\documentclass[useAMS,usenatbib]{mn2e}
\pdfoutput=1 

\usepackage{graphicx}
\usepackage{amssymb}
\usepackage{natbib}
\usepackage{longtable}

\def\deg {^{\circ} }

\title[The ATLAS 5.5 GHz survey of the Extended Chandra Deep Field South]{The ATLAS 5.5 GHz survey of the Extended Chandra Deep Field South: Catalogue, Source Counts and Spectral Indices}

\author[M. Huynh et al.] {M.T.~Huynh, $^1$\thanks{E-mail: minh.huynh@uwa.edu.au}
A.M.~Hopkins, $^2$ E.~Lenc, $^{3,6}$ M.Y.~Mao, $^{3,4}$ E.~Middelberg,$^5$ \newauthor R.P.~Norris,$^3$ K.E.~Randall, $^{3,6}$ \\
$^1$ International Center for Radio Astronomy Research, M468, University of Western Australia, Crawley, WA 6009, Australia \\
$^2$ Australian Astronomical Observatory, PO Box 296, Epping, NSW 1710, Australia \\
$^3$ CSIRO Astronomy and Space Science (CASS), PO Box 76, Epping, NSW 1710, Australia \\
$^4$ School of Mathematics and Physics, University of Tasmania, Private Bag 27, Hobart, Tasmania 7001, Australia \\
$^5$ Astronomisches Institut der Ruhr-Universit¬at Bochum, Universit¬atsstr. 150, 44801 Bochum, Germany \\
$^6$ Sydney Institute for Astronomy, School of Physics, The University of Sydney, NSW 2006, Australia }

\begin{document}

\maketitle

\label{firstpage}

\begin{abstract}

Star forming galaxies are thought to dominate the sub-mJy radio population, but
recent work has shown that low luminosity AGN can still make a significant contribution to the faint radio source population. Spectral indices are an important tool for understanding the emission mechanism of the faint radio sources.
We have observed the extended Chandra Deep Field South at 5.5 GHz using a mosaic of 42 pointings with the Australia Telescope Compact Array (ATCA). 
Our image reaches an almost uniform sensitivity of $\sim$12 $\mu$Jy rms over 0.25 deg$^2$ with a restoring beam of 4.9 $\times$ 2.0 arcsec, making it one of the deepest 6cm surveys to date. 
We present the 5.5 GHz catalogue and source counts from this field. 
We take advantage of the large amounts of ancillary data in this field to study the 1.4 to 5.5 GHz spectral indices of the sub-mJy population. 
For the full 5.5 GHz selected sample we find a flat median spectral index, $\alpha_{\rm med} = -0.40$, which is consistent with previous results. However, the spectral index appears to steepen at the faintest flux density levels ($S_{5.5 GHz} < 0.1$ mJy), where $\alpha_{\rm med} = -0.68$.
We performed stacking analysis of the faint 1.4 GHz selected sample (40 $<$ $S_{1.4 GHz}$ $<$  200 $\mu$Jy) and also find a steep average spectral index, $\alpha = -0.8$, consistent with synchrotron emission. We find a weak trend of steepening spectral index with redshift. Several young AGN candidates are identified using spectral indices, suggesting Gigahertz Peaked Spectrum (GPS)  sources are as common in the mJy population as they are at Jy levels. 

\end{abstract}


\section{Introduction}

One of the most fundamental issues in astrophysics is when and how stars, galaxies and black holes form, and how they evolved with cosmic time. The black hole mass and bulge relation \citep{kormendy1995, magorrian1998} suggests there is a link between active galactic nuclei (AGN) and star formation, hence studying the interaction between stars, galaxies and AGN phenomena is crucial for understanding galaxy formation in the early universe and how those galaxies evolve to the objects we see today. Radio emission can be produced by both AGN and star-forming processes, and thus radio wavelengths provide a unique window to study the cosmic evolution of these two important processes.

Bright radio sources ($> 100$ mJy) are associated with AGN activity (e.g. \citealp{condon1984, georgakakis1999}), but the Euclidean-normalized radio source counts flatten below about 1 mJy and this cannot be explained by a population of radio-loud AGNs. A non-evolving population of local ($z < 0.1$) low-luminosity radio galaxies \cite{wall1986}, strongly evolving normal spirals \citep{condon1984, condon1989}, and star forming galaxies \citep{windhorst1985, rowan-robinson1993} have all been suggested to explain this new population. The faint radio source counts have been successfully modelled by star forming galaxies \citep{seymour2004, huynh2005}. The most commonly accepted paradigm has been that the sub-mJy population is largely made up of starforming galaxies. However a growing number of studies are finding that low luminosity AGN, both radio-loud and radio-quiet, still make a significant contribution to the sub-mJy population \citep{jarvis2004, huynh2008, seymour2008, smolcic2008, padovani2009, padovani2011}. 

There is conflicting evidence on the nature and properties of faint radio sources, and in particular their spectral index properties and whether there is a flattening of the average spectral indices for faint radio sources (e.g. \citealp{randall2012}). \cite{prandoni2006} found sources with 1.4 GHz flux densities less than a few mJy had an average 1.4 GHz to 5 GHz spectral index flatter than that of brighter radio sources. \cite{owen2009} found a flattening of the average 325 MHz to 1.4 GHz  spectral index below $S_{1.4GHz} < 10$ mJy, but spectral indices appeared to steepen again at the faintest flux densities. In the Lockman Hole, the deepest radio field to date, no flattening of the 610 MHz to 1.4 GHz spectral indices was observed for  $S_{1.4GHz} > 0.1$ mJy \citep{ibar2009}.

Galaxies dominated by star formation processes are expected to have a spectral index of $\alpha = -0.8$ ($S \propto \nu^\alpha$, \citealp{condon1992}), consistent with synchrotron emission from electrons accelerated by supernovae. Although thermal Bremsstrahlung (free-free) emission found in HII regions can have a flatter spectral index, a flat or inverted spectrum is usually attributed to the superposition of different self-absorbed components of varying sizes at the base of the radio jet of a radio-loud AGN. A flattening in the average spectral index would imply there is a population of sources at sub-mJy flux densities with flat or inverted spectra, which are likely to be AGN.  Studying the spectral index properties of the faint radio population is important for understanding the $z-\alpha$ relation, which is used to identify the highest redshift radio sources (e.g. \citealp{debreuck2004, klamer2006}), identifying young radio AGN such as Gigahertz Peaked Spectrum sources and Compact Steep Spectrum sources \citep{odea1998, randall2011}, and determining whether star formation or AGN processes are responsible for the radio emission in these sources. 

Here we present 5.5 GHz observations by the Australia Telescope Large Area Survey (ATLAS, \citealp{norris2006}) team of the extended Chandra Deep Field South which are well-matched in area to the extensive multiwavelength data in the region. This survey is important as it is the largest 6cm survey deeper than 100 $\mu$Jy. We present the 5.5 GHz source catalogue and counts, and discuss the spectral index properties of the ATLAS radio sources. The paper is organized as follows: In Section 2 we describe the observations and data reduction. We discuss the source extraction and present the source catalogue in Section 3. Source counts are derived from the catalogue and presented in Section 4. We perform a detailed spectral index ($\alpha^{5.5 GHz}_{1.4GHz}$) analysis, and identify ultra-steep and Gigahertz Peaked Spectrum sources in Section 5.  Concluding remarks are given in Section 6. We assume a Hubble constant of 71 km s$^{-1}$ Mpc$^{-1}$, $\Omega_{\rm M}$ = 0.27 and $\Omega_{\Lambda}$ = 0.73 throughout this paper. 

\section{The Observations}

\subsection{Observing Strategy}

\begin{figure*} 
   \centering
   \includegraphics[width=0.95\columnwidth]{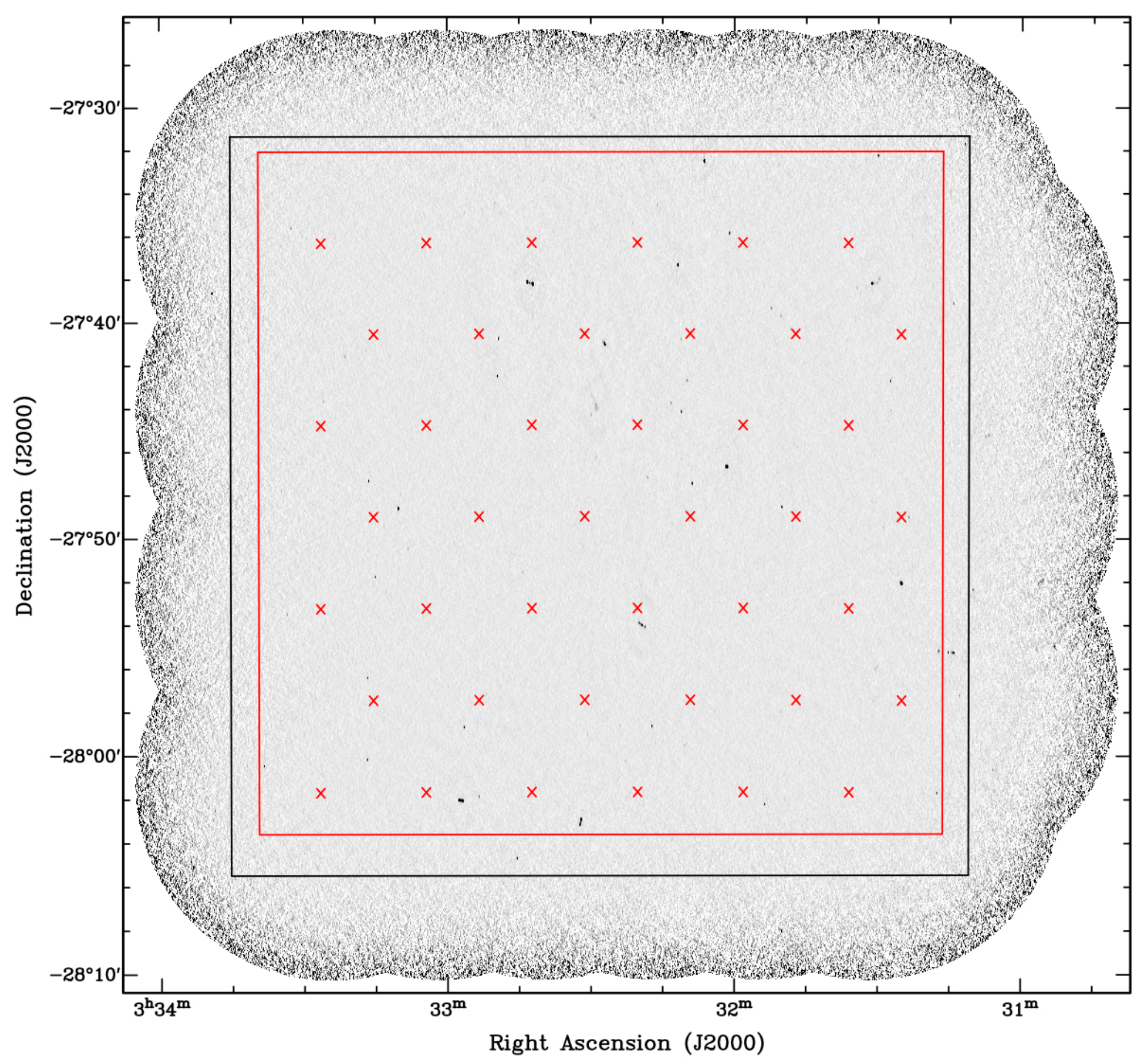}
   \includegraphics[width=0.85\columnwidth]{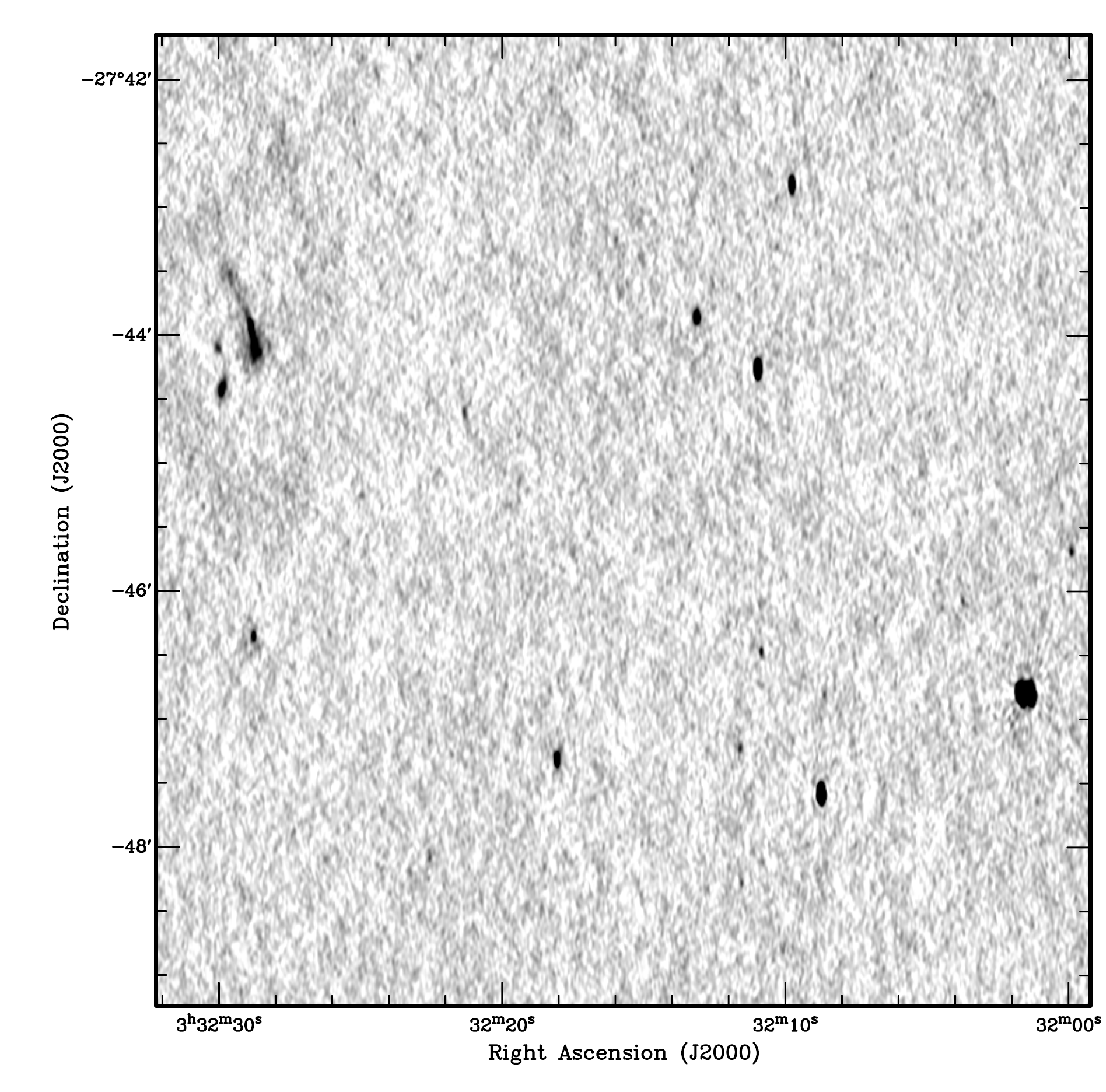}
   \caption{LEFT: Greyscale of the full eCDFS 5.5 GHz mosaic. The 42 mosaic pointing centers are marked by crosses. The inner red rectangle marks the area covered by the MUSYC optical imaging \citep{taylor2009} and the outer black rectangle marks the area covered by Very Large Array (VLA) 1.4 GHz imaging \citep{miller2008}. RIGHT: A cutout of a central portion of the mosaic to show more detail.}
   \label{fig:eCDFS55}
   \end{figure*}

We observed the extended Chandra Deep Field South at 6cm in two runs using the new Compact Array Broadband Backend (CABB; \citealp{wilson2011}) with the full CABB 2048 MHz bandwidth centred at 5.5 GHz. At 6cm the primary beam of the ATCA observations is $\sim$10 arcmin (FWHM). A total of 42 pointings in a hexagonal layout (Figure \ref{fig:eCDFS55}) was used to uniformly image the full 30 $\times$ 30 arcmin eCDFS area. Each pointing is separated by about 5 arcmin, which is approximately 0.5 FWHM of the primary beam at 6cm.  The resulting mosaic is centred on the original GOODS field with approximate RA and Dec (J2000) of 3h32m20s and $-27\deg48\arcmin34\arcsec$.

The first run in August 2009 consisted of an initial set of 12 and 8 hour observations. A further 144 hours was allocated in January 2010. The ATCA was in the 6D and 6A configurations in August 2009 and January 2010, respectively, providing a maximum baseline of 6km. The mosaic pointings were cycled through rapidly with each pointing observed for only 1 minute in each pass. 
Taking into account overheads such as calibrator scans, telescope drive-time and down time from inclement weather, the total effective integration time of the mosaic is about 115 hours, or 2.7 hours per pointing.
The secondary calibrator, 0347-279, was observed in the middle of the mosaic cycle for gain and phase calibration. During the August 2009 observations, secondary calibrator observations were separated by about 20 minutes. The January 2010 observing run had uncharacteristically poor phase stability and so the phase calibrator was observed about every 10 minutes instead. Primary flux density calibration was performed through observations of the source PKS B1934-638, which is the standard primary calibrator for ATCA observations ($S_{5.5GHz}$ = 5.1 Jy\footnote{See ATCA Calibrator list: http://www.narrabri.atnf.csiro.au/\\calibrators/}). 

\subsection{Data Reduction}

We used the Multichannel Image Reconstruction, Image Analysis and Display (MIRIAD) software package \citep{sault1995} to reduce the CABB data. The latest version of MIRIAD was used to ensure tasks were updated to deal with data from this new back end. The MIRIAD task {\em atlod} was run with the ``birdie" option to remove self-inteference from the 640 MHz clock harmonics and 100 channels at both band edges, where the bandpass response is poor.  

The 42 pointings were individually reduced and imaged. Automated flagging was performed with the task {\em mirflag}\footnote{http://www.atnf.csiro.au/people/Emil.Lenc/tools/Tools/\\Mirflag\_Plotvis.html} using an rms cutoff mode. In this mode RFI was identified if the rms in a channel exceeds the rms in a good channel by a set factor, or cutoff. 
This cutoff was 5$\sigma$ for the bright calibrators and 3$\sigma$ for the field observations. To make sure the flagging was successful the data were inspected visually using the 3D visualisation tool {\em plotvis\footnotemark[2]} and any remaining RFI was manually excised. We explored a range of robust weights in the imaging stage (Briggs 1995) as well as uniform and natural weighting. We found a robust weight of 1 resulted in a good compromise between sensitivity and resolution. An image 2300 $\times$ 2300 pixels in size with 0.5 arcsec pixels was generated for each pointing to ensure recovery of the full field of the primary beam. Cleaning was performed using multi-frequency cleaning task {\em mfclean} with the clean region set to 8 arcmin from the center, which extends beyond the 10\% response level of the primary beam and hence encompasses almost all the area of interest. One iteration of self-calibration was performed for each pointing and the images were restored with a beam of 4.9 $\times$ 2.0 arcsec, the average synthesized beam of the 42 pointings. The MIRIAD task {\em linmos} was then used to stitch together the individual pointings to create a single mosaic map, shown in Figure \ref{fig:eCDFS55}. 

The CABB bandwidth is so large that we also split the 2000 channels to make four sub-band images of approximately 500 channels each. The central frequencies of these sub-band images are 4.80, 5.25, 5.72 and 6.19 GHz. We use these images to examine the SED of our sources and identify Gigahertz Peaked Spectrum sources in Section 5.3. A high resolution image with a beam of 2.9 $\times$ 1.0 arcsec was also generated with super-uniform weighting, and this is used to examine radio morphology \citep{Dehghan2011}.

\subsection{Image Analysis}

\begin{figure*}[h]
\centering
\includegraphics[width=0.95\columnwidth]{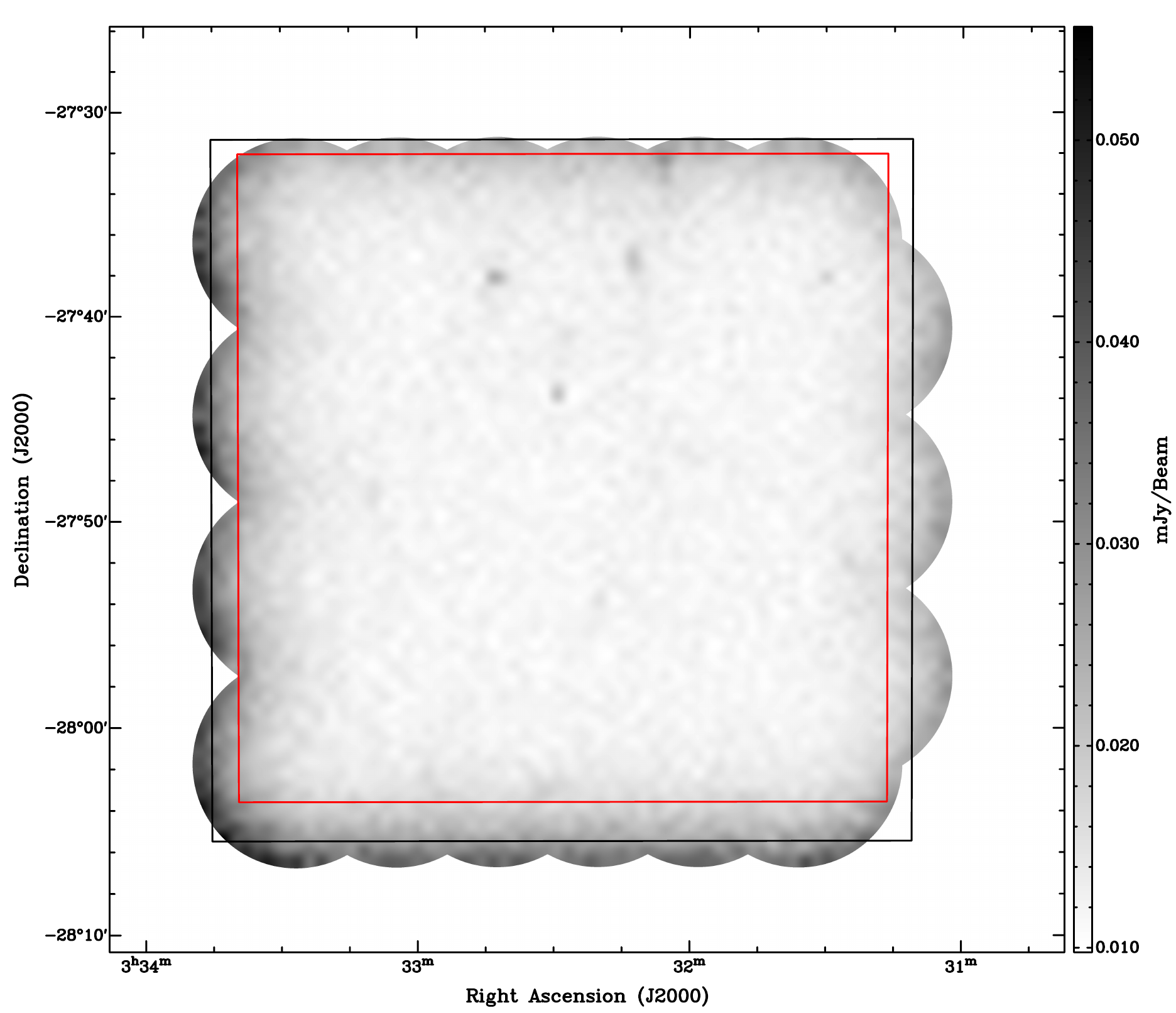}
\hspace{5mm}
\includegraphics[width=0.95\columnwidth]{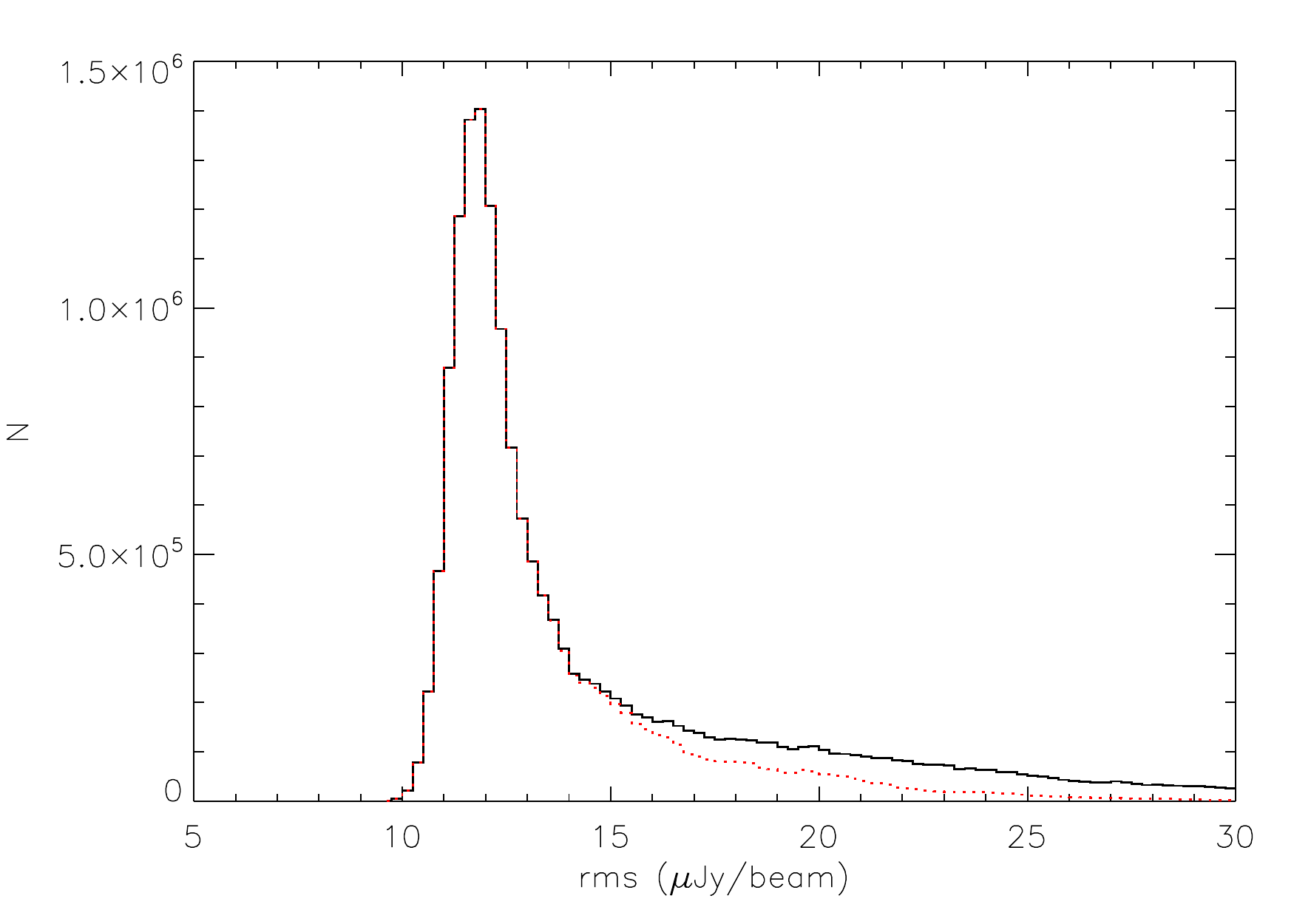}
\caption{LEFT: Greyscale of the noise image from SExtractor, generated using a mesh-size of $10  \times 10$ synthesized beams. The inner red rectangle marks the area covered by the MUSYC optical imaging \protect\citep{taylor2009} and the outer black rectangle marks the area covered by VLA 1.4 GHz imaging \protect\citep{miller2008}. RIGHT: The pixel distribution of the SExtractor noise image. Solid black line indicates the distribution for full noise image on the left, and the red dotted line indicates the distribution in the MUSYC imaging area. The peak of the distribution is at $\sim$ 12 $\mu$Jy. The high noise tail is mostly at the edges of the mosaic, where the primary beam response is lower. } 
\label{fig:noiseim}
\end{figure*}

The primary beam correction at the high and low end of the 2 GHz band differs by about 50\% at the half-power point of the primary beam for 5.5 GHz.
This may affect the recovered flux density for sources at the edge of a pointing since {\em linmos} uses a single primary beam correction. 
To check this effect we ran simulations for sources with spectral indices varying from $\alpha$ = 1 to $\alpha$ = $-2$ ($S \propto \nu^\alpha$). We find the flux density of a flat spectrum source is recovered to within 1\% in the 2 GHz band at the half-power of the primary beam. The flux density is recovered to $\sim$ 3\% for a canonical synchrotron spectrum of $\alpha$ = $-0.8$.  It is only for the steepest spectrum sources ($\alpha$ = $-2$) that the flux density is over-estimated, by $\sim$ 7\%, at the half-power point of the primary beam, and we don't expect these sources to be common. We therefore conclude that as long as the source lies within the half-power point (i.e. within 0.5 FWHM of a pointing center) this effect is minimal.

We examined the noise characteristics of the mosaic using the Sextractor software package \citep{bertin1996}. While SExtractor was developed for the analysis of optical images, it is known to work well on radio images also \citep{bondi2003, huynh2005,  prandoni2006}. The choice of mesh-size is important. A mesh-size which is too small would result in the estimation being affected by individual sources, while larger mesh sizes can miss systematic small-scale variations in the noise. Previous surveys have found that mesh-sizes with widths of 8 to 12 times the synthesised beam, produce good results for noise estimation in deep radio continuum surveys \citep{huynh2005, schinnerer2007, schinnerer2010}. A more detailed analysis of simulated radio continuum images confirms that a mesh-size of 10 to 20 synthesized beams produces optimum background noise estimates \citep{huynh2012}.  To estimate the local background noise variations we therefore used a mesh-size of $\sim$ $10 \times 10$ synthesized beams ($67 \times 67$ pixels). The SExtractor generated noise image is shown in Figure \ref{fig:noiseim} (left), limited to areas within 5 arcmin, $\sim$0.5 FWHM of the primary beam. We find that the mosaic has an rms noise of $\sim$11.9 $\mu$Jy/beam, but with a large tail at higher noise levels (Figure \ref{fig:noiseim}, right). This tail is the result of (i) higher noise regions around extended sources that aren't cleaned perfectly and (ii) the greater noise at the edge of the mosaic, which are furthest away from any one pointing and therefore have the greatest primary beam correction. 

The effective integration time of the mosaic is approximately 115 hours. According to the online ATCA sensitivity calculator, ATCA should reach 6 $\mu$Jy/beam rms in 12 hours at 5.5 GHz with the new CABB upgrade. This
corresponds to a sensitivity of 6 $\times$ $\sqrt{42}/1.4$ = 28 $\mu$Jy/beam rms for a nyquist-sampled 42 pointing mosaic in 12 hours. We therefore should have achieved $\frac{30}{\sqrt{115/12}} \sim 9.7$ $\mu$Jy/beam. The sensitivity we achieve is within $\sim$ 20\% of this value estimated from the ATCA sensitivity calculator. 

\section{Source Extraction}

There are many radio source identification and measurement tools in common use, including MIRIAD/AIPS Gaussian
fitting routines {\em imsad}, {\em sad} and {\em vsad}, {\em sfind} \citep{hopkins2002}, Duchamp \citep{whiting2012} and SExtractor \citep{bertin1996}.
Most of these use a simple S/N thresholding technique whereby a source with a peak flux density, or pixel value, above a set threshold (usually some multiple of the local noise) is  deemed to be a true source.  The false-discovery rate (FDR) approach is a robust statistical procedure which compares the distribution of image pixels to that of an image of equal size containing only noise, and the user threshold then places a limit on the fraction of identified sources which may be false, based simply on the statistics of the noise distribution \citep{miller2001}. The FDR method has been implemented for source measurement in radio images in the MIRIAD task {\em sfind} \citep{hopkins2002}. The FDR method was found to be more reliable than simple S/N threshold methods for parameters which return similar completeness \citep{huynh2012}, so we therefore use {\em sfind} for the source detection. 

We restricted the search area to within 5 arcmin ($\sim$ 0.5 FWHM of the primary beam) of the outer mosaic pointing to minimise primary beam effects, and ran {\em sfind} with `rmsbox' set to 10 synthesized beams and `alpha' set to 1, which are parameters found by \cite{huynh2012} to return reasonable noise estimates, completeness and reliability.  
Setting the `alpha' parameter to 1 would return a list of sources which is 99\% reliable for the case of a perfect image with pure Gaussian noise.  
{\em sfind} returned an initial list of 143 sources for inspection. Each source was then individually characterised as both a point source and Gaussian using the MIRIAD task {\em imfit}.

\begin{figure*}
\includegraphics[width=3.cm]{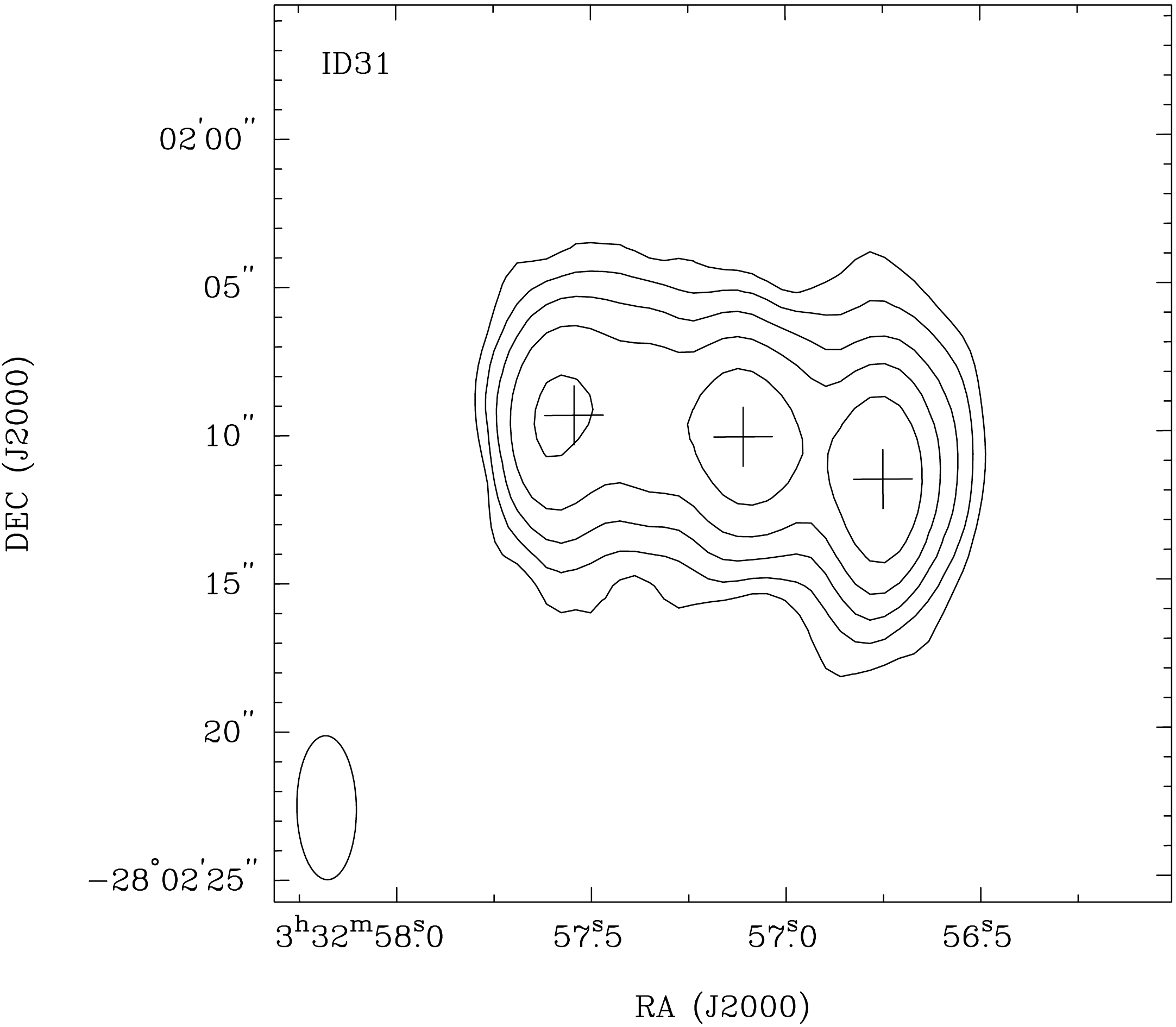}
\includegraphics[width=3.cm]{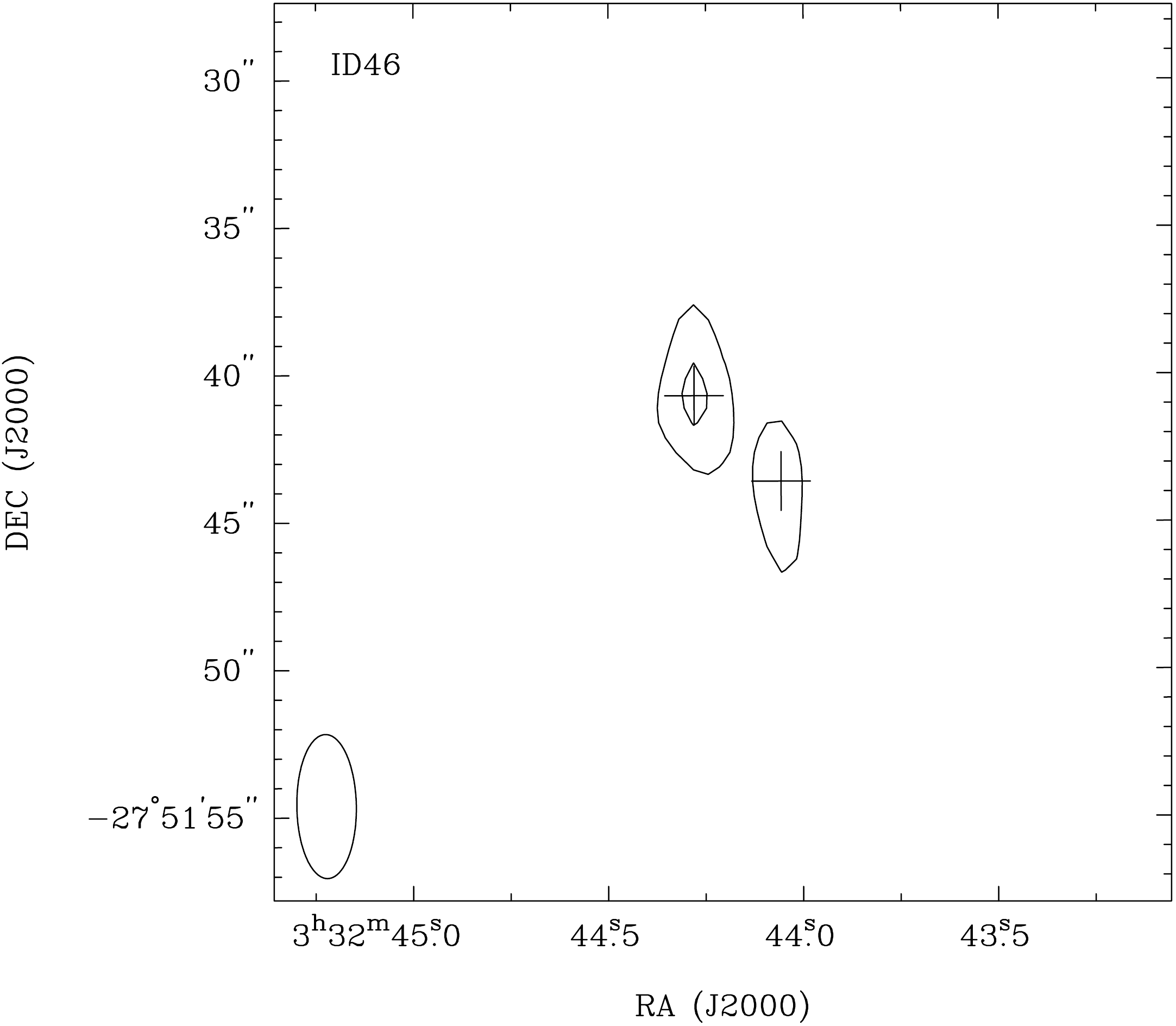}
\includegraphics[width=3.cm]{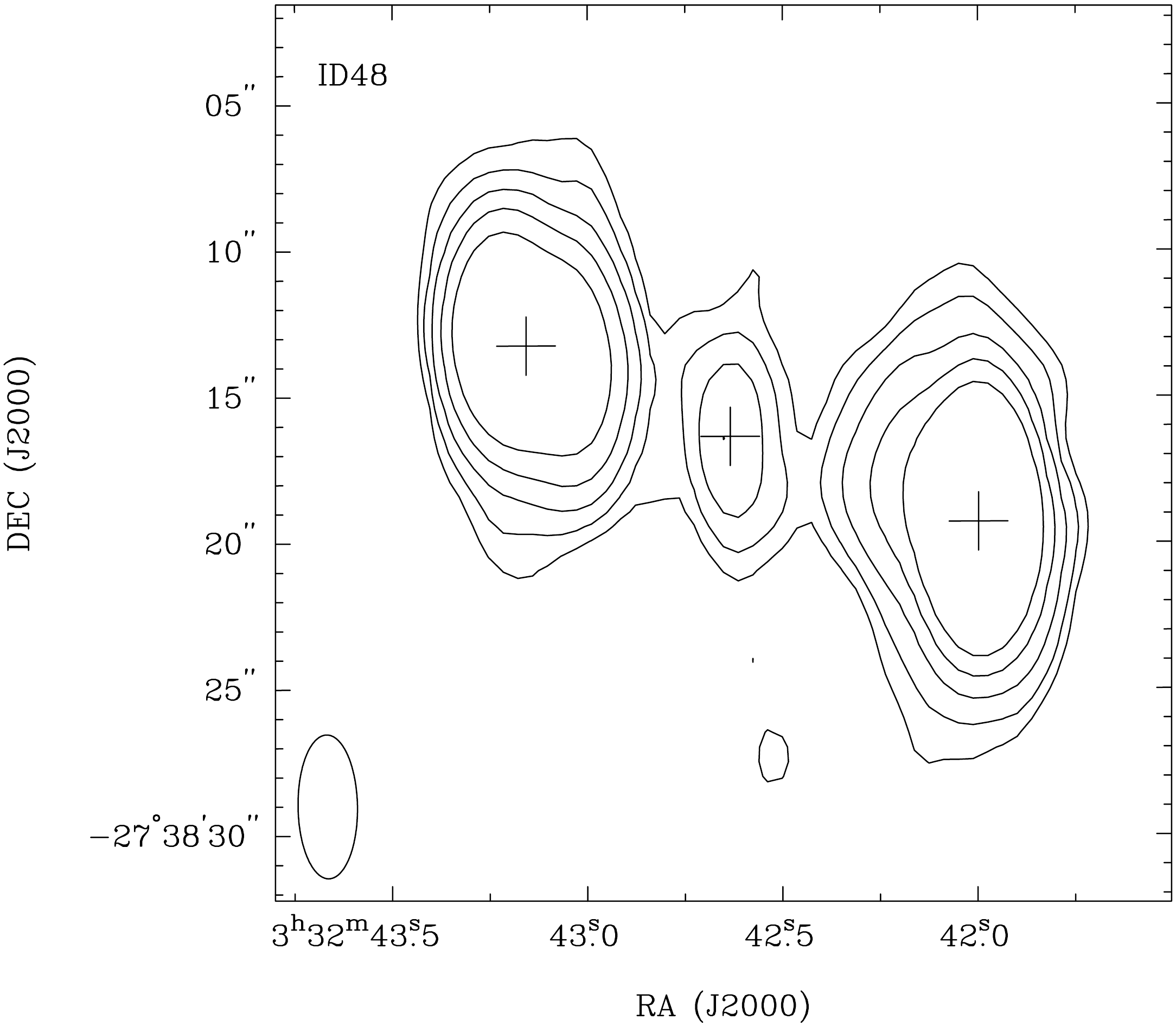}
\includegraphics[width=3.cm]{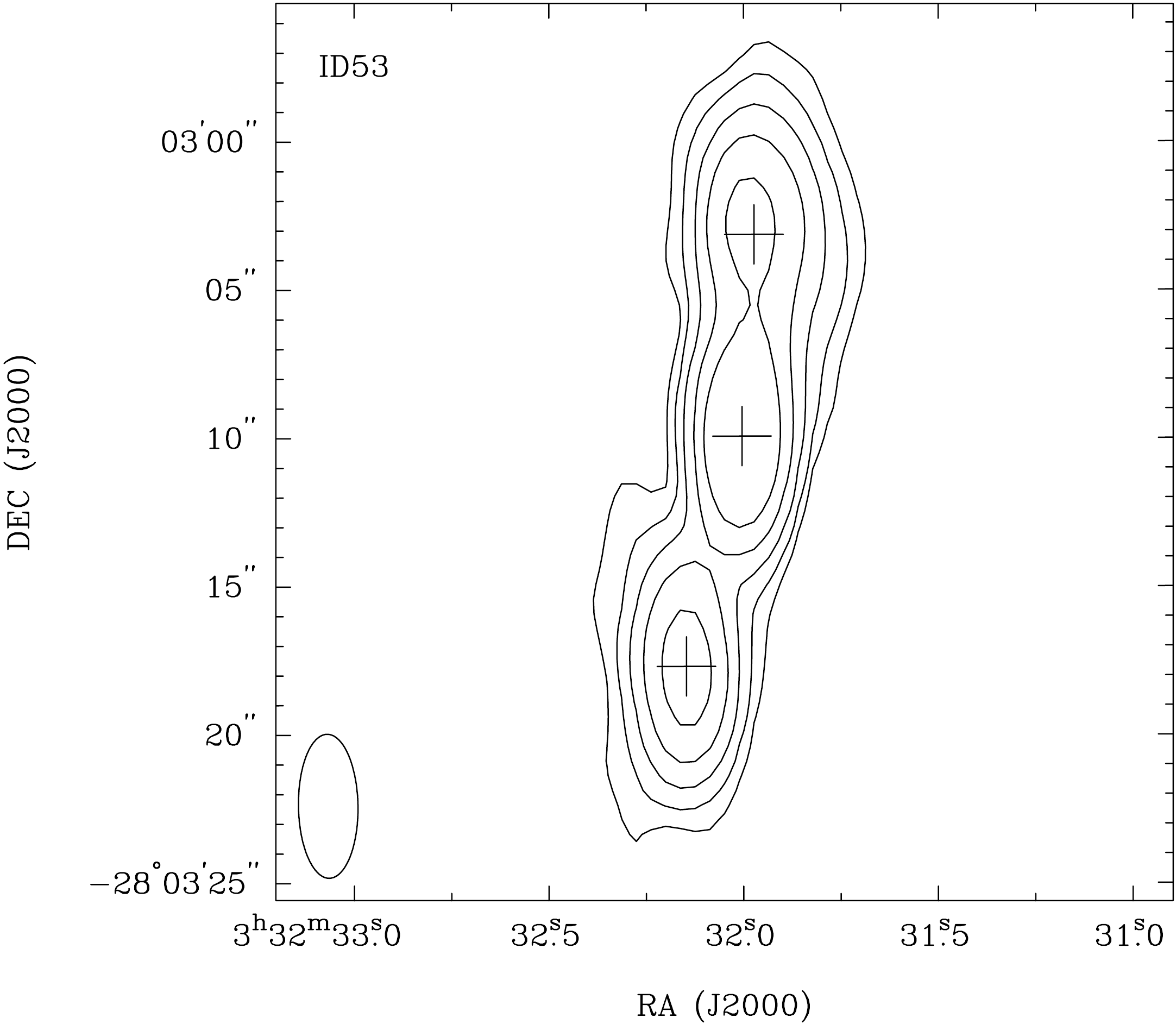}
\includegraphics[width=3.cm]{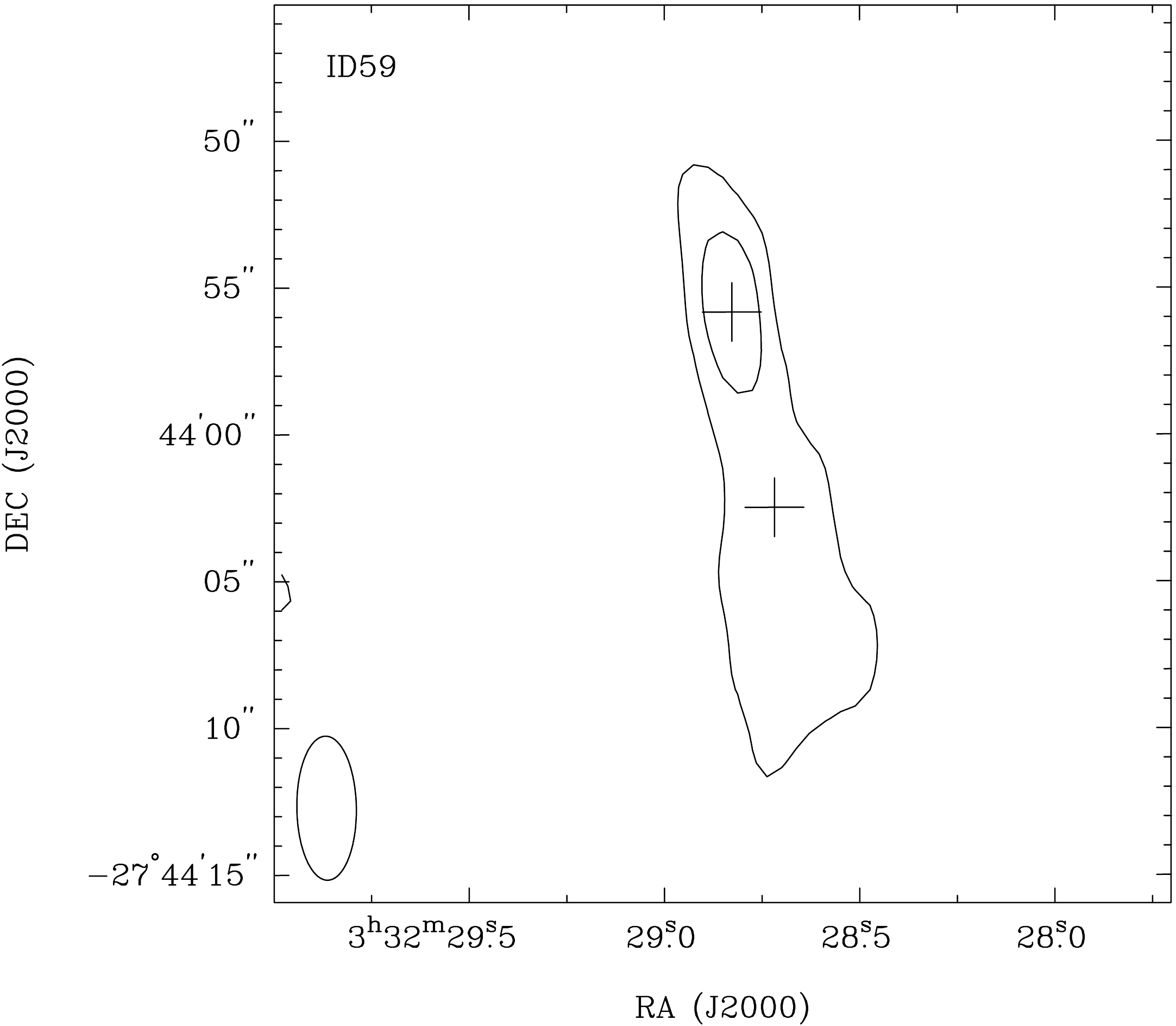}
\includegraphics[width=3.cm]{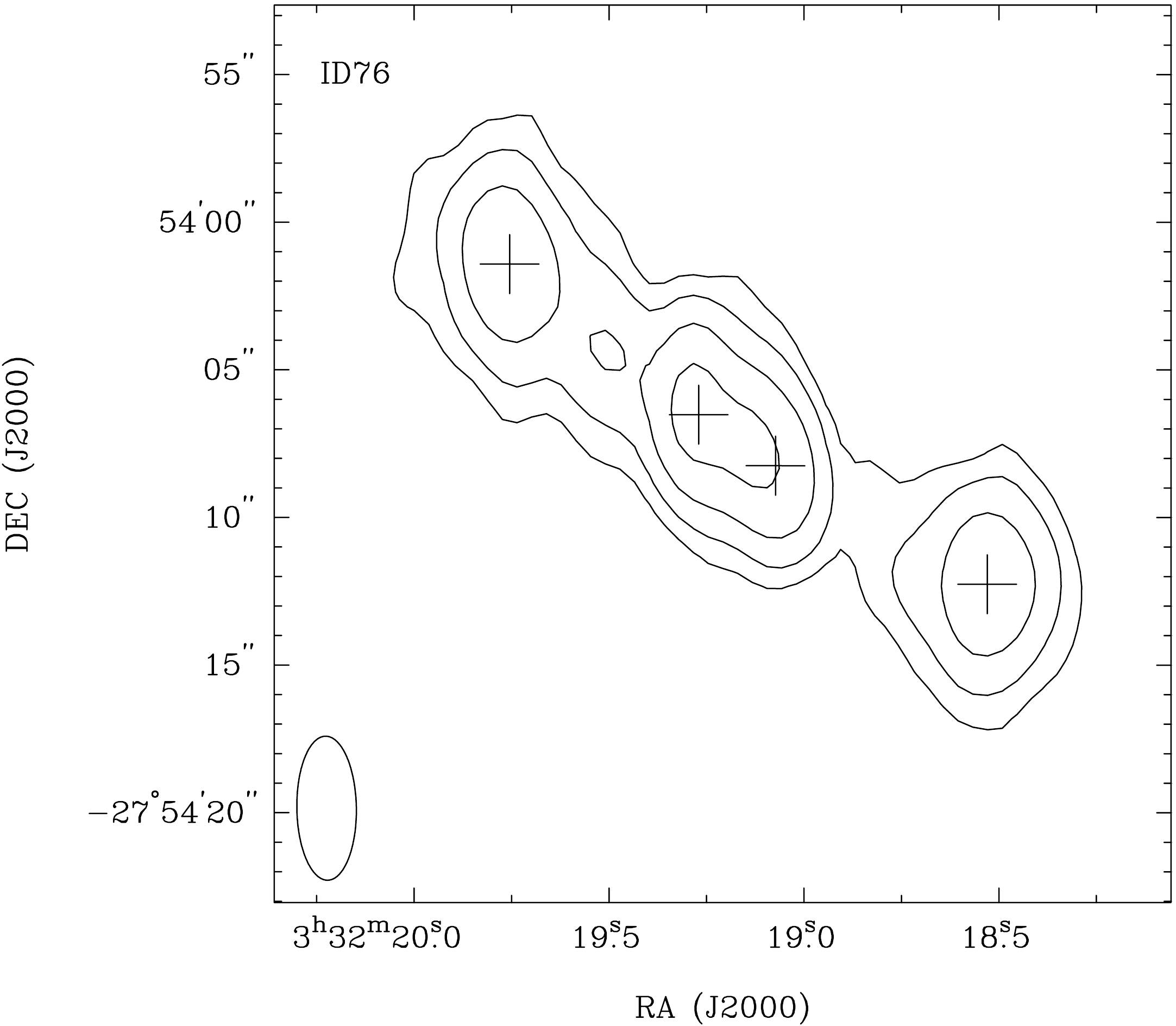}
\includegraphics[width=3.cm]{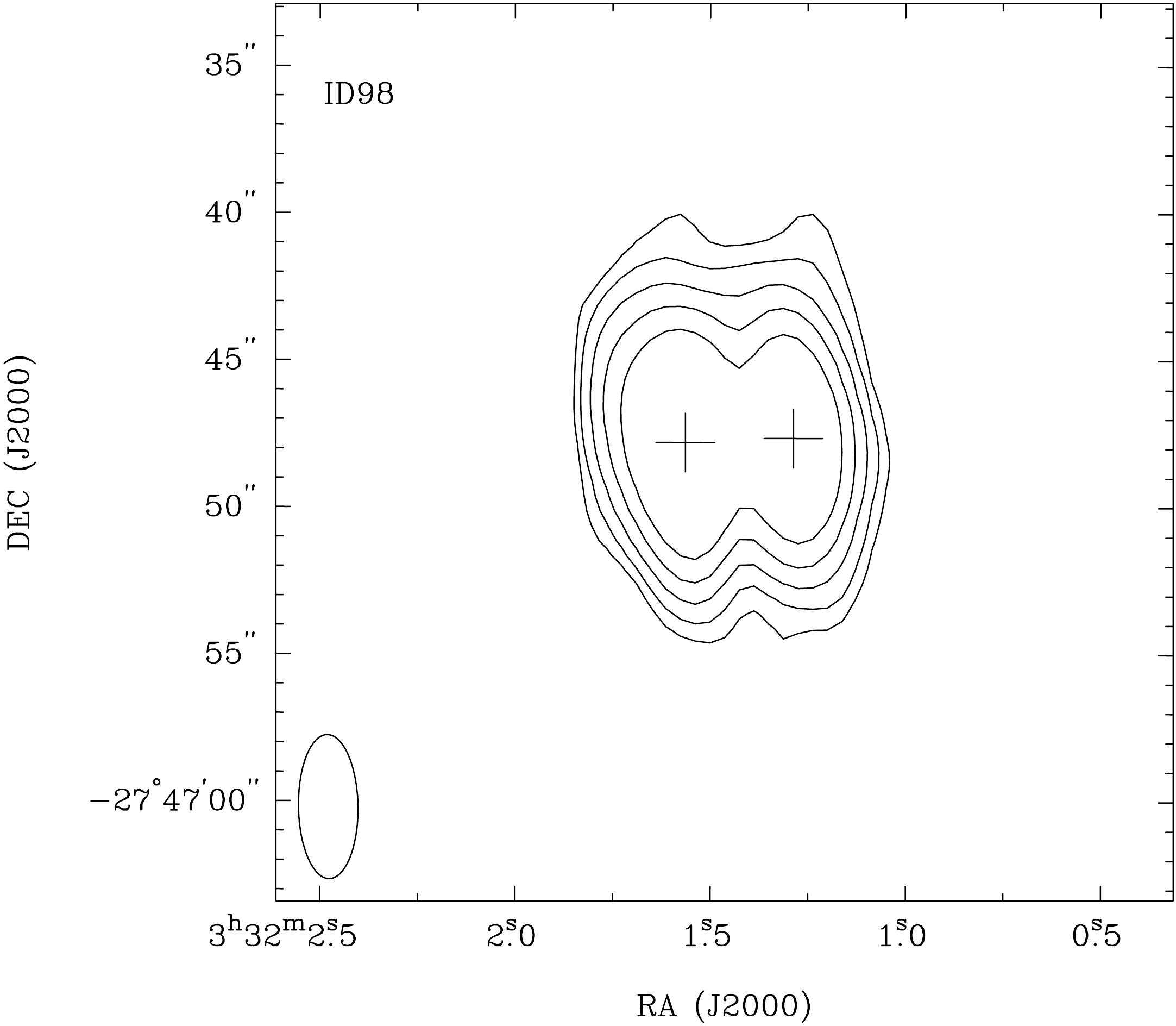}
\includegraphics[width=3.cm]{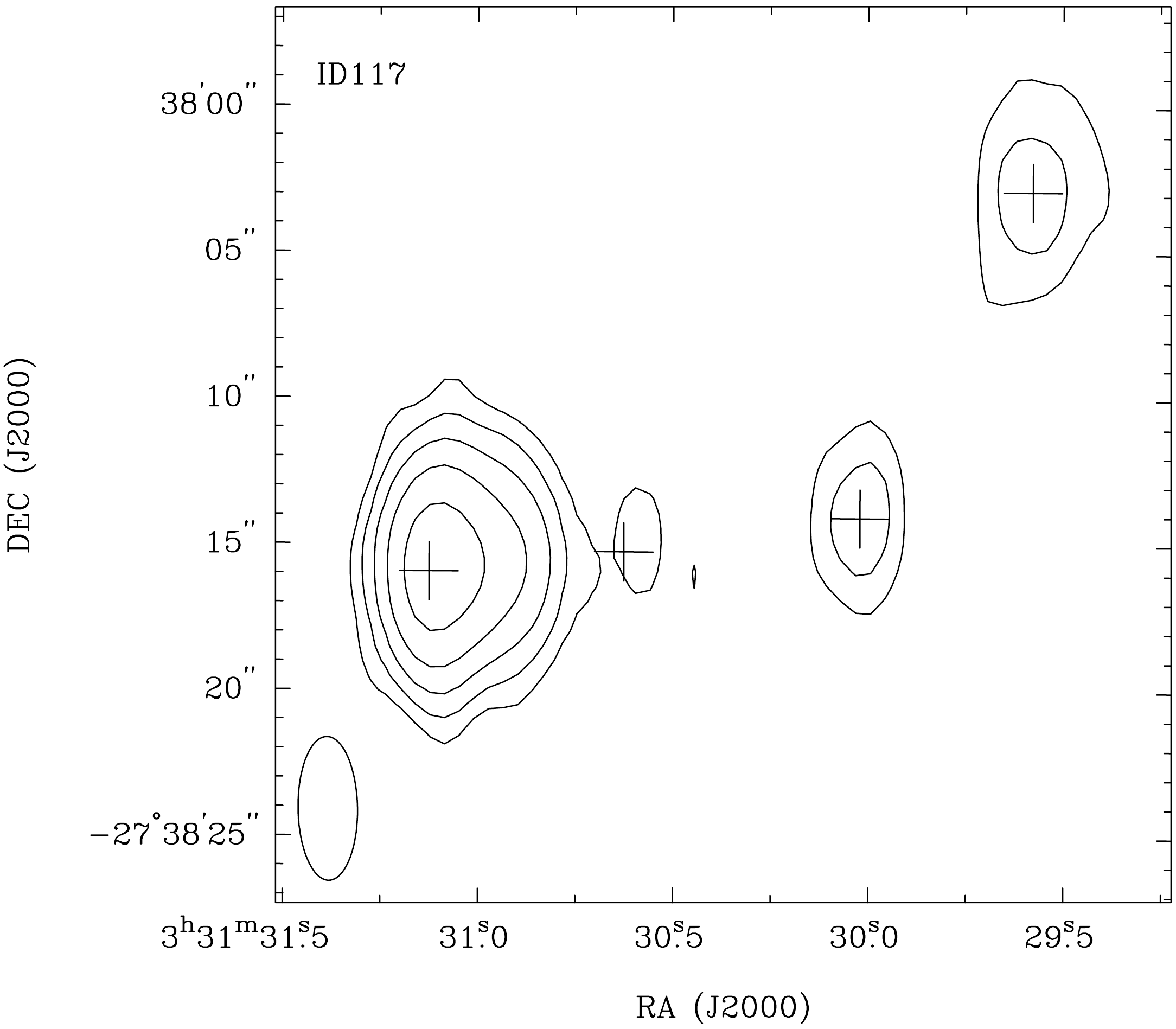}
\includegraphics[width=3.cm]{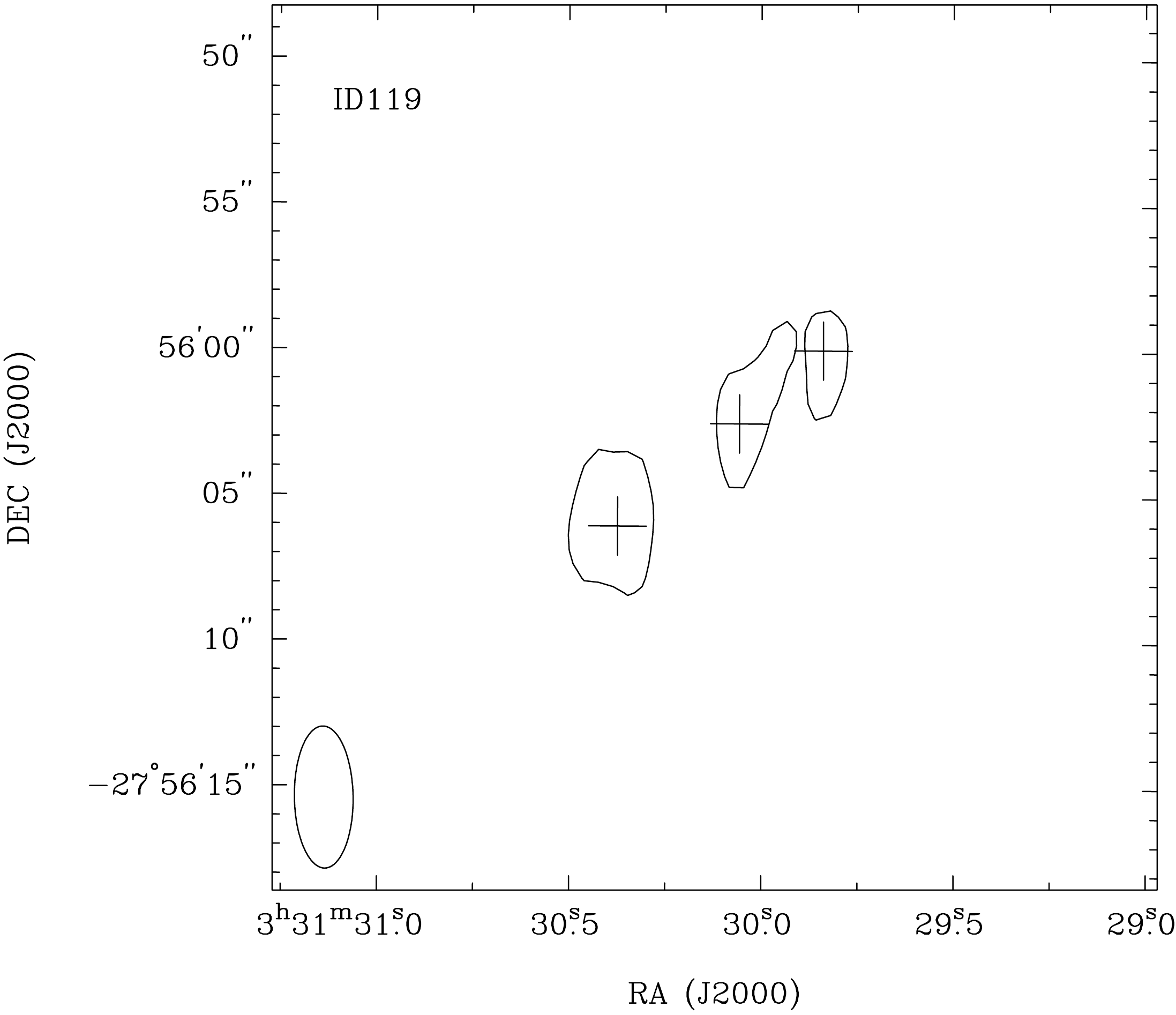}
 \includegraphics[width=3.cm]{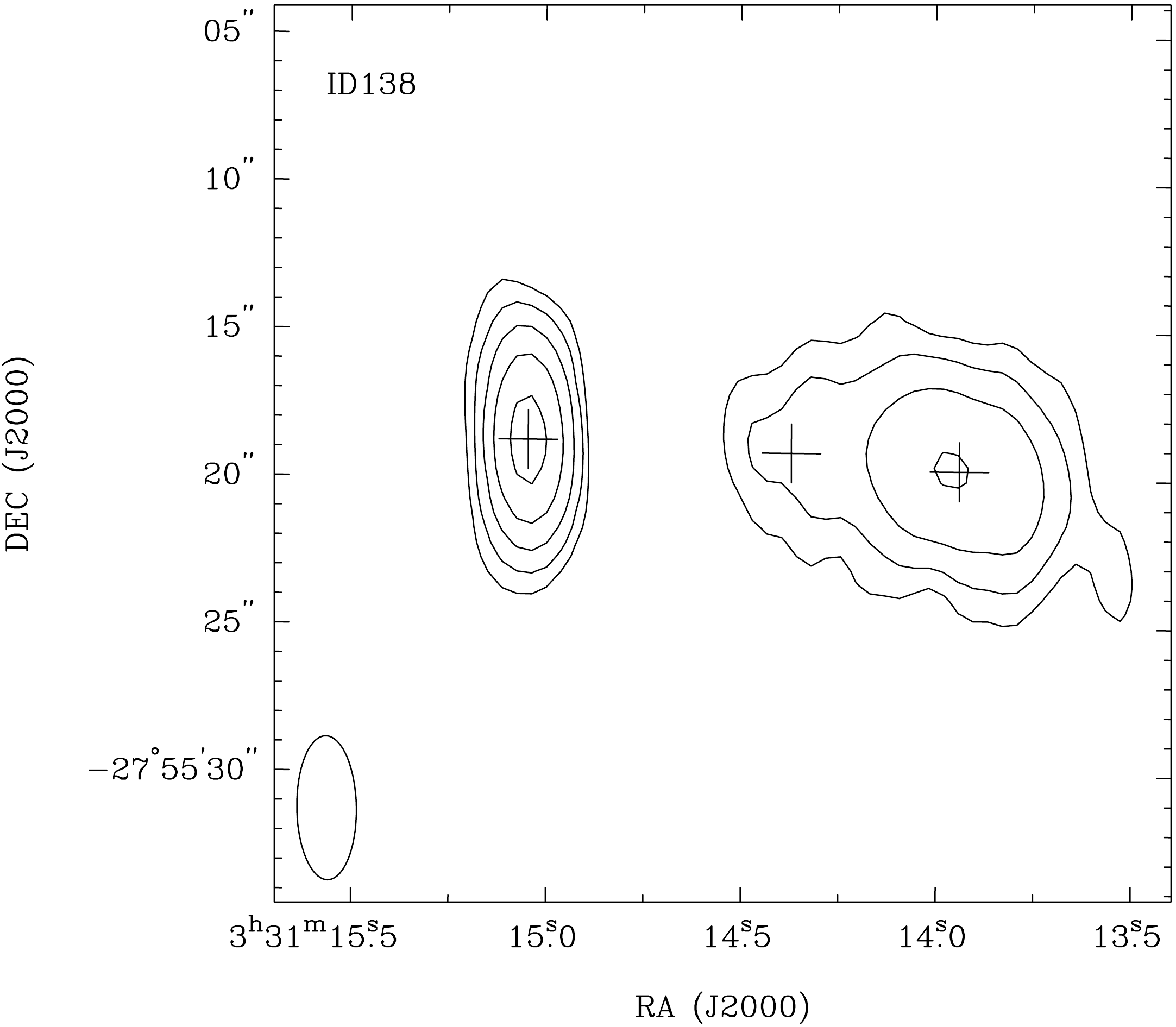}
\caption{$30\arcsec \times 30\arcsec$ contour images of the multiple sources in the catalogue. The contour levels are set at 5, 10, 20, 40 and 80 times the local noise level. The synthesized beam is shown in the bottom left corner. Crosses mark the positions of the catalogued components.}
 \label{fig:muilticomp}
\end{figure*}

Ten radio sources show classical core-lobe AGN morphology, and hence are clearly components of a single source (see Figure \ref{fig:muilticomp}). 
These sources were fitted as multiple Gaussians with  {\em imfit} and the multiple components are listed individually in the final catalogue. In total there are 123 sources and 142 source components in the final source list.

\subsection{Deconvolution}
\label{sec:deconv}

The ratio of the integrated flux to the peak flux is a direct measure of the extension of a radio source:
\begin{equation}
 S_{\rm tot}/S_{\rm peak} = \theta_{\rm min} \theta_{\rm max} / b_{\rm min} b_{\rm max}  
 \end{equation}
where $\theta_{\rm min}$ and $\theta_{\rm min}$  are the source FWHM axes and $b_{\rm min}$ and $b_{\rm max}$ are the synthesized beam FWHM axes. This ratio can be used to determine whether a source is resolved.  

\begin{figure} 
   \centering
   \includegraphics[width=0.95\columnwidth]{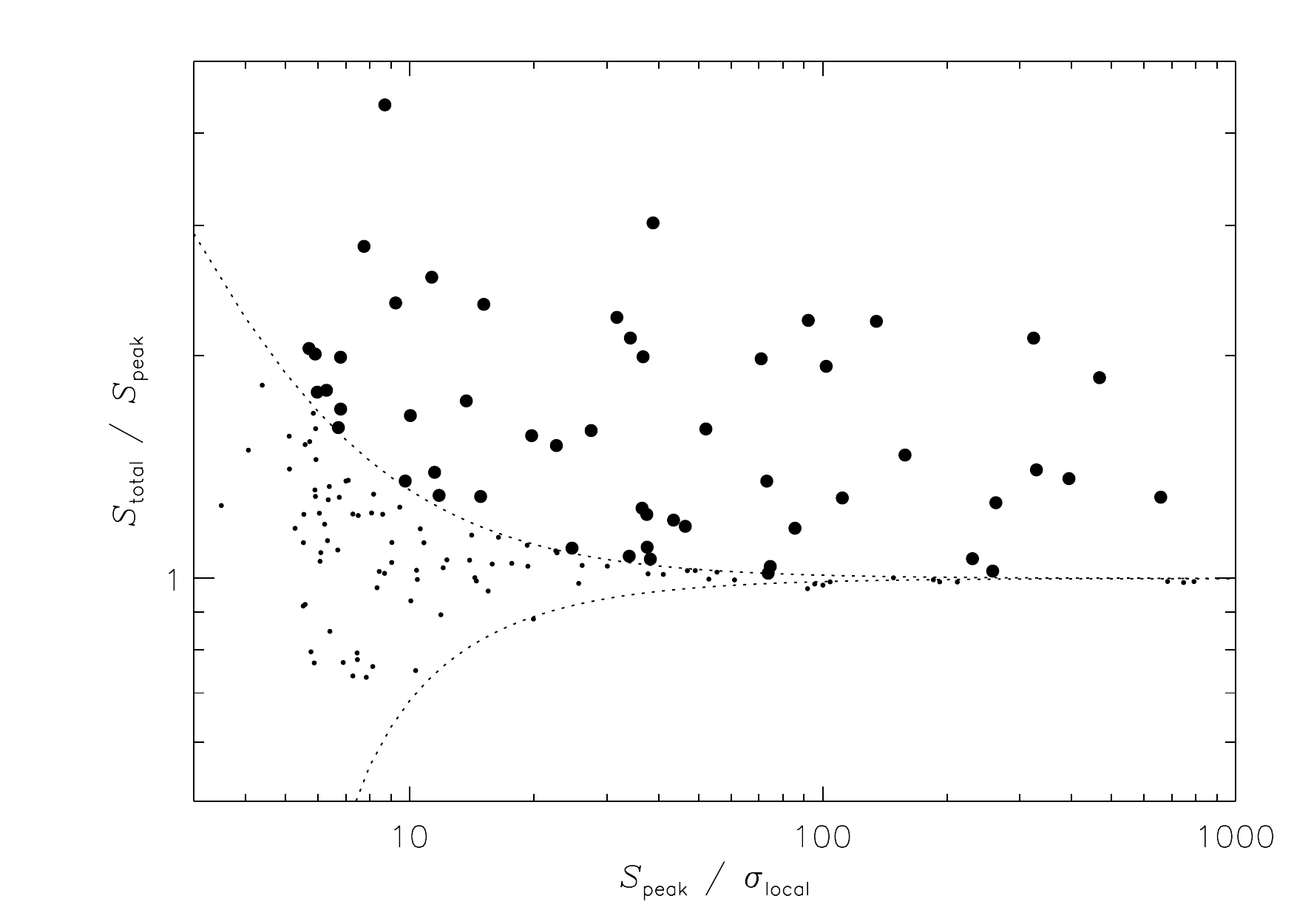}
   \caption{The ratio of integrated ($S_{\rm tot}$) flux density to peak flux density ($S_{\rm peak}$) as a function of source signal to noise ($S_{\rm peak}/\sigma$). The dotted line shows the upper and lower envelopes of the flux ratio distribution that contains 90\% of the unresolved sources. The large dots indicate sources which are deconvolved successfully and considered resolved.}
   \label{fig:deconv}
   \end{figure}

An analysis similar to \cite{prandoni2006} and \cite{huynh2005} was performed to determine which sources are resolved. Whether a source is successfully deconvolved depends on the S/N ratio of the source and not just the image beam-size. Using the Gaussian fits, we examined the ratio of integrated flux density to peak flux density as a function of source signal to noise (Figure \ref{fig:deconv}). Assuming the sources with $S_{\rm tot}/S_{\rm peak} < 1$ are due to noise then the upper envelope is defined as 
\begin{equation}
 S_{\rm tot}/S_{\rm peak} = 1 + 10 / (1 + [S_{\rm peak}/\sigma])^{1.5} .
 \end{equation}
This envelope is found by determining a lower curve that contains 90\% of the $S_{\rm tot}/S_{\rm peak} < 1$ sources (see Figure \ref{fig:deconv}) and then mirroring it on the 
$S_{\rm tot}/S_{\rm peak} > 1$ side. Only 47/143 (33\%) sources lie above the upper envelope and are considered to be successfully deconvolved (i.e. resolved).

\subsection{The Source Catalogue}

The source catalogue is reported in Table \ref{tab:cat}.  Point-source measurements are given for sources which are not successfully deconvolved. The fitted peak, integrated source flux density and deconvolved source sizes from the Gaussian fits are given for the resolved, or successfully deconvolved, sources. Absolute calibration errors dominate for high S/N sources, but internal fitting errors shown in Table \ref{tab:cat} dominate for the majority of sources, which are low S/N. 

Column (1) - ID. A letter, such as `a', `b', etc., indicates a component of a multiple source. 

Column (2) - Source IAU name

Columns (3) and (4) - Source position: Right Ascension and Declination (J2000)

Column (5) - Peak source flux density ($\mu$Jy)

Column (6) - Uncertainty in peak source flux density ($\mu$Jy)

Column (7) - Integrated flux density ($\mu$Jy). Zero indicates source is not successfully deconvolved and hence no integrated flux density is given. 

Column (8) - Resolved flag. Zero indicates source is not successfully deconvolved. 

Column (9) - Deconvolved major axis (arcsec). Zero indicates source is not successfully deconvolved. 

Column (10) - Deconvolved minor axis (arcsec). Zero indicates source is not successfully deconvolved. 

Column (11) - Deconvolved position angle (degrees), measured from north through east. Zero indicates source is not successfully deconvolved. 

Column(12) - Local noise level, rms, in $\mu$Jy. 



\begin{table*}
\caption{The ATLAS 5.5 GHz catalogue}  
{\scriptsize
\begin{tabular}{llllrrrrrrrr}  
\hline
ID & IAU name & RA & Dec  & $S_{peak}$ & d$S_{peak}$ & $S_{int}$ & Decon & Decon& Decon & Decon & $\sigma_{local}$ \\  
     &                   &   (J2000) &  (J2000)   & ($\mu$Jy)  & ($\mu$Jy)  & ($\mu$Jy) & Flag & Bmajor & Bminor & PA & \\ \hline
      3 &    ATCDFS5 J033348.75-280232.8 & 3:33:48.75 & -28:02:32.8 &      354.8 &      69.6 &        0.0 & 0 &   0.00 &   0.00 &    0.0 &       55.6 \\
      4 &    ATCDFS5 J033346.59-273405.4 & 3:33:46.59 & -27:34:05.4 &      181.5 &      37.0 &        0.0 & 0 &   0.00 &   0.00 &    0.0 &       46.9 \\
      5 &    ATCDFS5 J033341.32-273809.5 & 3:33:41.32 & -27:38:09.5 &      216.9 &      22.2 &        0.0 & 0 &   0.00 &   0.00 &    0.0 &       36.8 \\
      6 &    ATCDFS5 J033338.35-280031.0 & 3:33:38.35 & -28:00:31.0 &      536.0 &      20.0 &        0.0 & 0 &   0.00 &   0.00 &    0.0 &       27.2 \\
      7 &    ATCDFS5 J033334.59-274751.2 & 3:33:34.59 & -27:47:51.2 &      148.1 &      16.5 &        0.0 & 0 &   0.00 &   0.00 &    0.0 &       23.6 \\
      8 &    ATCDFS5 J033333.43-275332.8 & 3:33:33.43 & -27:53:32.8 &      467.7 &      14.1 &        0.0 & 0 &   0.00 &   0.00 &    0.0 &       18.4 \\
      9 &    ATCDFS5 J033332.57-273539.2 & 3:33:32.57 & -27:35:39.2 &      460.6 &      21.8 &        0.0 & 0 &   0.00 &   0.00 &    0.0 &       19.5 \\
     10 &    ATCDFS5 J033327.53-275726.6 & 3:33:27.53 & -27:57:26.6 &      109.4 &      15.1 &        0.0 & 0 &   0.00 &   0.00 &    0.0 &       16.5 \\
     11 &    ATCDFS5 J033325.85-274343.2 & 3:33:25.85 & -27:43:43.2 &      186.2 &      19.9 &        0.0 & 0 &   0.00 &   0.00 &    0.0 &       15.0 \\
     12 &    ATCDFS5 J033322.75-275500.1 & 3:33:22.75 & -27:55:00.1 &      100.9 &      17.1 &        0.0 & 0 &   0.00 &   0.00 &    0.0 &       15.1 \\
     13 &    ATCDFS5 J033321.31-274138.6 & 3:33:21.31 & -27:41:38.6 &      267.8 &      18.4 &        0.0 & 0 &   0.00 &   0.00 &    0.0 &       13.4 \\
     14 &    ATCDFS5 J033318.71-274940.0 & 3:33:18.71 & -27:49:40.0 &       72.1 &      12.2 &        0.0 & 0 &   0.00 &   0.00 &    0.0 &       13.7 \\
     15 &    ATCDFS5 J033318.30-273440.4 & 3:33:18.30 & -27:34:40.4 &       96.3 &      12.9 &        0.0 & 0 &   0.00 &   0.00 &    0.0 &       14.2 \\
     16 &    ATCDFS5 J033316.96-274121.7 & 3:33:16.96 & -27:41:21.7 &       78.3 &      14.1 &      125.2 & 1 &   2.82 &   1.12 &  -50.1 &       11.7 \\
     17 &    ATCDFS5 J033316.76-280016.2 & 3:33:16.76 & -28:00:16.2 &     1291.5 &      12.6 &        0.0 & 0 &   0.00 &   0.00 &    0.0 &       13.0 \\
     18 &    ATCDFS5 J033316.73-275630.4 & 3:33:16.73 & -27:56:30.4 &      775.2 &      16.5 &        0.0 & 0 &   0.00 &   0.00 &    0.0 &       12.7 \\
     19 &    ATCDFS5 J033316.35-274725.1 & 3:33:16.35 & -27:47:25.1 &     1262.2 &      15.7 &        0.0 & 0 &   0.00 &   0.00 &    0.0 &       13.3 \\
     20 &    ATCDFS5 J033314.98-275151.3 & 3:33:14.98 & -27:51:51.3 &      688.9 &      16.0 &        0.0 & 0 &   0.00 &   0.00 &    0.0 &       12.3 \\
     21 &    ATCDFS5 J033314.84-280432.2 & 3:33:14.84 & -28:04:32.2 &      241.4 &      15.8 &        0.0 & 0 &   0.00 &   0.00 &    0.0 &       23.2 \\
     22 &    ATCDFS5 J033313.12-274930.3 & 3:33:13.12 & -27:49:30.3 &      148.6 &      13.5 &        0.0 & 0 &   0.00 &   0.00 &    0.0 &       13.3 \\
     23 &    ATCDFS5 J033312.64-275232.1 & 3:33:12.64 & -27:52:32.1 &       76.5 &      10.7 &        0.0 & 0 &   0.00 &   0.00 &    0.0 &       12.4 \\
     24 &    ATCDFS5 J033311.81-274138.7 & 3:33:11.81 & -27:41:38.7 &      101.7 &      12.5 &        0.0 & 0 &   0.00 &   0.00 &    0.0 &       11.7 \\
     25 &    ATCDFS5 J033310.19-274842.2 & 3:33:10.19 & -27:48:42.2 &     9857.3 &      70.8 &        0.0 & 0 &   0.00 &   0.00 &    0.0 &       13.3 \\
     26 &    ATCDFS5 J033309.70-274801.7 & 3:33:09.70 & -27:48:01.7 &       90.8 &      15.0 &      162.1 & 1 &   4.48 &   1.35 &   23.2 &       15.2 \\
     27 &    ATCDFS5 J033308.17-275033.3 & 3:33:08.17 & -27:50:33.3 &      466.2 &      12.9 &        0.0 & 0 &   0.00 &   0.00 &    0.0 &       12.3 \\
     28 &    ATCDFS5 J033304.45-273802.2 & 3:33:04.45 & -27:38:02.2 &       62.1 &      10.7 &        0.0 & 0 &   0.00 &   0.00 &    0.0 &       12.0 \\
     29 &    ATCDFS5 J033303.73-273611.3 & 3:33:03.73 & -27:36:11.3 &      308.4 &      14.6 &        0.0 & 0 &   0.00 &   0.00 &    0.0 &       11.6 \\
     30 &    ATCDFS5 J033301.80-273636.4 & 3:33:01.80 & -27:36:36.4 &       62.5 &      11.4 &        0.0 & 0 &   0.00 &   0.00 &    0.0 &       12.2 \\
   31a &    ATCDFS5 J033257.54-280209.3 & 3:32:57.54 & -28:02:09.3 &     1304.6 &      81.4 &     2913.3 & 1 &   3.57 &   2.49 &  -66.0 &       14.1 \\
   31b &    ATCDFS5 J033257.10-280210.1 & 3:32:57.10 & -28:02:10.1 &     1918.0 &      85.5 &     4270.2 & 1 &   3.92 &   1.15 &   82.5 &       14.2 \\
   31c &    ATCDFS5 J033256.75-280211.5 & 3:32:56.75 & -28:02:11.5 &     2274.0 &      10.1 &     3339.0 & 1 &   2.56 &   1.61 &  -16.8 &       14.4 \\
     34 &    ATCDFS5 J033256.47-275848.4 & 3:32:56.47 & -27:58:48.4 &      932.5 &      21.1 &      967.8 & 1 &   1.10 &   0.31 &    6.6 &       12.5 \\
     35 &    ATCDFS5 J033256.27-273500.8 & 3:32:56.27 & -27:35:00.8 &      109.7 &      13.5 &        0.0 & 0 &   0.00 &   0.00 &    0.0 &       13.5 \\
     36 &    ATCDFS5 J033253.33-280159.3 & 3:32:53.33 & -28:01:59.3 &      539.2 &      20.9 &      658.1 & 1 &   2.99 &   0.59 &    1.3 &       14.4 \\
     37 &    ATCDFS5 J033252.07-274425.6 & 3:32:52.07 & -27:44:25.6 &      195.0 &      17.3 &        0.0 & 0 &   0.00 &   0.00 &    0.0 &       11.2 \\
     38 &    ATCDFS5 J033251.82-274436.7 & 3:32:51.82 & -27:44:36.7 &       77.8 &      15.5 &        0.0 & 0 &   0.00 &   0.00 &    0.0 &       11.7 \\
     39 &    ATCDFS5 J033251.84-275716.9 & 3:32:51.84 & -27:57:16.9 &       68.5 &      16.2 &        0.0 & 0 &   0.00 &   0.00 &    0.0 &       11.7 \\
     40 &    ATCDFS5 J033249.95-273432.6 & 3:32:49.95 & -27:34:32.6 &      152.2 &      21.2 &        0.0 & 0 &   0.00 &   0.00 &    0.0 &       13.8 \\
     41 &    ATCDFS5 J033249.43-274235.4 & 3:32:49.43 & -27:42:35.4 &      825.3 &      15.2 &        0.0 & 0 &   0.00 &   0.00 &    0.0 &       11.1 \\
     42 &    ATCDFS5 J033249.19-274050.8 & 3:32:49.19 & -27:40:50.8 &     2196.4 &      40.4 &        0.0 & 0 &   0.00 &   0.00 &    0.0 &       11.5 \\
     43 &    ATCDFS5 J033249.30-275844.4 & 3:32:49.30 & -27:58:44.4 &       70.4 &      12.9 &        0.0 & 0 &   0.00 &   0.00 &    0.0 &       12.5 \\
     44 &    ATCDFS5 J033247.89-274232.4 & 3:32:47.89 & -27:42:32.4 &       74.7 &      16.4 &      134.2 & 1 &   4.02 &   1.87 &  -11.4 &       11.9 \\
     45 &    ATCDFS5 J033245.37-280450.2 & 3:32:45.37 & -28:04:50.2 &      642.7 &     60.4 &     1018.4 & 1 &   3.16 &   0.98 &  -39.3 &       23.4 \\
   46a &    ATCDFS5 J033244.28-275140.7 & 3:32:44.28 & -27:51:40.7 &      120.6 &      17.7 &      140.8 & 1 &   1.50 &   0.99 &   -9.0 &       11.4 \\
   46b &    ATCDFS5 J033244.05-275143.6 & 3:32:44.05 & -27:51:43.6 &       77.2 &      17.9 &      130.8 & 1 &   6.53 &   0.34 &   -0.9 &       11.4 \\
   48a &    ATCDFS5 J033243.15-273813.2 & 3:32:43.15 & -27:38:13.2 &     4401.6 &     266.0 &     9300.8 & 1 &   3.72 &   1.62 &   66.2 &       13.6 \\
   48b &    ATCDFS5 J033242.63-273816.3 & 3:32:42.63 & -27:38:16.3 &      563.1 &      46.9 &      675.1 & 1 &   2.60 &   0.62 &   10.6 &       13.6 \\
   48c &    ATCDFS5 J033241.99-273819.2 & 3:32:41.99 & -27:38:19.2 &    10344.0 &     422.0 &    13320.0 & 1 &   1.82 &   1.31 &   21.3 &       15.7 \\
     50 &    ATCDFS5 J033242.00-273949.5 & 3:32:42.00 & -27:39:49.5 &      132.3 &      14.0 &        0.0 & 0 &   0.00 &   0.00 &    0.0 &       12.7 \\
     51 &    ATCDFS5 J033241.62-280128.2 & 3:32:41.62 & -28:01:28.2 &      110.3 &      13.3 &        0.0 & 0 &   0.00 &   0.00 &    0.0 &       14.7 \\
     52 &    ATCDFS5 J033237.74-275000.7 & 3:32:37.74 & -27:50:00.7 &       70.9 &      17.1 &        0.0 & 0 &   0.00 &   0.00 &    0.0 &       11.0 \\
   53a &    ATCDFS5 J033232.14-280317.7 & 3:32:32.14 & -28:03:17.7 &     2204.8 &      82.4 &     2832.4 & 1 &   2.43 &   1.14 &   -2.8 &       19.8 \\
   53b &    ATCDFS5 J033231.97-280303.1 & 3:32:31.97 & -28:03:03.1 &     1944.7 &     135.0 &     3764.0 & 1 &   4.96 &   1.86 &    2.0 &       19.1 \\
   53c &    ATCDFS5 J033232.00-280309.9 & 3:32:32.00 & -28:03:09.9 &     4564.2 &     159.0 &        0.0 & 0 &   0.00 &   0.00 &    0.0 &       19.1 \\
     56 &    ATCDFS5 J033231.55-275029.0 & 3:32:31.55 & -27:50:29.0 &      105.8 &      16.6 &        0.0 & 0 &   0.00 &   0.00 &    0.0 &       11.6 \\
     57 &    ATCDFS5 J033230.56-275911.2 & 3:32:30.56 & -27:59:11.2 &      112.2 &      9.1 &      186.4 & 1 &   5.96 &   0.53 &    7.9 &       11.2 \\
     58 &    ATCDFS5 J033229.86-274424.7 & 3:32:29.86 & -27:44:24.7 &      175.3 &      10.4 &      411.5 & 1 &   5.58 &   2.19 &  -17.4 &       11.6 \\
   59a &    ATCDFS5 J033228.82-274355.8 & 3:32:28.82 & -27:43:55.8 &      269.8 &      37.5 &      468.9 & 1 &   5.29 &   1.10 &   12.5 &       19.6 \\
   59b &    ATCDFS5 J033228.71-274402.4 & 3:32:28.71 & -27:44:02.4 &      215.5 &      41.1 &      670.0 & 1 &   0.00 &   0.00 &    0.0 &       17.5 \\
     60 &    ATCDFS5 J033228.74-274620.6 & 3:32:28.74 & -27:46:20.6 &      184.9 &      12.8 &        0.0 & 0 &   0.00 &   0.00 &    0.0 &       11.4 \\
     64 &    ATCDFS5 J033226.97-274107.0 & 3:32:26.97 & -27:41:07.0 &     4807.2 &     131.7 &     6740.4 & 1 &   3.42 &   0.90 &   17.5 &       14.6 \\
     66 &    ATCDFS5 J033224.30-280114.4 & 3:32:24.30 & -28:01:14.4 &      131.0 &      13.9 &        0.0 & 0 &   0.00 &   0.00 &    0.0 &       11.8 \\
     67 &    ATCDFS5 J033223.82-275845.3 & 3:32:23.82 & -27:58:45.3 &       97.8 &      13.8 &        0.0 & 0 &   0.00 &   0.00 &    0.0 &       12.5 \\
     68 &    ATCDFS5 J033223.69-273648.5 & 3:32:23.69 & -27:36:48.5 &       85.4 &      13.8 &        0.0 & 0 &   0.00 &   0.00 &    0.0 &       12.1 \\
     69 &    ATCDFS5 J033222.60-280023.5 & 3:32:22.60 & -28:00:23.5 &       89.0 &      11.5 &      177.1 & 1 &   4.34 &   1.54 &  -34.8 &       13.1 \\
     70 &    ATCDFS5 J033222.51-274804.5 & 3:32:22.51 & -27:48:04.5 &       69.0 &      11.0 &        0.0 & 0 &   0.00 &   0.00 &    0.0 &       11.9 \\
     71 &    ATCDFS5 J033221.73-280152.2 & 3:32:21.73 & -28:01:52.2 &       94.1 &      17.1 &        0.0 & 0 &   0.00 &   0.00 &    0.0 &       14.8 \\
     72 &    ATCDFS5 J033221.27-274435.8 & 3:32:21.27 & -27:44:35.8 &       86.3 &      12.3 &        0.0 & 0 &   0.00 &   0.00 &    0.0 &       12.6 \\
     73 &    ATCDFS5 J033221.07-273530.5 & 3:32:21.07 & -27:35:30.5 &      120.8 &      14.2 &        0.0 & 0 &   0.00 &   0.00 &    0.0 &       13.0 \\
     75 &    ATCDFS5 J033219.79-274123.1 & 3:32:19.79 & -27:41:23.1 &       80.2 &      11.6 &        0.0 & 0 &   0.00 &   0.00 &    0.0 &       12.0 \\
   76a &    ATCDFS5 J033219.75-275401.4 & 3:32:19.75 & -27:54:01.4 &      475.1 &      23.7 &      947.2 & 1 &   3.33 &   2.30 &   47.4 &       13.0 \\
   76b &    ATCDFS5 J033219.27-275406.5 & 3:32:19.27 & -27:54:06.5 &      659.2 &      78.3 &     1085.0 & 1 &   2.99 &   0.65 &   57.1 &       12.6 \\
   76c &    ATCDFS5 J033219.07-275408.2 & 3:32:19.07 & -27:54:08.2 &      460.0 &     133.0 &      572.4 & 1 &   1.81 &   0.47 &   49.5 &       12.6 \\
   76d &    ATCDFS5 J033218.53-275412.2 & 3:32:18.53 & -27:54:12.2 &      430.7 &      19.7 &      910.3 & 1 &   3.14 &   2.75 &  -76.9 &       12.6 \\
     78 &    ATCDFS5 J033218.01-274718.6 & 3:32:18.01 & -27:47:18.6 &      442.9 &      20.0 &        0.0 & 0 &   0.00 &   0.00 &    0.0 &       11.3 \\
     79 &    ATCDFS5 J033217.05-275846.5 & 3:32:17.05 & -27:58:46.5 &     1733.3 &      13.6 &        0.0 & 0 &   0.00 &   0.00 &    0.0 &       11.7 \\
     80 &    ATCDFS5 J033215.95-273438.5 & 3:32:15.95 & -27:34:38.5 &      213.8 &      14.1 &        0.0 & 0 &   0.00 &   0.00 &    0.0 &       14.2 \\
     81 &    ATCDFS5 J033214.84-275640.1 & 3:32:14.84 & -27:56:40.1 &       93.2 &      11.3 &        0.0 & 0 &   0.00 &   0.00 &    0.0 &       11.5 \\
     82 &    ATCDFS5 J033213.89-275000.9 & 3:32:13.89 & -27:50:00.9 &      101.8 &      15.9 &        0.0 & 0 &   0.00 &   0.00 &    0.0 &       11.6 \\
     83 &    ATCDFS5 J033213.48-274953.2 & 3:32:13.48 & -27:49:53.2 &       99.7 &      10.2 &        0.0 & 0 &   0.00 &   0.00 &    0.0 &       11.7 \\

\hline

\end{tabular}
}
\label{tab:cat}
\end{table*}

\setcounter{table}{0}
\begin{table*}
\caption{continued}  
{\scriptsize
\begin{tabular}{llllrrrrrrrr}  
\hline
ID & IAU name & RA & Dec & $S_{peak}$ & d$S_{peak}$ & $S_{int}$ & Decon & Decon& Decon & Decon & $\sigma_{local}$ \\  
     &                   &    (J2000)    &       (J2000)                 & ($\mu$Jy)  & ($\mu$Jy)  & ($\mu$Jy) & Flag & Bmajor & Bminor & PA & \\ \hline

     84 &    ATCDFS5 J033213.08-274350.9 & 3:32:13.08 & -27:43:50.9 &      283.7 &      10.0 &      429.1 & 1 &   2.38 &   1.86 &    5.2 &       12.5 \\
     85 &    ATCDFS5 J033211.65-273726.2 & 3:32:11.65 & -27:37:26.2 &    12248.0 &      85.0 &        0.0 & 0 &   0.00 &   0.00 &    0.0 &       15.6 \\
     86 &    ATCDFS5 J033211.55-274713.2 & 3:32:11.55 & -27:47:13.2 &       95.4 &      14.4 &        0.0 & 0 &   0.00 &   0.00 &    0.0 &       11.9 \\
     87 &    ATCDFS5 J033210.92-274415.2 & 3:32:10.92 & -27:44:15.2 &     2095.0 &      16.5 &        0.0 & 0 &   0.00 &   0.00 &    0.0 &       11.3 \\
     88 &    ATCDFS5 J033210.99-274053.7 & 3:32:10.99 & -27:40:53.7 &      229.9 &      18.0 &        0.0 & 0 &   0.00 &   0.00 &    0.0 &       12.4 \\
     89 &    ATCDFS5 J033210.79-274628.0 & 3:32:10.79 & -27:46:28.0 &      104.6 &      14.7 &        0.0 & 0 &   0.00 &   0.00 &    0.0 &       11.6 \\
     90 &    ATCDFS5 J033210.16-275938.4 & 3:32:10.16 & -27:59:38.4 &      157.7 &      21.4 &        0.0 & 0 &   0.00 &   0.00 &    0.0 &       12.0 \\
     91 &    ATCDFS5 J033209.80-275932.4 & 3:32:09.80 & -27:59:32.4 &       69.2 &      19.2 &        0.0 & 0 &   0.00 &   0.00 &    0.0 &       13.1 \\
     92 &    ATCDFS5 J033209.71-274248.3 & 3:32:09.71 & -27:42:48.3 &      616.0 &      14.8 &        0.0 & 0 &   0.00 &   0.00 &    0.0 &       11.7 \\
     93 &    ATCDFS5 J033208.67-274734.6 & 3:32:08.67 & -27:47:34.6 &     3294.6 &      40.2 &     3371.2 & 1 &   0.90 &   0.18 &   10.9 &       12.8 \\
     94 &    ATCDFS5 J033206.10-273235.7 & 3:32:06.10 & -27:32:35.7 &    14121.0 &     146.9 &        0.0 & 0 &   0.00 &   0.00 &    0.0 &       20.8 \\
     95 &    ATCDFS5 J033204.68-280057.2 & 3:32:04.68 & -28:00:57.2 &       69.3 &      17.2 &        0.0 & 0 &   0.00 &   0.00 &    0.0 &       14.3 \\
     96 &    ATCDFS5 J033203.88-275805.3 & 3:32:03.88 & -27:58:05.3 &      111.4 &       9.5 &        0.0 & 0 &   0.00 &   0.00 &    0.0 &       11.7 \\
     97 &    ATCDFS5 J033203.66-274604.2 & 3:32:03.66 & -27:46:04.2 &       66.5 &      15.0 &        0.0 & 0 &   0.00 &   0.00 &    0.0 &       11.8 \\
   98a &    ATCDFS5 J033201.56-274647.7 & 3:32:01.56 & -27:46:47.7 &     4835.1 &     115.0 &     6597.2 & 1 &   2.09 &   1.02 &   49.7 &       12.3 \\
   98b &    ATCDFS5 J033201.28-274647.6 & 3:32:01.28 & -27:46:47.6 &     3463.2 &     120.0 &     4383.8 & 1 &   2.32 &   0.66 &   30.7 &       13.2 \\
    100 &    ATCDFS5 J033200.84-273556.9 & 3:32:00.84 & -27:35:56.9 &     2394.1 &      18.8 &        0.0 & 0 &   0.00 &   0.00 &    0.0 &       11.4 \\
    101 &    ATCDFS5 J033159.83-274540.4 & 3:31:59.83 & -27:45:40.4 &      107.5 &       8.7 &        0.0 & 0 &   0.00 &   0.00 &    0.0 &       11.6 \\
    102 &    ATCDFS5 J033155.00-274411.0 & 3:31:55.00 & -27:44:11.0 &       64.2 &      12.6 &        0.0 & 0 &   0.00 &   0.00 &    0.0 &       11.2 \\
    103 &    ATCDFS5 J033153.42-280221.2 & 3:31:53.42 & -28:02:21.2 &      787.8 &      15.9 &        0.0 & 0 &   0.00 &   0.00 &    0.0 &       15.9 \\
    104 &    ATCDFS5 J033152.12-273926.5 & 3:31:52.12 & -27:39:26.5 &      555.4 &      17.4 &        0.0 & 0 &   0.00 &   0.00 &    0.0 &       11.7 \\
    105 &    ATCDFS5 J033150.78-274703.9 & 3:31:50.78 & -27:47:03.9 &      119.5 &       8.8 &        0.0 & 0 &   0.00 &   0.00 &    0.0 &       12.3 \\
    106 &    ATCDFS5 J033150.13-273948.3 & 3:31:50.13 & -27:39:48.3 &      222.8 &      16.9 &      347.5 & 1 &   3.78 &   1.34 &   14.4 &       11.3 \\
    107 &    ATCDFS5 J033150.02-275806.4 & 3:31:50.02 & -27:58:06.4 &      160.9 &      12.7 &        0.0 & 0 &   0.00 &   0.00 &    0.0 &       11.2 \\
    108 &    ATCDFS5 J033149.88-274838.9 & 3:31:49.88 & -27:48:38.9 &      847.8 &      38.2 &     1147.7 & 1 &   1.77 &   1.12 &   76.7 &       11.6 \\
    109 &    ATCDFS5 J033148.72-273312.2 & 3:31:48.72 & -27:33:12.2 &      110.4 &      22.2 &        0.0 & 0 &   0.00 &   0.00 &    0.0 &       16.1 \\
    110 &    ATCDFS5 J033147.37-274542.5 & 3:31:47.37 & -27:45:42.5 &      116.6 &       6.7 &        0.0 & 0 &   0.00 &   0.00 &    0.0 &       11.1 \\
    111 &    ATCDFS5 J033146.59-275734.7 & 3:31:46.59 & -27:57:34.7 &      128.8 &      13.6 &      179.2 & 1 &   3.73 &   0.22 &   21.2 &       11.2 \\
    112 &    ATCDFS5 J033146.62-274553.3 & 3:31:46.62 & -27:45:53.3 &       62.4 &      10.1 &        0.0 & 0 &   0.00 &   0.00 &    0.0 &       12.0 \\
    113 &    ATCDFS5 J033146.10-280026.4 & 3:31:46.10 & -28:00:26.4 &      136.8 &      12.0 &        0.0 & 0 &   0.00 &   0.00 &    0.0 &       11.2 \\
    114 &    ATCDFS5 J033144.02-273836.5 & 3:31:44.02 & -27:38:36.5 &       79.2 &      13.5 &        0.0 & 0 &   0.00 &   0.00 &    0.0 &       12.1 \\
    115 &    ATCDFS5 J033140.05-273648.0 & 3:31:40.05 & -27:36:48.0 &       77.5 &      11.7 &        0.0 & 0 &   0.00 &   0.00 &    0.0 &       11.6 \\
    116 &    ATCDFS5 J033134.22-273828.7 & 3:31:34.22 & -27:38:28.7 &      273.7 &      20.3 &        0.0 & 0 &   0.00 &   0.00 &    0.0 &       11.8 \\
  117a &    ATCDFS5 J033131.12-273815.9 & 3:31:31.12 & -27:38:15.9 &     1731.7 &     107.0 &     3260.0 & 1 &   2.98 &   1.76 &  -78.0 &       13.9 \\
  117b &    ATCDFS5 J033130.62-273815.2 & 3:31:30.62 & -27:38:15.2 &      109.6 &      12.0 &      167.5 & 1 &   5.24 &   1.90 &   -1.0 &       14.1 \\
  117c &    ATCDFS5 J033130.02-273814.0 & 3:31:30.02 & -27:38:14.0 &      240.3 &      12.6 &      277.5 & 1 &   1.75 &   0.60 &  -62.4 &       14.5 \\
  117d &    ATCDFS5 J033129.57-273802.8 & 3:31:29.57 & -27:38:02.8 &      255.5 &      15.0 &      441.3 & 1 &   3.96 &   3.01 &  -24.1 &       20.3 \\
    118 &    ATCDFS5 J033130.74-275734.9 & 3:31:30.74 & -27:57:34.9 &      163.5 &       9.6 &        0.0 & 0 &   0.00 &   0.00 &    0.0 &       10.8 \\
  119a &    ATCDFS5 J033130.37-275606.0 & 3:31:30.37 & -27:56:06.0 &       96.7 &      12.5 &      272.0 & 1 &   4.69 &   3.33 &   38.1 &       12.5 \\
  119b &    ATCDFS5 J033130.05-275602.5 & 3:31:30.05 & -27:56:02.5 &       96.7 &      12.9 &        0.0 & 0 &   0.00 &   0.00 &    0.0 &       12.9 \\
  119c &    ATCDFS5 J033129.83-275559.9 & 3:31:29.83 & -27:55:59.9 &       78.6 &      13.9 &      160.9 & 1 &   5.59 &   1.82 &    3.8 &       13.9 \\
    124 &    ATCDFS5 J033129.77-273218.4 & 3:31:29.77 & -27:32:18.4 &     1670.9 &      36.3 &        0.0 & 0 &   0.00 &   0.00 &    0.0 &       18.5 \\
    126 &    ATCDFS5 J033128.58-274934.7 & 3:31:28.58 & -27:49:34.7 &      174.6 &       9.6 &        0.0 & 0 &   0.00 &   0.00 &    0.0 &       12.1 \\
    127 &    ATCDFS5 J033127.20-274247.2 & 3:31:27.20 & -27:42:47.2 &      603.2 &      20.1 &      709.5 & 1 &   2.15 &   0.73 &  -11.2 &       13.0 \\
    128 &    ATCDFS5 J033127.04-275958.6 & 3:31:27.04 & -27:59:58.6 &       88.4 &      19.4 &        0.0 & 0 &   0.00 &   0.00 &    0.0 &       12.6 \\
    129 &    ATCDFS5 J033127.05-274409.7 & 3:31:27.05 & -27:44:09.7 &      186.6 &       9.1 &        0.0 & 0 &   0.00 &   0.00 &    0.0 &       13.0 \\
    130 &    ATCDFS5 J033126.78-274237.1 & 3:31:26.78 & -27:42:37.1 &       98.0 &      14.1 &        0.0 & 0 &   0.00 &   0.00 &    0.0 &       13.6 \\
    131 &    ATCDFS5 J033124.90-275208.0 & 3:31:24.90 & -27:52:08.0 &     6481.8 &     179.7 &    12109.0 & 1 &   3.44 &   1.05 &   59.9 &       13.9 \\
    132 &    ATCDFS5 J033123.30-274905.8 & 3:31:23.30 & -27:49:05.8 &      560.7 &      13.1 &        0.0 & 0 &   0.00 &   0.00 &    0.0 &       13.6 \\
    133 &    ATCDFS5 J033120.13-273901.3 & 3:31:20.13 & -27:39:01.3 &       88.3 &      10.3 &        0.0 & 0 &   0.00 &   0.00 &    0.0 &       14.2 \\
    134 &    ATCDFS5 J033118.73-274902.3 & 3:31:18.73 & -27:49:02.3 &       99.9 &      13.9 &        0.0 & 0 &   0.00 &   0.00 &    0.0 &       14.2 \\
    135 &    ATCDFS5 J033117.34-280147.0 & 3:31:17.34 & -28:01:47.0 &      672.5 &      31.5 &        0.0 & 0 &   0.00 &   0.00 &    0.0 &       17.1 \\
    136 &    ATCDFS5 J033117.04-275515.2 & 3:31:17.04 & -27:55:15.2 &      474.2 &      31.9 &     1068.9 & 1 &   4.38 &   2.62 &  -17.8 &       14.9 \\
    137 &    ATCDFS5 J033115.99-274443.1 & 3:31:15.99 & -27:44:43.1 &      389.5 &      20.4 &      428.0 & 1 &   1.47 &   0.41 &  -24.4 &       15.8 \\
  138a &    ATCDFS5 J033115.04-275518.7 & 3:31:15.04 & -27:55:18.7 &     1512.6 &      24.4 &        0.0 & 0 &   0.00 &   0.00 &    0.0 &       14.6 \\
  138b &    ATCDFS5 J033114.37-275519.1 & 3:31:14.37 & -27:55:19.1 &      159.4 &      12.3 &      407.2 & 1 &   4.15 &   3.15 &   41.7 &       20.0 \\
  138c &    ATCDFS5 J033113.94-275519.7 & 3:31:13.94 & -27:55:19.7 &      606.6 &      15.6 &     1834.9 & 1 &   5.37 &   2.10 &   74.8 &       15.6 \\
    140 &    ATCDFS5 J033114.45-275545.9 & 3:31:14.45 & -27:55:45.9 &       98.8 &      16.1 &        0.0 & 0 &   0.00 &   0.00 &    0.0 &       14.9 \\
    142 &    ATCDFS5 J033113.95-273910.2 & 3:31:13.95 & -27:39:10.2 &      501.2 &      26.9 &      537.0 & 1 &   0.77 &   0.33 &   80.8 &       14.8 \\
    143 &    ATCDFS5 J033112.59-275718.1 & 3:31:12.59 & -27:57:18.1 &      269.8 &      13.1 &        0.0 & 0 &   0.00 &   0.00 &    0.0 &       15.0 \\
    144 &    ATCDFS5 J033109.80-275225.1 & 3:31:09.80 & -27:52:25.1 &      695.3 &      27.9 &        0.0 & 0 &   0.00 &   0.00 &    0.0 &       22.7 \\
    145 &    ATCDFS5 J033109.19-274954.8 & 3:31:09.19 & -27:49:54.8 &      119.8 &      22.9 &        0.0 & 0 &   0.00 &   0.00 &    0.0 &       19.5 \\
    
\hline

\end{tabular}
}
\label{tab:cat2}
\end{table*}

\subsection{Comparison with VLA survey}

The CDFS area has been observed with the Very Large Array (VLA) at 4.9 GHz. Four VLA pointings were used to cover a region of approximately 20 $\times$ 20 arcmin in size, with depths varying from 7$\mu$Jy/beam rms at the pointing centers to  50 $\mu$Jy/beam rms at the edges \citep{kellermann2008}. The resolution of the VLA image is approximately 3.5 arcsec, which is almost the same size as the synthesized beam of our ATCA imaging. 

\begin{figure*}
\includegraphics[width=0.9\columnwidth]{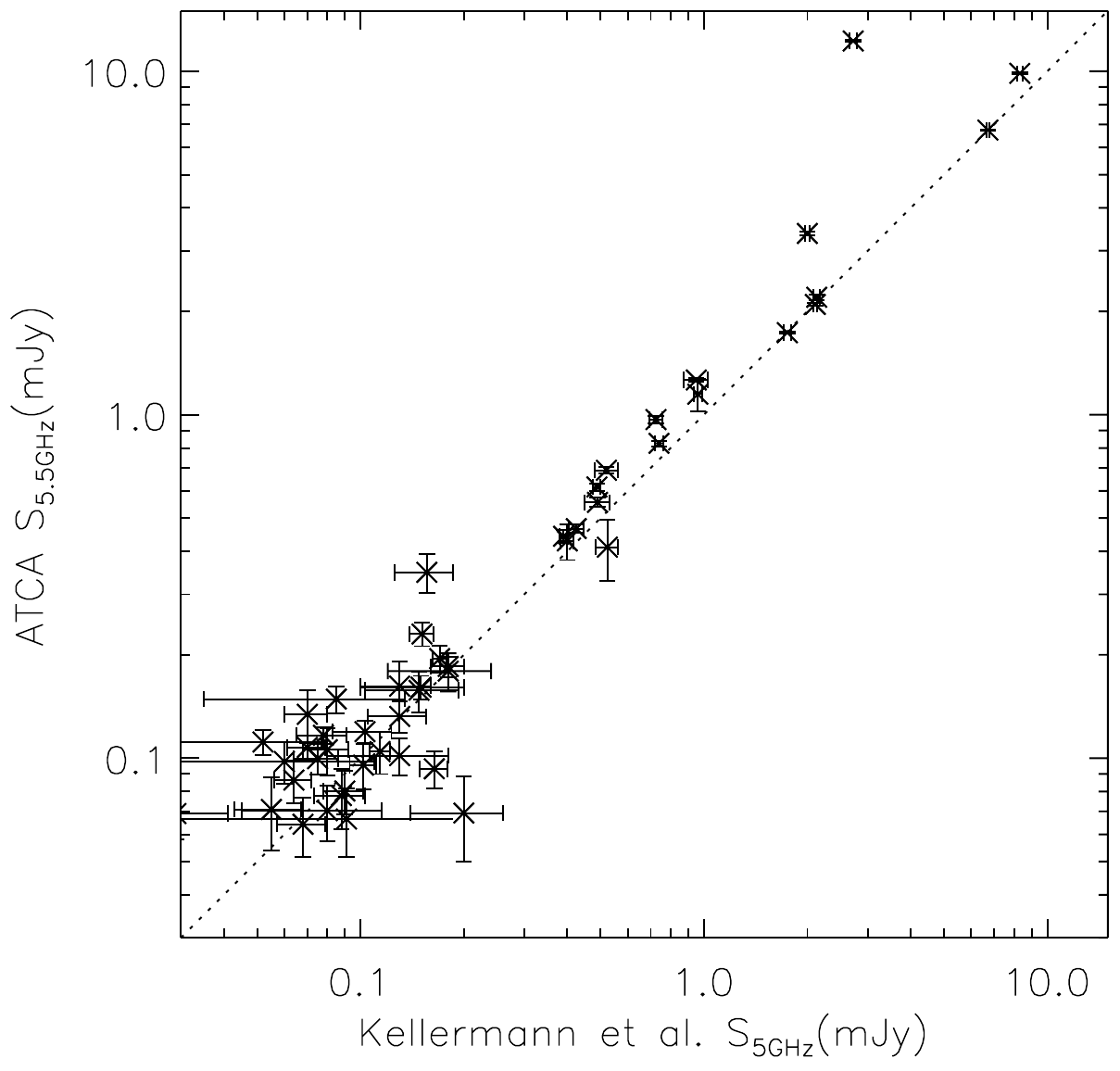}
\hspace{5mm}
\includegraphics[width=0.9\columnwidth]{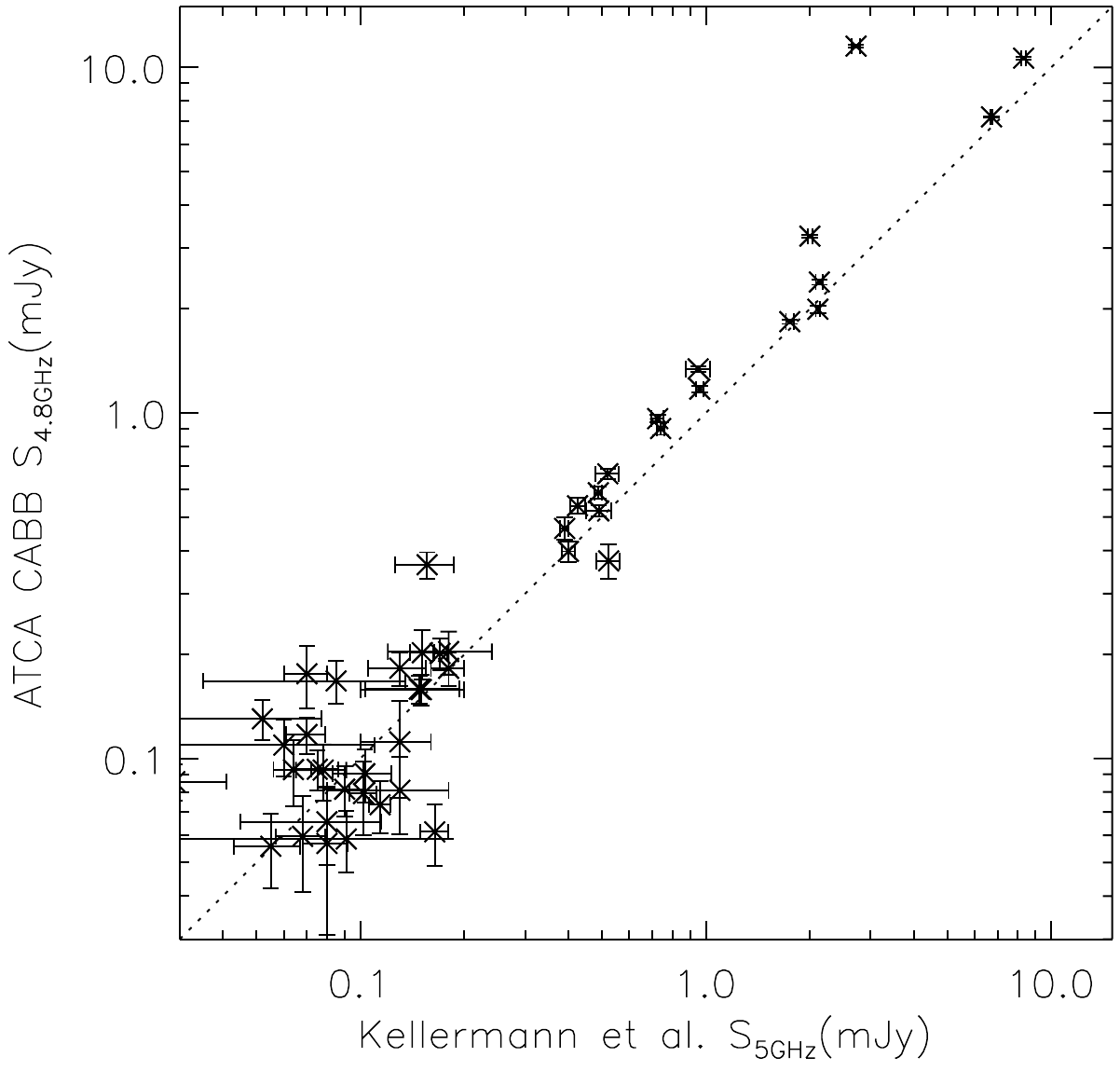}
\caption{LEFT: Comparison of the flux densities in Table \ref{tab:cat} with corresponding measurements in \protect\cite{kellermann2008}. The dotted line denotes ATCA flux density = VLA flux density. The ATCA values appear greater (by $\sim$20\%) than the VLA ones, but there is considerable scatter at the lowest flux densities. RIGHT: Same as the left, but the comparison is with ATCA flux densities measured in the 4.8 GHz sub-band image.} 
\label{fig:fluxcomp}
\end{figure*}

We compared the single component sources in our catalogue with those in the VLA catalogue. We find 49 of our sources have a VLA 4.9 GHz flux density measurement from \cite{kellermann2008}. A straight comparison of the flux densities (Figure \ref{fig:fluxcomp}, left) suggests that ATCA flux densities are higher than the VLA flux densities, and the ATCA/VLA flux density ratio has a mean of 1.28 $\pm$ 0.09 and median of 1.14. However one source with $S_{5.5 GHz} \sim 12.2$ mJy is a clear outlier. Subsequent analysis finds this source is an AGN with QSO-like optical colours and hence variability could be an issue with this source. Excluding this outlier, the ATCA/VLA flux density ratio has a mean of 1.21 $\pm$ 0.06 and median of 1.13. For a spectral index of $\alpha = -0.8$ we expect the ATCA flux densities to be about 8\% less than the VLA measurements, if the VLA and ATCA are calibrated on the same scale. To minimise spectral index effects we compare the VLA flux densities to those of our sources as measured in the ATCA 4.8 GHz sub-band image (Figure \ref{fig:fluxcomp}, right). We find that the ATCA/VLA ratio, excluding the outlying source, has a mean of 1.23 $\pm$ 0.08 and median of 1.18. The ATCA flux densities are therefore about 20\% greater than the ones measured by the VLA. 

Discrepancies in measured flux densities can result from the details of the fitting algorithm used in the source extraction, but this is expected to affect only faint sources. The higher ATCA flux density is clear in the strong sources with 0.3  $< S_{5.5 GHz} < 1$ mJy, not just in the faint sources, which suggests that choice of fitting algorithm is not the cause of the differences.  A possible discrepancy between the VLA and ATCA flux density scales has been noted in the past \citep{norris2006, kellermann2008}, but with ATCA flux densities reported to be lower than VLA ones by 5 to 20\%. However, those comparisons were done at 1.4 GHz, and so it is possible that the calibrator scales assumed can lead to different flux density ratio comparison results between 1.4 GHz and 5 GHz.

\subsection{Completeness and Flux Boosting}
\label{sec:sims}

\begin{figure*}
\includegraphics[width=0.9\columnwidth]{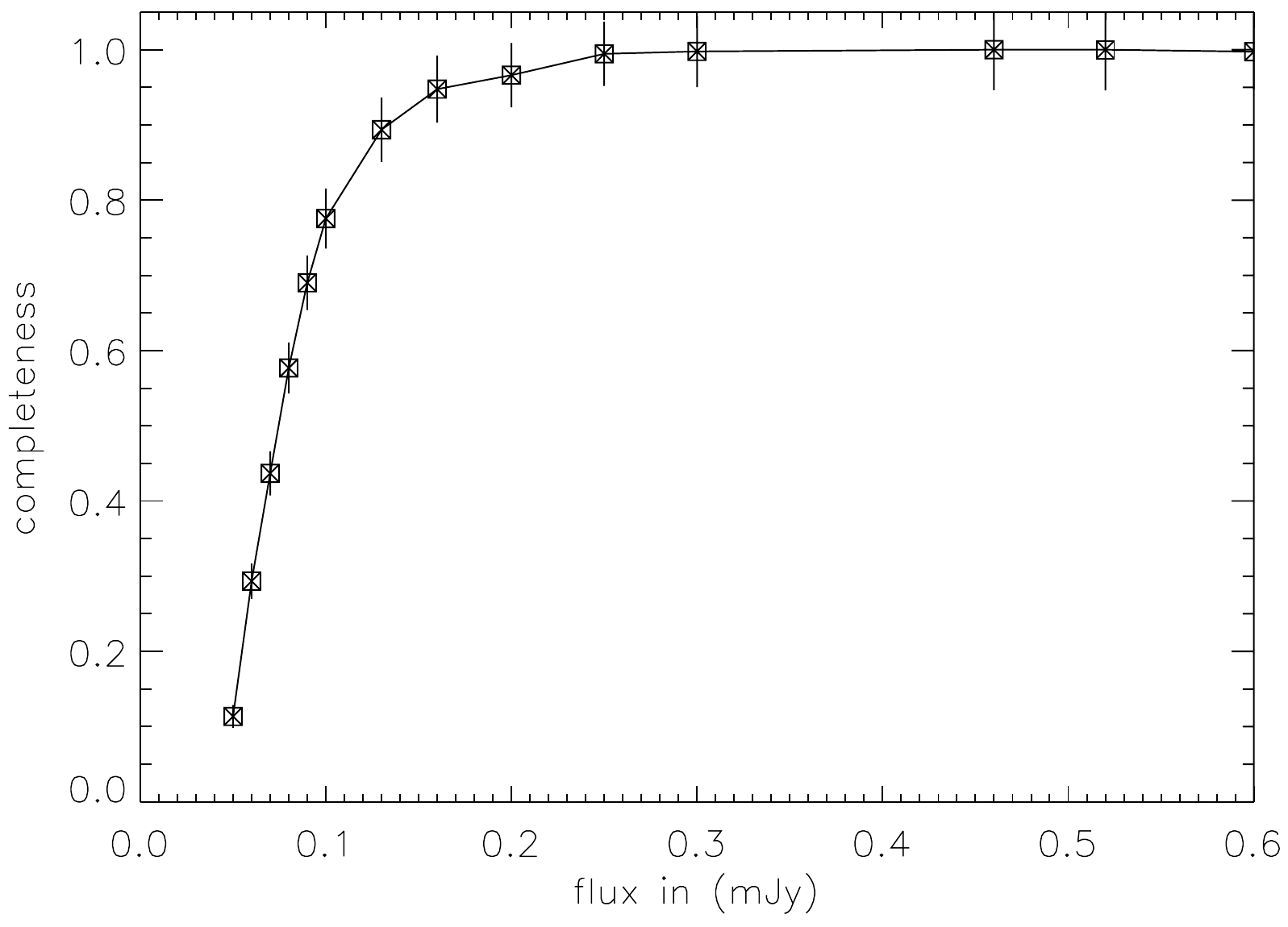}
\hspace{5mm}
\includegraphics[width=0.9\columnwidth]{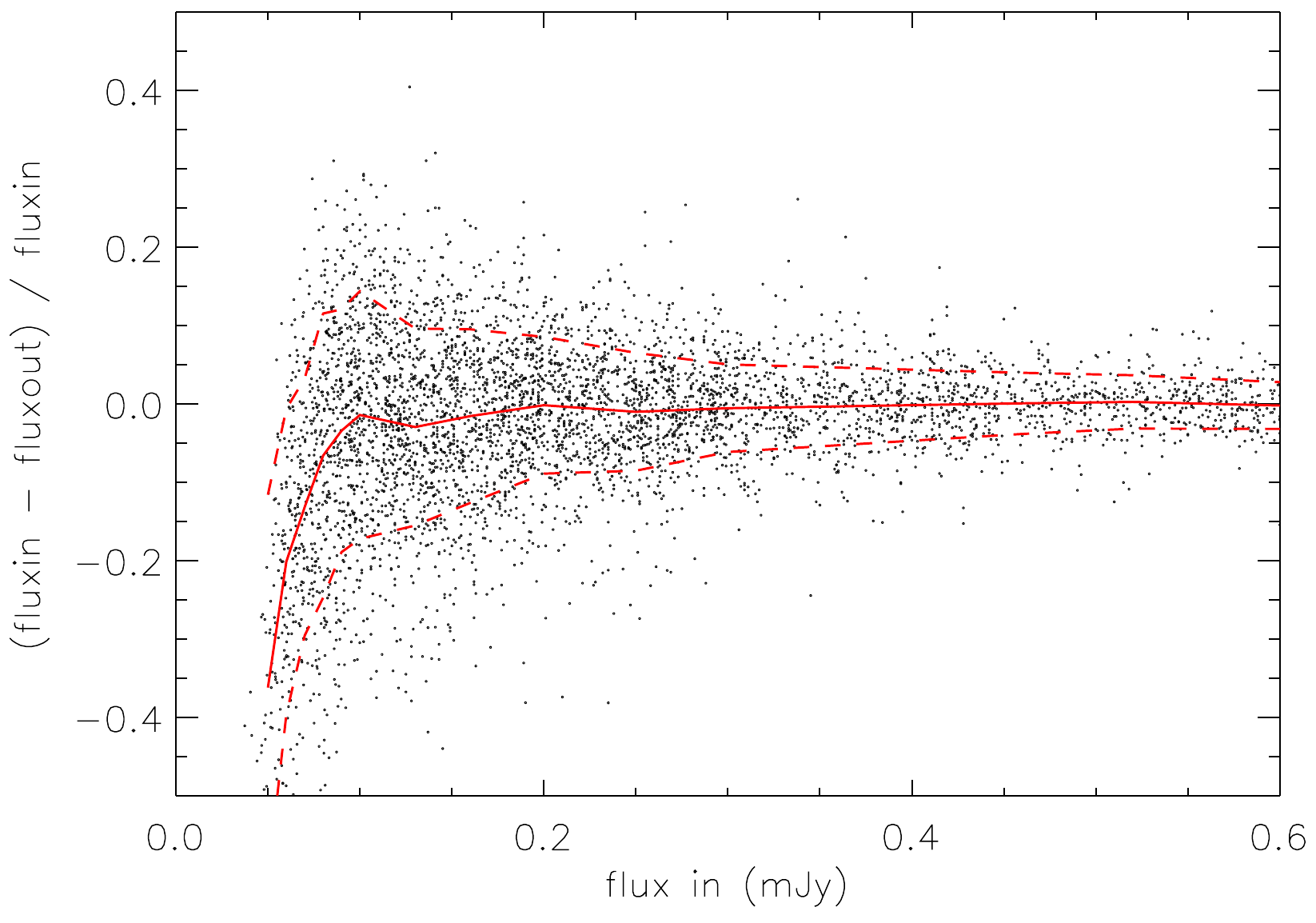}
\caption{LEFT: Completeness as a function of input flux density, as derived from the Monte-Carlo simulations. Completeness is the number of extracted sources divided by number of input sources. RIGHT: The distribution of (input flux density
 - output flux density)/input flux density as a function of output flux density for the simulated sources. The solid red line is the median of the simulation and  the dashed lines mark the 1 sigma upper and lower bounds. The effect of flux boosting at the faint end is dramatically illustrated by the rapid downturn to negative values below about 0.1 mJy.}
\label{fig:comp-boost}

\end{figure*}

The completeness of the source catalog was estimated by simulations. Monte-Carlo simulations were performed by injecting 8000 sources at random locations of the map and extracting them using the same technique as adopted for the production of the catalog. While completeness levels vary significantly across the image, becoming worse at the edges of the mosaic due to primary beam effects, we can recover the overall completeness level of the generated catalog by injecting sources over the full image. A single source was injected per simulation, and the input flux density varied from 30 to 3000 $\mu$Jy to sample the full range of interest. The completeness as a function of flux density is shown in Figure \ref{fig:comp-boost} (left).  The completeness rises steeply from about 10\% at 50 $\mu$Jy to approximately 90\% at 130 $\mu$Jy. The 50\% completeness level occurs at approximately 75 $\mu$Jy. 

Source detection algorithms that rely on finding a peak above a local noise background can lead to flux boosting. This is because sources that lie on a noise peak have their flux density increased and therefore have a higher probability of being detected, while sources which lie on a noise trough have decreased flux densities and may be excluded altogether. This effect is most pronounced in the faintest flux density bins, i.e. in sources with a SNR close to the limit of the catalog. The degree of flux boosting can be estimated by examining the output to input flux density of the simulations (Figure \ref{fig:comp-boost}, right). We find that flux densities are boosted by about 5\% to 1\% for measured flux densities of 100 to 200 $\mu$Jy, on average. The flux boosting is negligible for sources with true flux densities brighter than 200 $\mu$Jy.

\begin{figure*}
\includegraphics[width=0.9\columnwidth]{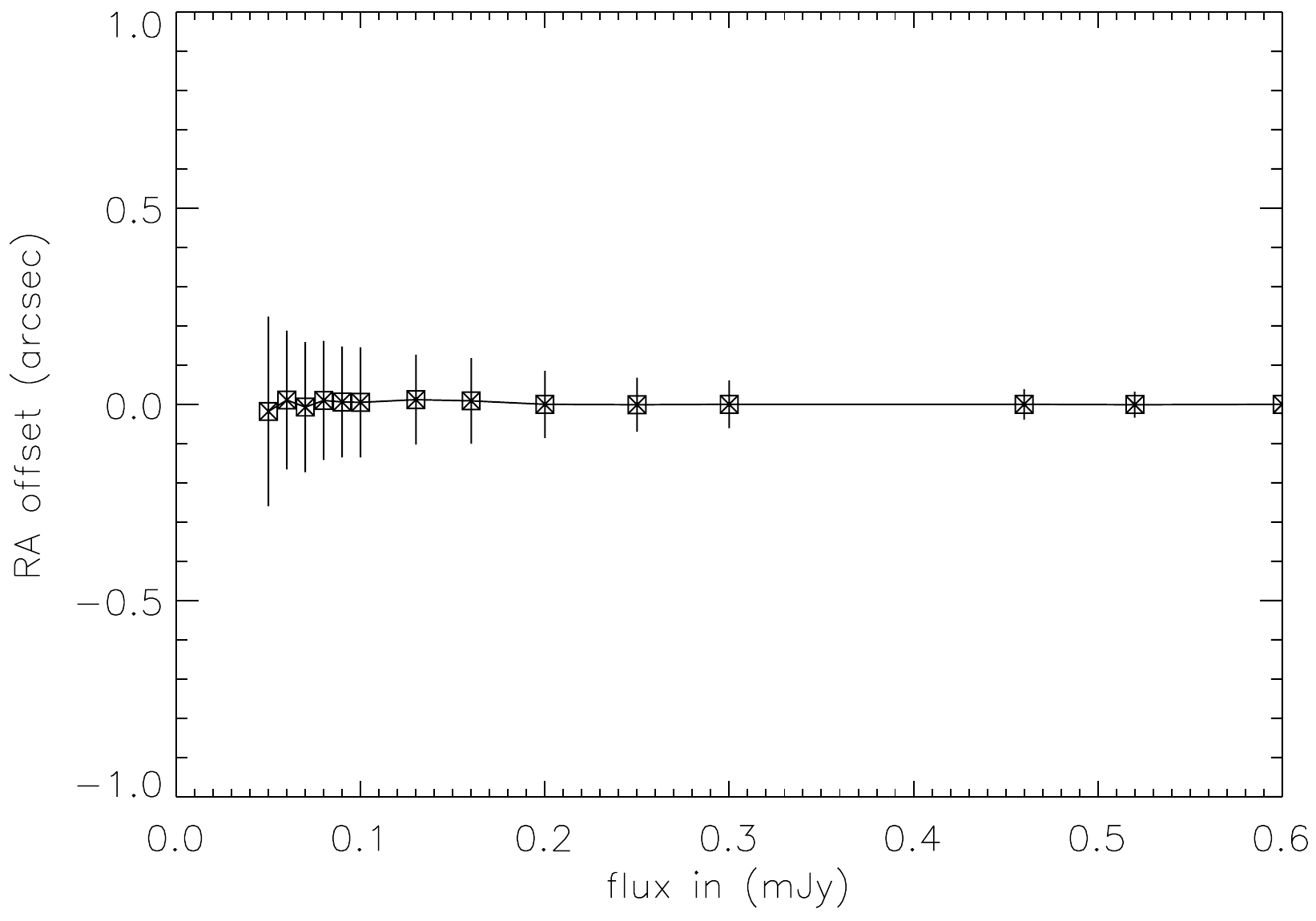}
\hspace{5mm}
\includegraphics[width=0.9\columnwidth]{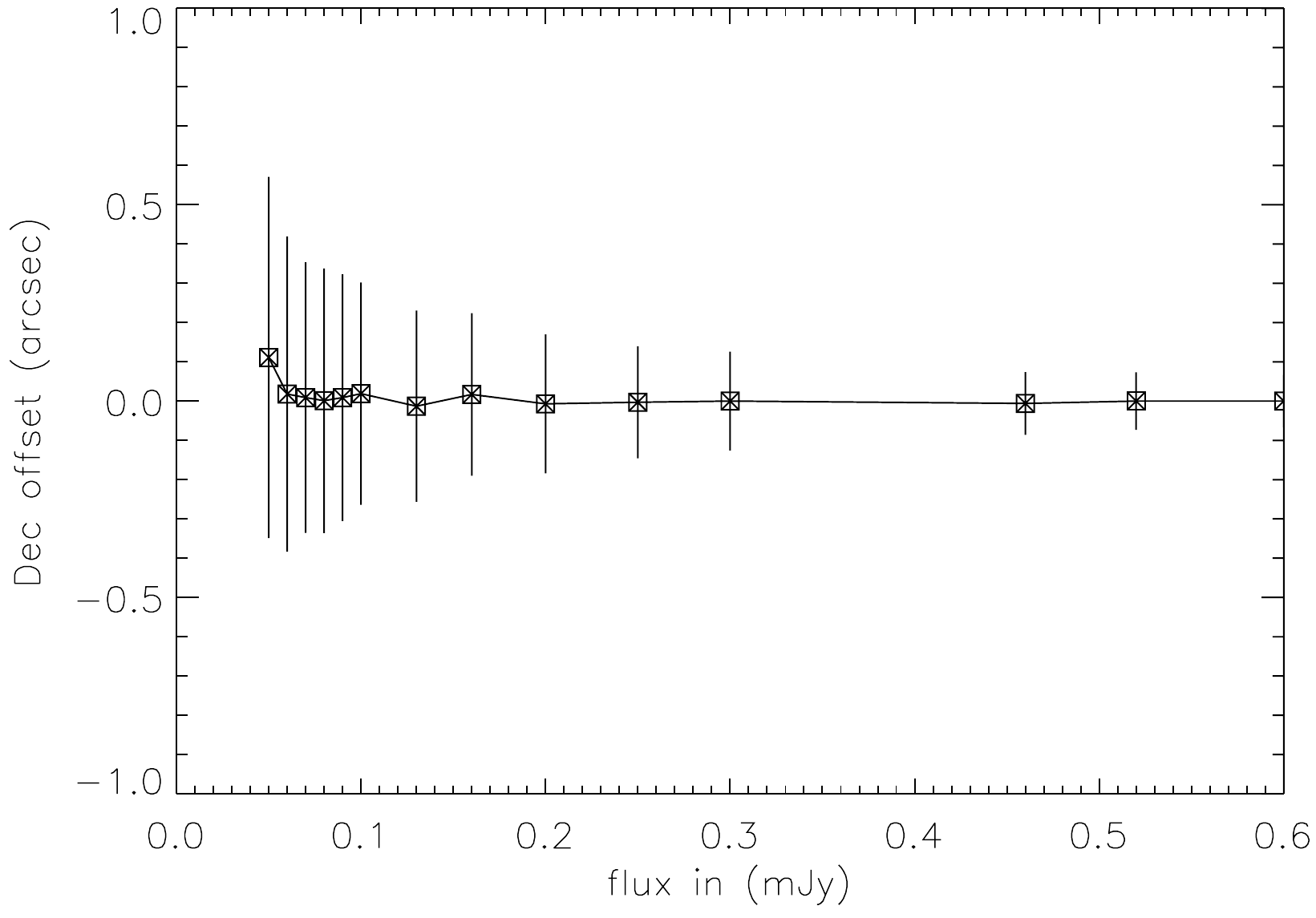}
\caption{The offset in RA (left) and Dec (right) between the recovered positions
 of sources in the simulation and the true input positions, as a function of input flux density. The error bars mark the 1 sigma uncertainty in the position as 
a function of input flux density. }
\label{fig:delradeldec}
\end{figure*}

The simulations can also be used to gauge the positional accuracy of the catalog by comparing input and output positions. The median of the RA and Dec offsets are plotted for various input flux density bins in Figure \ref{fig:delradeldec}. The positional accuracy can be estimated from the standard deviation in the offsets of each of these bins. We find that at the faintest levels (50 $\mu$Jy) the fitted RA and Dec uncertainties are approximately 0.24 and 0.46 arcsec, respectively. The total positional accuracy is $\sim$0.3 arcsec or better for sources that are 0.1 mJy and brighter. 

\subsection{Source Size and Resolution Bias}

Weak extended sources with large total integrated flux densities may have peak flux densities that fall below the detection threshold. To derive source counts which are complete in terms of total flux density the so-called resolution bias must be determined. 
We follow the formalism of  \cite{prandoni2001}  and \cite{huynh2005} in calculating the resolution bias. 

\begin{figure}
\includegraphics[width=0.95\columnwidth]{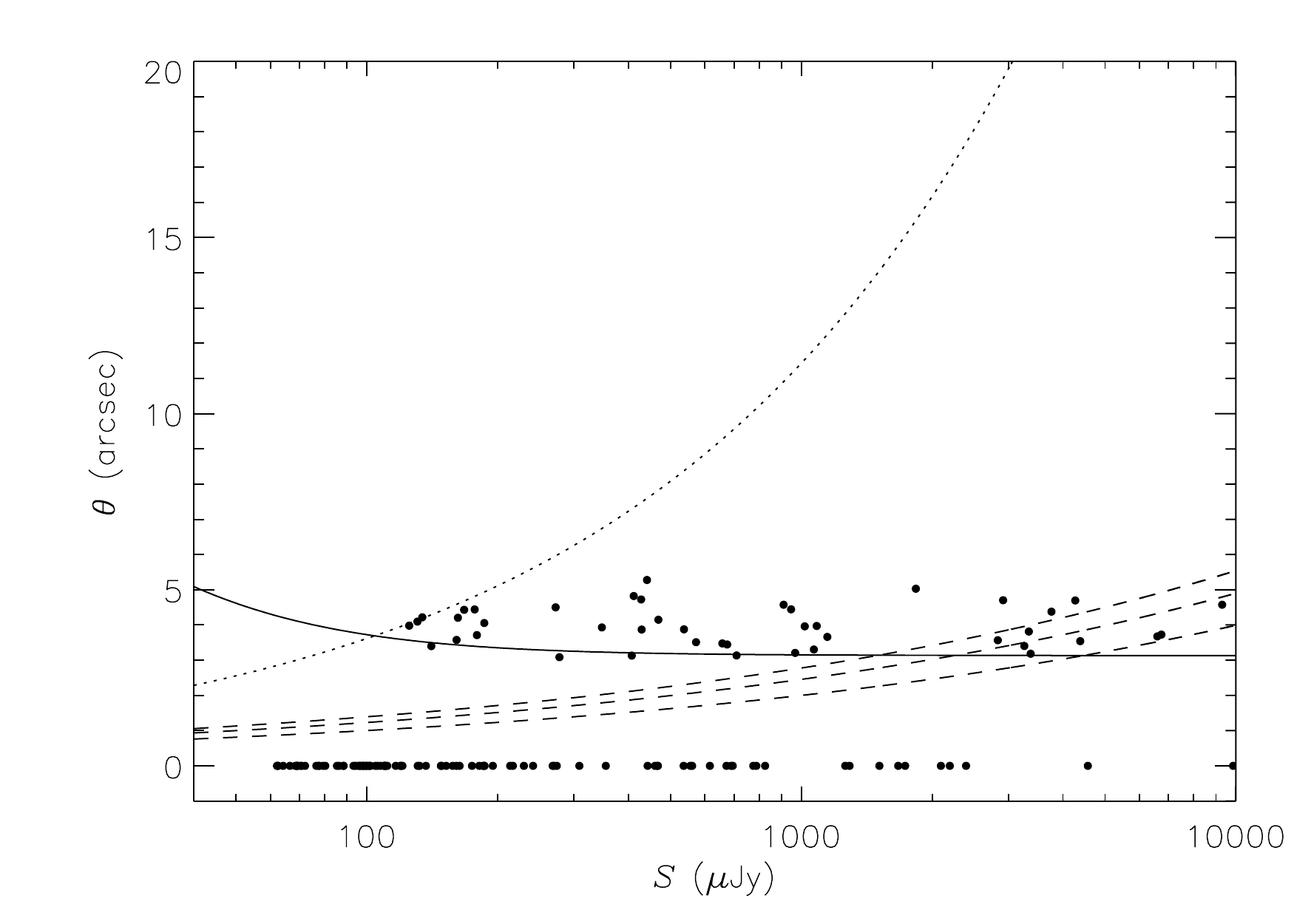}
\caption{The fitted angular size as a function of total flux density. Sources fitted or deconvolved as a point source are plotted with an angular size of zero. The solid line indicates the minimum angular size ($\theta_{\rm min}$) of sources in the survey, below which deconvolution is not considered meaningful. The dotted line shows the maximum angular size ($\theta_{\rm max}$) above which the survey becomes incomplete due to resolution bias. The dashed lines indicate the median source sizes expected from the Windhorst et al. 1990 relation, as a function of flux density, for a spectral index of 0, -0.5 and -0.8 between 1.4 and 5.5 GHz. }
\label{fig:maxmintheta}
\end{figure}

The maximum size ($\theta_{\rm max}$) a source of total flux density $S_{\rm tot}$ can have is determined by
\begin{equation}
 S_{\rm tot}/\sigma_{\rm det} = \theta_{\rm max}^2 / b_{\rm min} b_{\rm max}  
 \end{equation}
where $b_{\rm min}$ and $b_{\rm max}$ are the synthesized beam FWHM axes and $\sigma_{\rm det}$ is the detection limit. As the {\sc sfind} detection limit varies across the image, we use the 50\% completeness level, as determined by the simulations of Section \ref{sec:sims}.  The resulting $\theta_{\rm max}$ as a function of total flux density is plotted in Figure \ref{fig:maxmintheta} (dotted line). We plot the angular sizes ($\theta$) of the catalogued sources as a function of total flux density in Figure \ref{fig:maxmintheta} also. The angular size $\theta$ is defined as the geometric mean of the fitted Gaussian major and minor axes. We find that the largest catalogued sources are in good agreement with the $\theta_{\rm max}$ function.

The minimum angular size ($\theta_{\rm min}$) is estimated from Equation 2, with $\sigma$ equal to the noise of the full image (12 $\mu$Jy). The resulting $\theta_{\rm min}$ as a function of total flux density is plotted in Figure \ref{fig:maxmintheta} (solid line). This $\theta_{\rm min}$ constraint is important at low flux density levels, where $\theta_{\rm max}$ becomes unphysical (smaller than a point source). Overplotted in Figure \ref{fig:maxmintheta} (dashed lines) is the expected median angular size obtained from Windhorst et al. 1990 relations for a 1.4 GHz sample: 
\[ \theta_{\rm med} = 2\arcsec S_{\rm 1.4 GHz}^{0.30}\]  where $S_{\rm 1.4 GHz}$ is in mJy. The extrapolation to 5.5 GHz was done assuming a spectral index of 0, -0.5 and -0.8 between 1.4 and 5.5 GHz. The integral angular size distribution, $h(\theta)$, is \citep{Windhorst1990}: \[ h(\theta) = \exp ^{-\ln 2 \, (\theta/\theta_{\rm med})^{0.62}}\;.\] At the bright end our source sizes are consistent with the \cite{Windhorst1990} relation, however most of the sources are unresolved and therefore we can not draw any conclusions about the full sample. 

\begin{figure*}
\includegraphics[width=0.9\columnwidth]{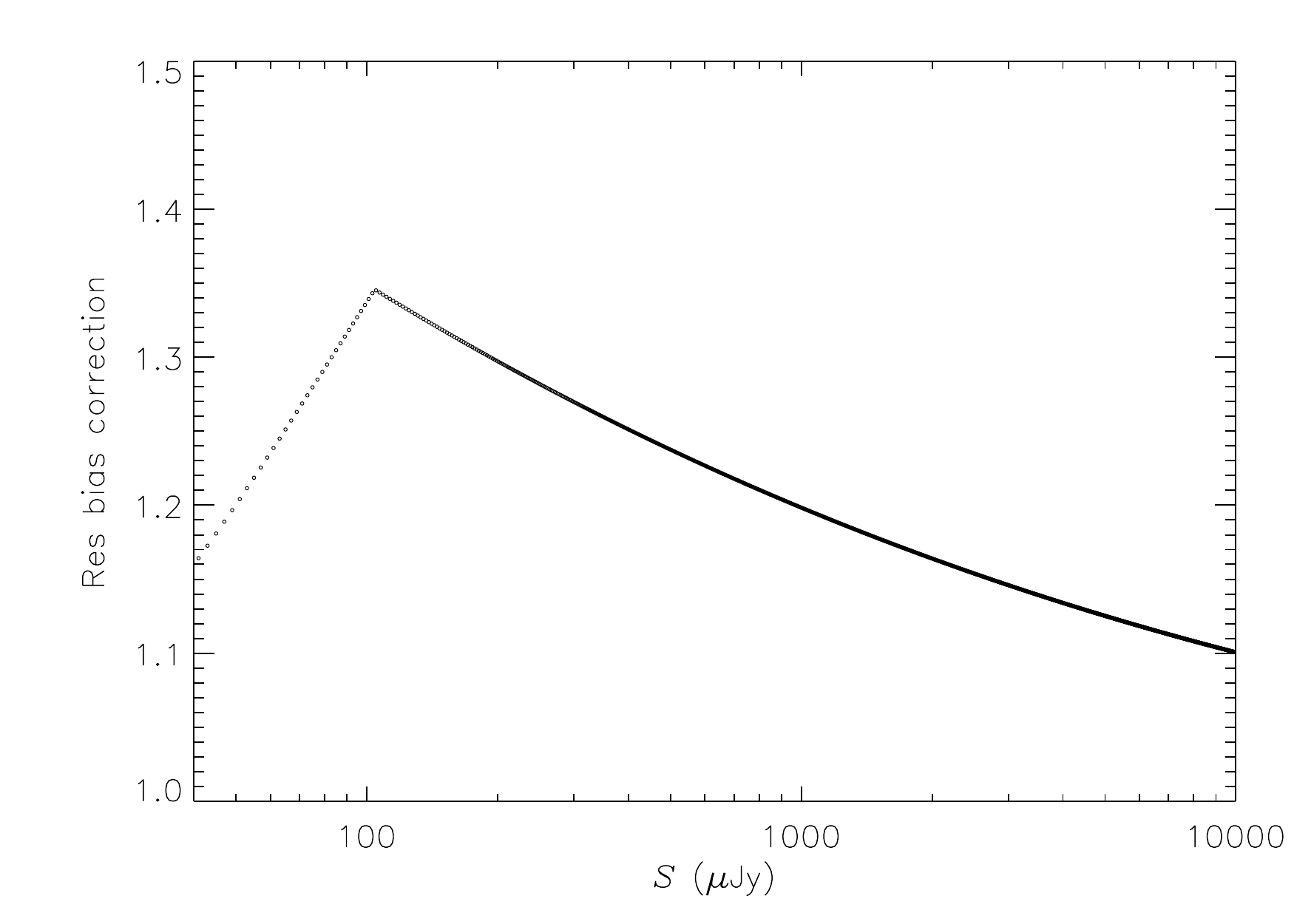}
\hspace{5mm}
\includegraphics[width=0.85\columnwidth]{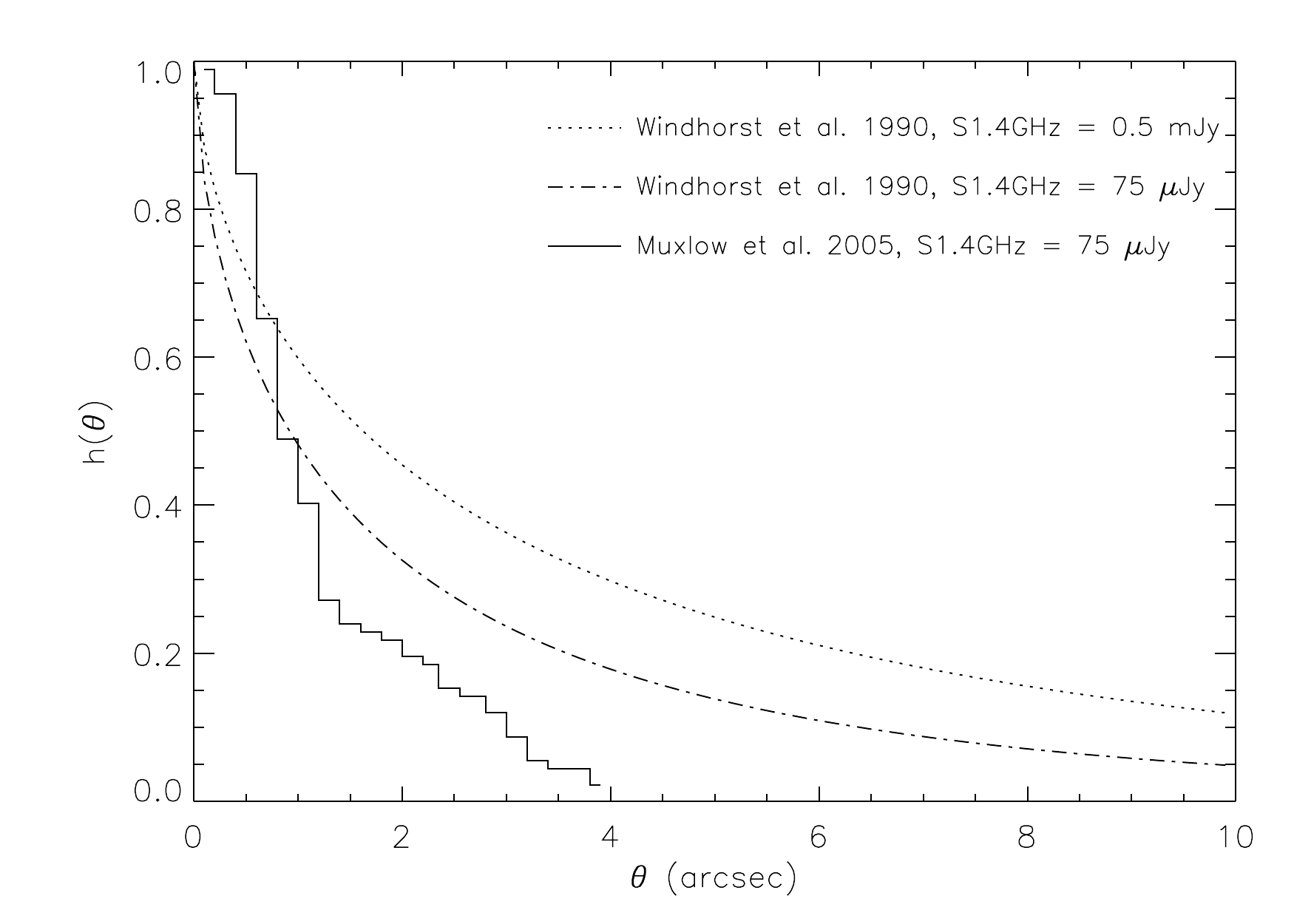}
\caption{LEFT: The resolution bias correction as a function of flux density, assuming the Windhorst et al. 1990 integral angular source size distribution. RIGHT: The integral angular size distribution from the Windhorst et al. 1990 relation (for flux densities as shown) and that observed in the Muxlow et al. 2005 sample. The Muxlow et al. 2005 sample has a median flux density of 75 $\mu$Jy.}
\label{fig:rescorr}
\end{figure*}

As outlined in \cite{prandoni2001}, the overall angular size limit, $\theta_{\rm lim} = {\rm max}(\theta_{\rm max},\theta_{\rm min})$ and the expected integral size distribution, $h(\theta)$, allows an estimation of the fraction of sources larger than the maximum detectable size, and hence missed by the survey. The resolution bias correction factor is then simply $\frac{1} {1 - h(\theta)}\;.$
This correction factor is plotted in Figure \ref{fig:rescorr} (left). It has a maximum of about 1.35 at a flux density of 100 $\mu$Jy, where the limiting overall angular size, $\theta_{\rm lim}$, becomes dominated by $\theta_{\rm min}$. \cite{Windhorst1990} estimate the uncertainty in the resolution bias correction to be about 10\%. The caveat to the resolution bias correction is that the integral angular size distribution at these low flux densities is not well known. High resolution MERLIN and VLA observations of 92 radio sources with $S_{1.4 GHz} > 40 \mu$Jy  \citep{muxlow2005} hint that the faint radio population may have an integral angular size distribution different to the brighter sample. The integral angular size distribution for the \cite{muxlow2005} sample differs from the \cite{Windhorst1990} relation by up to 20\%(Figure  \ref{fig:rescorr}, right). High resolution observations are likely to miss low surface brightness galaxies, but \cite{muxlow2005} estimate that they have missed only 10\% of the faint radio sources in their field.  If the \cite{muxlow2005} sizes are used the resolution correction factor near 100 $\mu$Jy would be approximately 5\%, and zero elsewhere. 

\section{Source Counts}

The differential radio source counts were constructed from the 6cm eCDFS catalogue of Section 3. In computing the source counts we used integrated flux densities for extended sources and the point source fit for all other sources.  The components of multiple sources were summed and counted as a single source. For comparison with with other 6cm studies, the source counts are normalised to a non-evolving Euclidean model. At 6cm the standard Euclidean integral counts are $N(> S_{\rm 6cm}) = 60 \times S_{\rm 6cm}^{-1.5} \; {\rm sr}^{-1}$, where $S_{\rm 6cm}$ is in Jy \citep{fomalont1991,ciliegi2003}.  The results are summarised in Table \ref{tab:srccount}, where for each bin we report the flux density interval, median flux density, the number of sources detected ($N$), the number of sources after completeness and resolution bias corrections have been applied ($N_C$), the differential source count  ($dN_{C}/dS$), and the normalised counts ($N_C/N_{\rm exp}$). The counts are normalised to $N_{\rm exp}$, the number expected in the bin from the standard Euclidean count. The Poissonian errors in the count are  $C N^{1/2}/N_{\rm exp}$, where $C$ is the total correction factor, $N_C/N$. The estimated total error in the counts is the Poissonian error with the resolution bias uncertainty (10\%) and completeness correction uncertainties (2 -- 4 \%) added in quadrature.

\begin{figure*}
\includegraphics[width=0.9\textwidth]{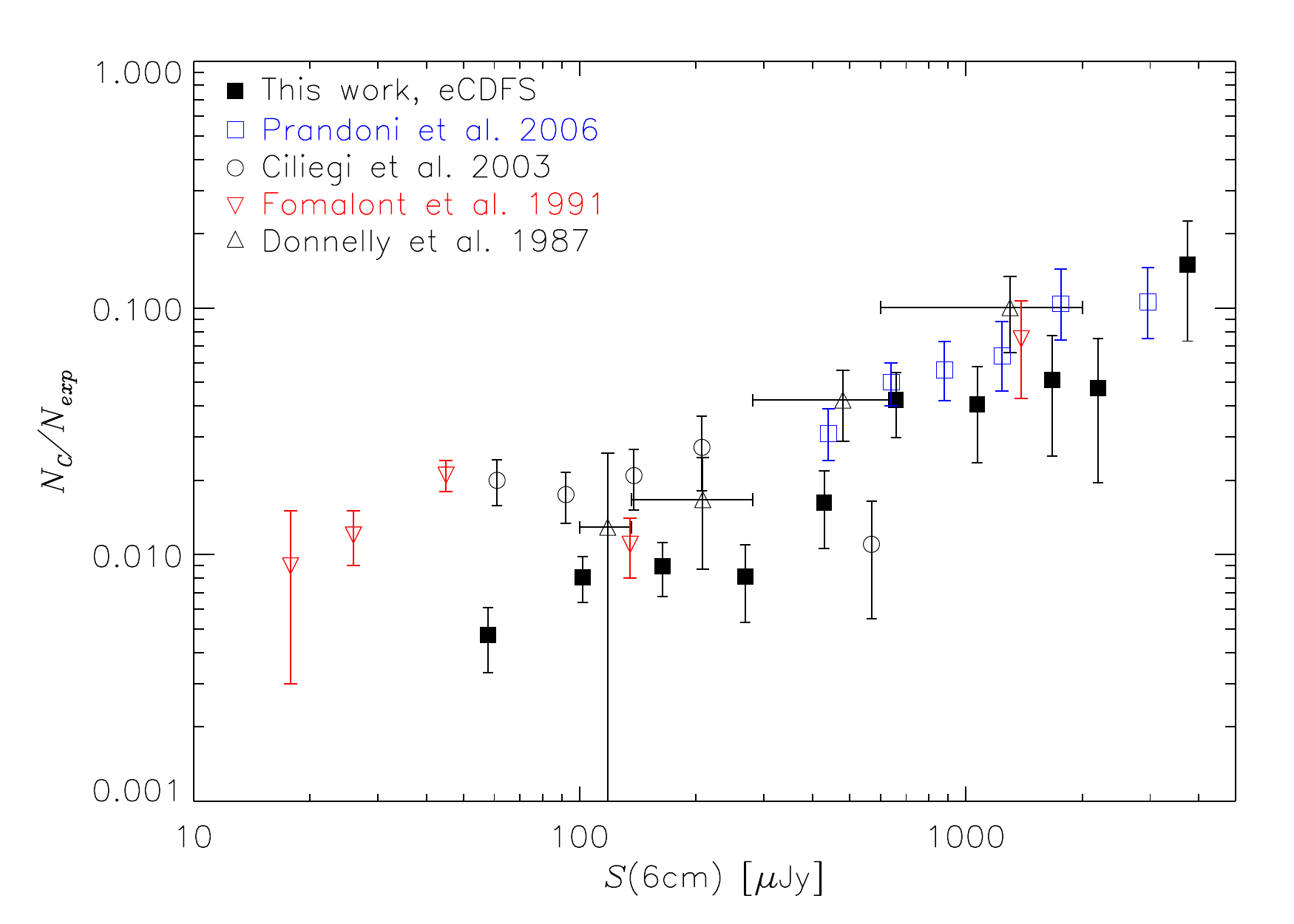}
\caption{Normalized 5.5 GHz differential source counts for different samples: Prandoni et al. (2006) (empty blue squares); Ciliegi et al. (2003) (empty circles); Fomalont et al. (1991) (red upside-down triangles); Donnelly et al. (1987) (empty triangles). The eCDFS 5.5 GHz source counts presented in this work (filled squares) are corrected for completeness and resolution bias as explained in the text. Vertical bars represent Poisson errors on the normalized counts.}
\label{fig:srccount}
\end{figure*}

Our results are compared with previous work in Figure \ref{fig:srccount}. In this Figure we only show source counts below 4 mJy. The counts are well known above this flux density \citep{altschuler1986, wrobel1990, prandoni2006}: there is an initial steep rise between 10 and 1Jy, an excess with respect to the Euclidean prediction between 1 Jy and 0.1 Jy and a steep slope in the counts for the flux density range 100 to 0.5 mJy. Some authors claim there is a flattening of the 6cm source counts at mJy levels (e.g. \citealp{ciliegi2003}), while others do not see the flattening down to 0.5 mJy \citep{prandoni2006}. 

Our source counts are consistent with other published 6cm source counts for $S_{\rm 6cm} > 0.3$ mJy. At fainter flux densities our source counts are significantly lower than \cite{ciliegi2003} counts by a factor of about two. Our counts are consistent with \cite{donnelly1987} and Fomalont et al. 1991, however the counts in the faintest bins ($S < 50\mu$Jy) from Fomalont et al. 1991 are about a factor of two higher than our counts in the lowest bin. Fomalont et al. 1991 catalogued sources to about 4$\sigma$ in their image, so it is likely that they have spurious sources in their faintest bins. We note that out survey area is almost 3 times greater than Ciliegi et al. 2003 (0.25 deg$^2$ versus 0.087 deg$^2$) and 5 times greater than Fomalont et al. 1991 (0.25 deg$^2$ versus 0.05 deg$^2$).
Using the \cite{driver2010} expression for quantifying cosmic variance, we find that our survey has a cosmic variance of about 35 to 90\% for bins of $dz = 0.1$ in size at $z < 1$. Similarly the Ciliegi et al. 2003 survey has a cosmic variance of 49 to 115\%. 
The difference in the counts at the faint end can therefore be attributed to cosmic variance.

Finally we note the 6cm surveys in the literature have a central frequency of 5 GHz and the difference of 0.5 GHz in the observing frequency will have an impact on the source counts. To estimate the counts at 5 GHz we applied spectral indices of 0, -0.4 and 0.8 to our catalogue and recalculated the source counts. We find that the source counts change by about 5\% -- 40\% but this does not account for the significantly lower counts seen in the faintest bins of our data compared to previous 6cm surveys. 

\begin{table}
\centering
\caption{The 5.5 GHz source counts.}
\begin{tabular}{ccrrcc} \hline
$\Delta S$ & $<$S$>$ & $N$ & $N_{C}$ & $dN_{C}/dS$ & $N_{C}/N_{exp}$ \\
($\mu$Jy) & ($\mu$Jy) & & &  (sr$^{-1}$ Jy$^{-1}$) &  ($\times 10^{-2}$)\\ \hline
50 -- 79 & 58 & 16 & 77.3 & $1.68  \times 10^{10}$ & 0.47 $\pm$ 0.14 \\
79 -- 126 & 102 & 30 & 51.0 & $6.98  \times 10^9$ & 0.81 $\pm$ 0.17 \\
126 -- 199 & 164 & 20 & 27.4 & $2.36  \times 10^9$ & 0.90 $\pm$ 0.22 \\
199 -- 315 & 268 & 9 & 11.4 & $6.23  \times 10^8$ & 0.81 $\pm$ 0.28 \\
315 -- 500 & 429 & 9 & 11.1 & $3.83  \times 10^8$ & 1.62 $\pm$ 0.56 \\
500 -- 792 & 658 & 13 & 15.8 & $3.43  \times 10^8$ & 4.24 $\pm$ 1.25 \\
792 -- 1256 & 1069 & 6 & 7.15 & $9.78 \times 10^7$ & 4.06 $\pm$  1.71 \\
1256 -- 1991 & 1671 & 4 & 4.67 & $4.03 \times 10^7$ & 5.11 $\pm$ 2.61 \\
1991 -- 3155 & 2196 & 3 & 3.46 & $1.88 \times 10^7$ & 4.73 $\pm$ 2.77 \\
3155 -- 5000 & 3755 & 4 & 4.54 & $1.56 \times 10^7$ & 14.97 $\pm$ 7.63 \\ \hline
\end{tabular}
\label{tab:srccount}
\end{table}

\section{Spectral Index Analysis}

\subsection{1.4 to 5.5 GHz Spectral Indices}

To investigate the spectral index properties of faint radio population we matched the 5.5 GHz catalogue to sources in the Very Large Array (VLA) 1.4 GHz survey of the extended Chandra Deep Field South \citep{miller2008}. This survey is a good match in the area covered, as seen in Figure 1. The depth of the \citep{miller2008} survey is about 8 $\mu$Jy rms, which should be sensitive enough to detect counterparts to all but the faintest 5.5 GHz sources with flat or inverted spectra. Importantly, the beam of the VLA observations is 2.8 $\times$ 1.6 $\arcsec$ beam, which is only a factor of $\sim$1.5 better than our observations. With similar resolutions these images have a similar surface brightness sensitivity, and thus the measured flux densities can be used directly for spectral index analyses. 

\begin{figure}
\includegraphics[width=0.95\columnwidth]{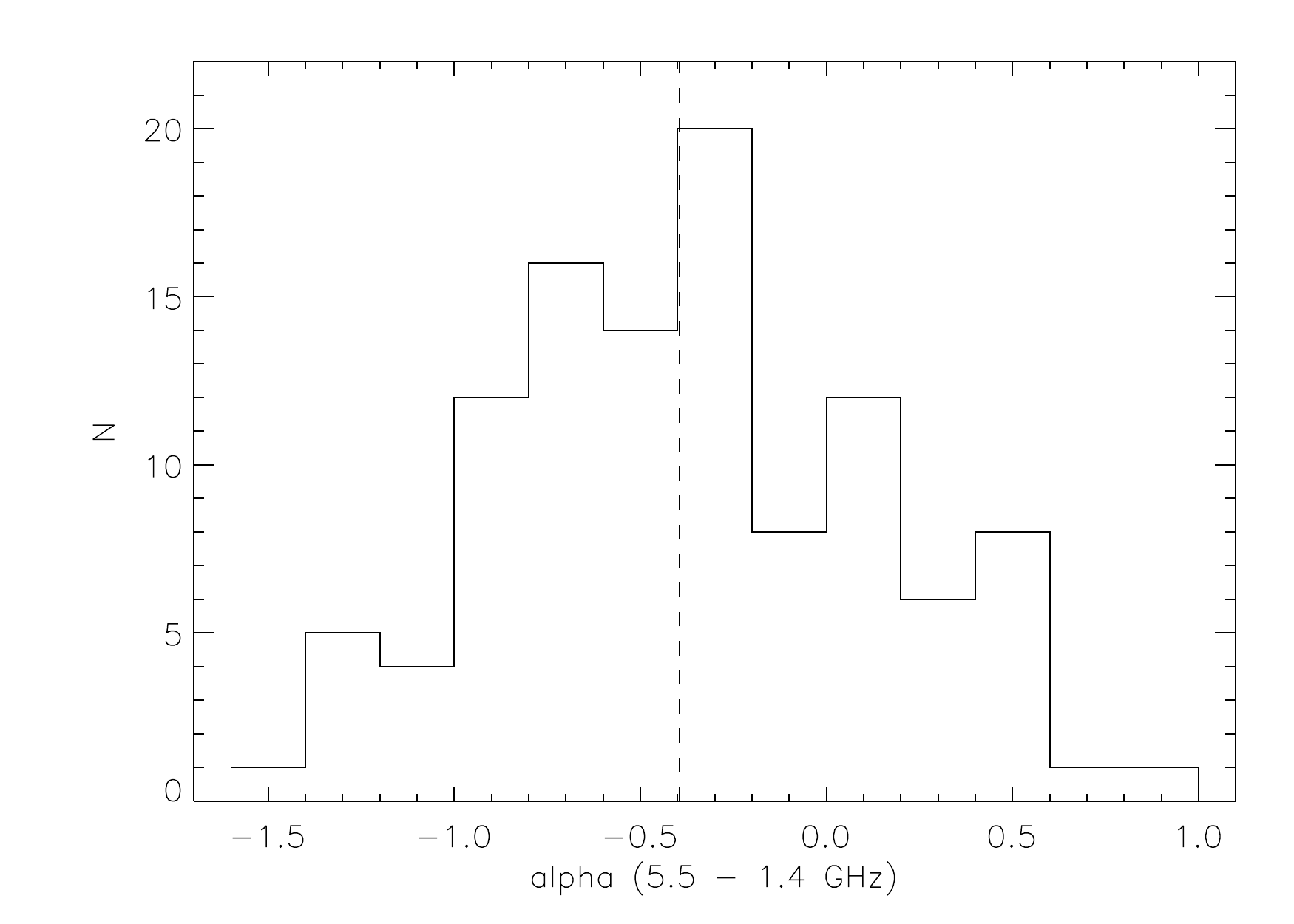}
\caption{Spectral index distribution for sources in the ATCA 6cm sample. The vertical dashed line indicates the median value of the sample ($\alpha_{med} = -0.40$).}
\label{fig:alphahist}
\end{figure}

\begin{figure*}
\includegraphics[width=0.95\columnwidth]{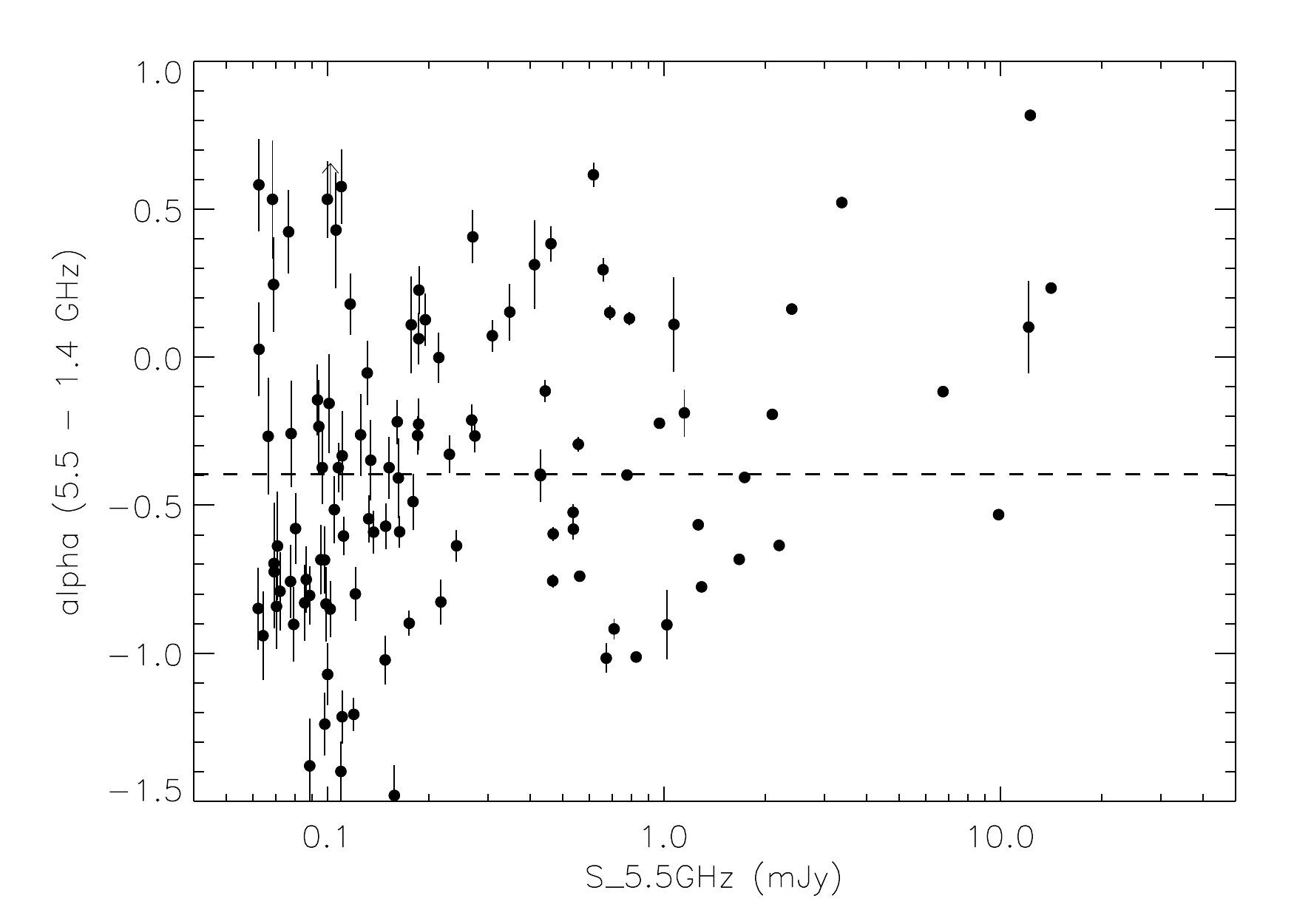}
\includegraphics[width=0.95\columnwidth]{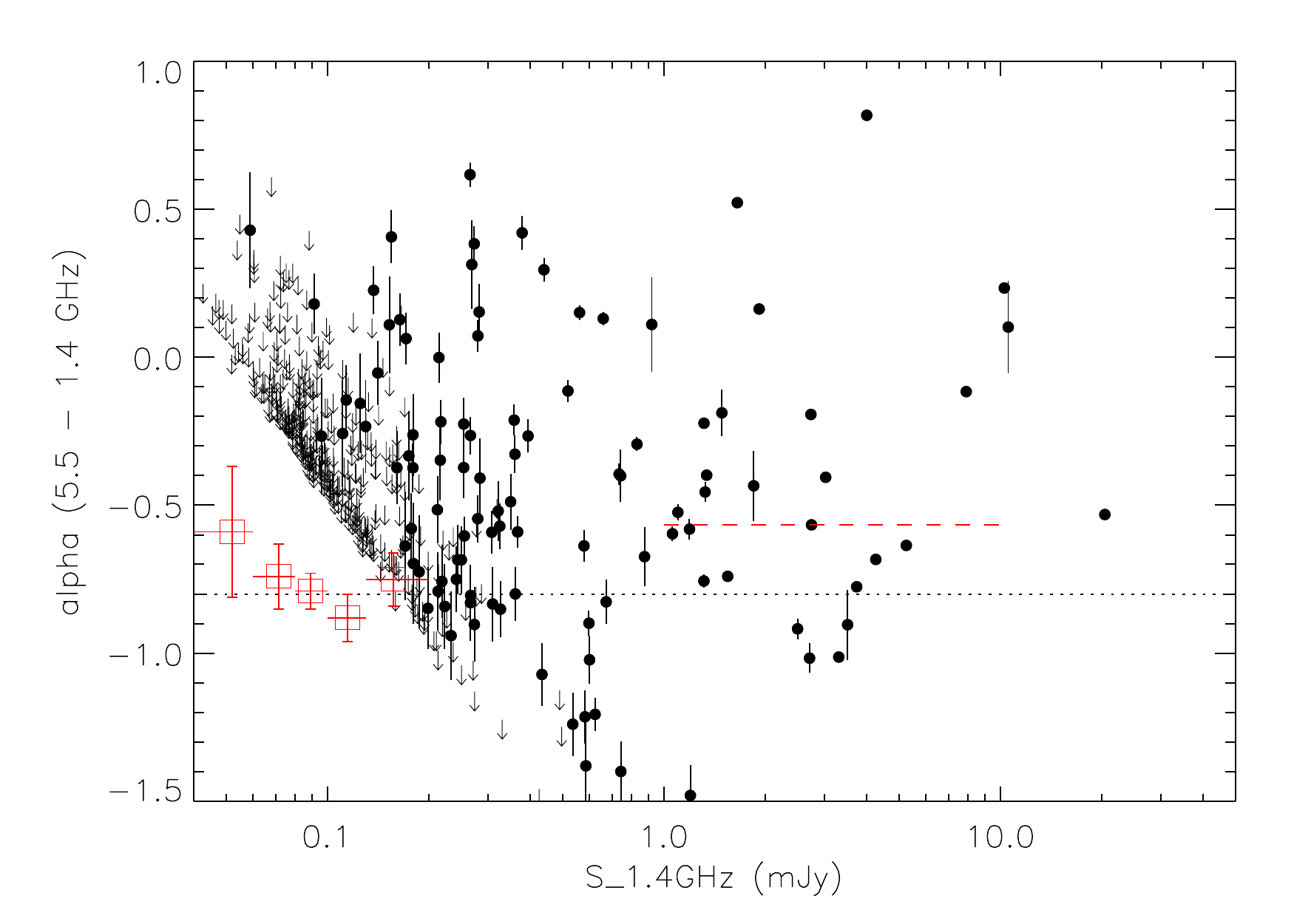}
\caption{LEFT: 1.4-5.5 GHz spectral index vs. 5.5 GHz flux density for the ATCA 5.5 GHz sample. Only sources catalogued at 5.5 GHz are shown. The dashed line indicates the median spectral index, $\alpha = -0.40$. RIGHT: 1.4-5.5 GHz spectral index vs. 1.4 GHz flux density for the \protect\cite{miller2008} 1.4 GHz sample. Only sources catalogued at 1.4 GHz are shown (black dots), and black arrows mark upper limits assuming a detection limit of 5$\sigma$ in the 5.5 GHz image. The average spectral indices from stacking analysis of the faint 1.4 GHz sources are shown as red squares. The dashed red line indicates the median spectral index of the $1 < S_{1.4 GHz} < 10$ mJy sample, $\alpha = -0.57$. The black dotted line denotes the canonical spectral index from synchrotron emission, $\alpha = -0.8$.}
\label{fig:alpha-s}
\end{figure*}

To begin the spectral index analysis we removed the multi-component sources from the analysis as their interpretation is complicated by the core-jet structure, leaving 113 individual 5.5 GHz sources for investigation. 101/113 sources have a 1.4 GHz match within 2 arcsec (FWHM of the synthesised beam of the VLA observations). The rest were inspected and we found four were out of the 1.4 GHz image area, and seven had a faint counterpart in the 1.4 GHz image that was below the Miller et al. (2008) catalogued threshold. The 1.4 flux density for these sources was measured manually with the MIRIAD task {\em imfit}. In summary 108/109 (99\%) of the 5.5 GHz sources in the 1.4 GHz image area have a 1.4 GHz counterpart, and hence a spectral index measurement.  
The source without a 1.4 GHz counterpart, ID82, has a $7\sigma$ limit of $S_{1.4 GHz} < 48\mu$Jy, and hence is inverted with $\alpha > 0.55$.
The median spectral index for the ATCA 5.5 GHz sample is $\alpha_{\rm med} = -0.40$ (see Figure \ref{fig:alphahist}), which is consistent with that of the deepest 6cm sample to date ($\alpha_{\rm med} = -0.38$ for $S_{5.5 GHz} > 16 \mu$Jy, Fomalont et al. 1991). Figure \ref{fig:alpha-s} (left) shows the spectral index as a function of 5.5 GHz flux density.  The median spectral index is $\alpha_{\rm med} = -0.35$ for $0.1 < S_{5.5 GHz} < 1$ mJy. This is consistent with published values for 6cm selected sources of similar flux density, for example Prandoni et al. 2006 who found $\alpha_{\rm med}=-0.4$ for $0.4 < S_{5GHz} < 4$ mJy and Donnelly et al. 1987 who found $\alpha_{\rm med} =-0.42$ for $0.4 < S_{5GHz} < 1.2$ mJy.

It has been claimed that spectral indices flatten from bright (mJy) to fainter (sub-mJy) flux densities. For example, Ciliegi et al. 2003 observed a flattening of the radio spectral indices in 6cm sources with $\alpha_{\rm med}=-0.37$ for $0.1 < S_{5GHz} < 0.2$ mJy and $\alpha_{\rm med} =-0.81$ for $S_{5GHz} > 0.2$ mJy. Our median spectral index for $0.1 < S_{5.5GHz} < 0.2$ mJy is $-0.37$, which is consistent with this result, however we do not have enough mJy sources to confirm the spectral flattening from the mJy to sub-mJy regime.
 \cite{owen2009} also observed the spectral flattening from mJy to sub-mJy flux densities, albeit between 325 MHz and 1.4 GHz, but they found that the radio spectra steepened again at the faintest flux densities ($S_{1.4GHz} < 0.1$ mJy). At the faintest flux density levels in our sample, $S_{5.5 GHz} < 0.1$ mJy, the median spectral index is $\alpha_{\rm med} = -0.68$ showing a trend to steeper spectra at the faintest flux densities. This is a direct confirmation of what \cite{owen2009} concluded through stacking analysis.

The fraction of flat or inverted sources  ($\alpha > -0.5$) is 55\% for the whole sample, consistent with the findings of a high proportion (50 -- 60\%) of flat or inverted sources in $\mu$Jy samples selected at similar frequencies \citep{fomalont1991, windhorst1993, prandoni2006}, but the fraction of flat or inverted sources drops to 11/30 (36\%) for $S_{5.5 GHz} < 0.1$ mJy.  The synchrotron-like median spectral index for $S_{5.5 GHz} < 0.1$ mJy is consistent with star-forming galaxies dominating the population at these faintest flux density levels. 

We also examined the spectral indices of the 1.4 GHz selected sample. Of the 514 1.4 GHz sources in the \cite{miller2008} observations, 3 are not in our 5.5 GHz catalogued region. We found 36 1.4 GHz sources that were associated with a multiple 6cm source and these were also removed from further spectral index analysis. 106/475 (22\%) of 1.4 GHz sources have a 6cm counterpart within 2 arcsec (Figure \ref{fig:alpha-s}, right). The {\em sfind} $\alpha$ parameter we used is best approximated by 5$\sigma$ for the noise characteristics of our image. Upper limits to the spectral index assuming a 5$\sigma$ detection limit at 5.5 GHz are thus plotted in Figure \ref{fig:alpha-s} (right) to illustrate where the spectral index information becomes incomplete. Figure \ref{fig:alpha-s} (right) shows that the 1.4 GHz sources start to be missed in the 5.5 GHz image for $S_{1.4 GHz}  \lesssim 0.5$ mJy. 

To derive median spectral indices of the full 1.4 GHz sample at faint flux densities we performed a stacking analysis in bins of 1.4 GHz flux density. The 6cm data was stacked at the positions of 1.4 GHz sources not detected in the 6cm image. The effective pixel value in the stack was taken to be the median value of the stacked positions, and the flux density determined by a point source fit to the signal at the centre of the stack. Offset stacks were generated by randomly offsetting each stacked position by up to 8 arcsec. The uncertainty in the flux density of the median stack is then estimated as the standard deviation of the flux density measured in 300 random offset stacks. The uncertainty from the point source fit is then added in quadrature to this estimate to derive a total uncertainty in the median stacked flux density. The results are summarised in Table \ref{tab:stackres}. As a check of our stacking technique, we also derived the noise-weighted flux densities from the stacks and found agreement (to within 9\%) with the median stacked flux density.

\begin{table*}
\caption{Results of stacking 1.4 GHz positions in the 5.5 GHz image. $N_{det}$ is the number of 1.4 GHz sources in each bin which are detected in the 5.5 GHz image. $N_{stack}$ is the number of 1.4 GHz sources not detected in the radio image and therefore the number stacked in this analysis, resulting in the stacked $S_{5.5 GHz}$ listed in Column 5. Stacked $\alpha$ is the spectral index for the stacked sample. Average $\alpha$ is the spectral index for all sources in the flux density bin, obtained by combining the detections with the stacked results.}
\centering
\begin{tabular}{lllllll} \hline
$S_{1.4 GHz} $ & $N_{det}$ & $N_{stack}$ & mean $S_{1.4 GHz}$ & stacked $S_{5.5 GHz}$ & stacked $\alpha$ & average $\alpha$ \\
($\mu$Jy)  &  & & ($\mu$Jy) & ($\mu$Jy) & & \\ \hline
$40 < S_{1.4 GHz} < 60$  & 1 & 19 & 52.0 $\pm$ 2.8 &  21.6 $\pm$ 6.4 & $-0.64 \pm 0.22$ & $-0.59 \pm 0.22$ \\ 
$60 < S_{1.4 GHz} < 80$  & 0 & 83 & 71.4 $\pm$ 1.6 &  25.8 $\pm$ 3.7 & $-0.74 \pm 0.11$ & $-0.74 \pm 0.11$ \\ 
$80 < S_{1.4 GHz} < 100$  & 2 & 102 & 89.0 $\pm$ 1.5 &  29.5 $\pm$ 2.2 & $-0.80 \pm 0.06$ & $-0.79 \pm 0.06$ \\ 
$100 < S_{1.4 GHz} < 130$  & 4 & 76 & 114.4 $\pm$ 1.9 &  32.4 $\pm$ 3.7 & $-0.92 \pm 0.08$ & $-0.88 \pm 0.08$ \\ 
$130 < S_{1.4 GHz} < 200$  & 15 & 66 & 156.7 $\pm$ 2.1 &  48.3 $\pm$ 5.2 & $-0.86 \pm 0.08$ & $-0.75 \pm 0.09$ \\ 
\hline
\end{tabular}
\label{tab:stackres}
\end{table*}

The median spectral index for $1 < S_{1.4 GHz} < 10$ mJy is $\alpha = -0.57$, consistent with the Prandoni et al. (2006) 1.4 GHz sample.
The stacking results show that the spectral index of the faint ($40 < S_{1.4 GHz} < 200$ $\mu$Jy) sources range from $\alpha = -0.64 \pm 0.22$ to $-0.92 \pm 0.08$, consistent with synchrotron emission from star forming galaxies. Again, this is consistent with Owen et al. 2009 who find radio sources are steep at the faintest flux density levels ($S_{1.4 GHz} < 100$ $\mu$Jy).
We find that 17/22 (77\%) of faint ($S_{1.4 GHz} < 200$ $\mu$Jy) sources {\it detected} at both frequencies are flat or inverted ($\alpha > -0.5$), but stacking analysis of the full sample of faint $\mu$Jy level 1.4 GHz sources shows they are steep on average and flat or inverted sources do not dominate at the faint flux density levels. 

Finally we note that the differences between VLA and ATCA flux densities (Section 3.3) implies a possible systematic effect on the measured spectral indices. If the ATCA 5.5 GHz flux densities are actually over-estimated then the true spectral indices are steeper by 0.13.

\subsection{Optical counterparts and redshifts}

Optical counterparts for our 5.5\,GHz sample were identified from the COMBO-17 (Wolf et al. 2004) and Multiwavelength Survey by Yale-Chile (MUSYC, \citealp{taylor2009})  imaging surveys of the eCDFS. Both surveys are about $30\arcmin \times 30\arcmin$ in size and hence well matched in area to our 5.5 GHz survey.
Spectroscopic redshifts are available for many sources in the eCDFS (e.g. \citealp{vanzella2008, balestra2010}; Mao et al.\ in prep) and we searched these catalogues for redshifts of the radio sources, adopting photometric redshifts only when spectroscopic redshifts are not available. We prioritize MUSYC photometric redshifts over COMBO-17,
due to the larger filter set used.
As in Section 5.1, we exclude the multiple sources from this analysis. We find 104/113 sources lie within MUSYC or COMBO-17 imaging, of which 97/104 (92\%) have counterparts within 2 arcsec ($\sim 0.67$ FWHM of the synthesized beam). Spectroscopic redshifts are available for 45 sources.
We adopt MUSYC photometric redshifts for 47 sources and COMBO-17 photometric redshifts for 5.

\begin{figure*}
\includegraphics[width=0.95\columnwidth]{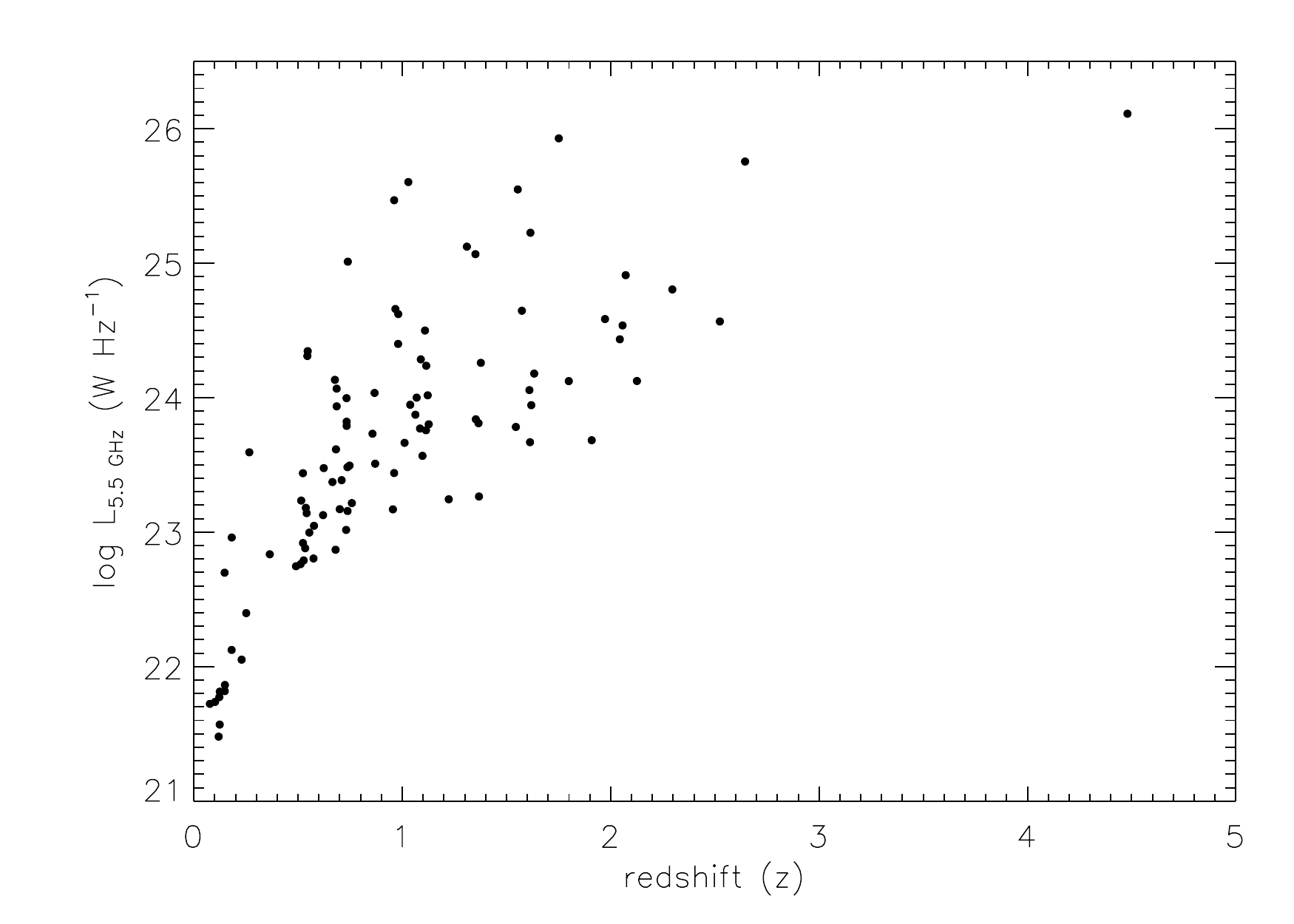}
\includegraphics[width=0.95\columnwidth]{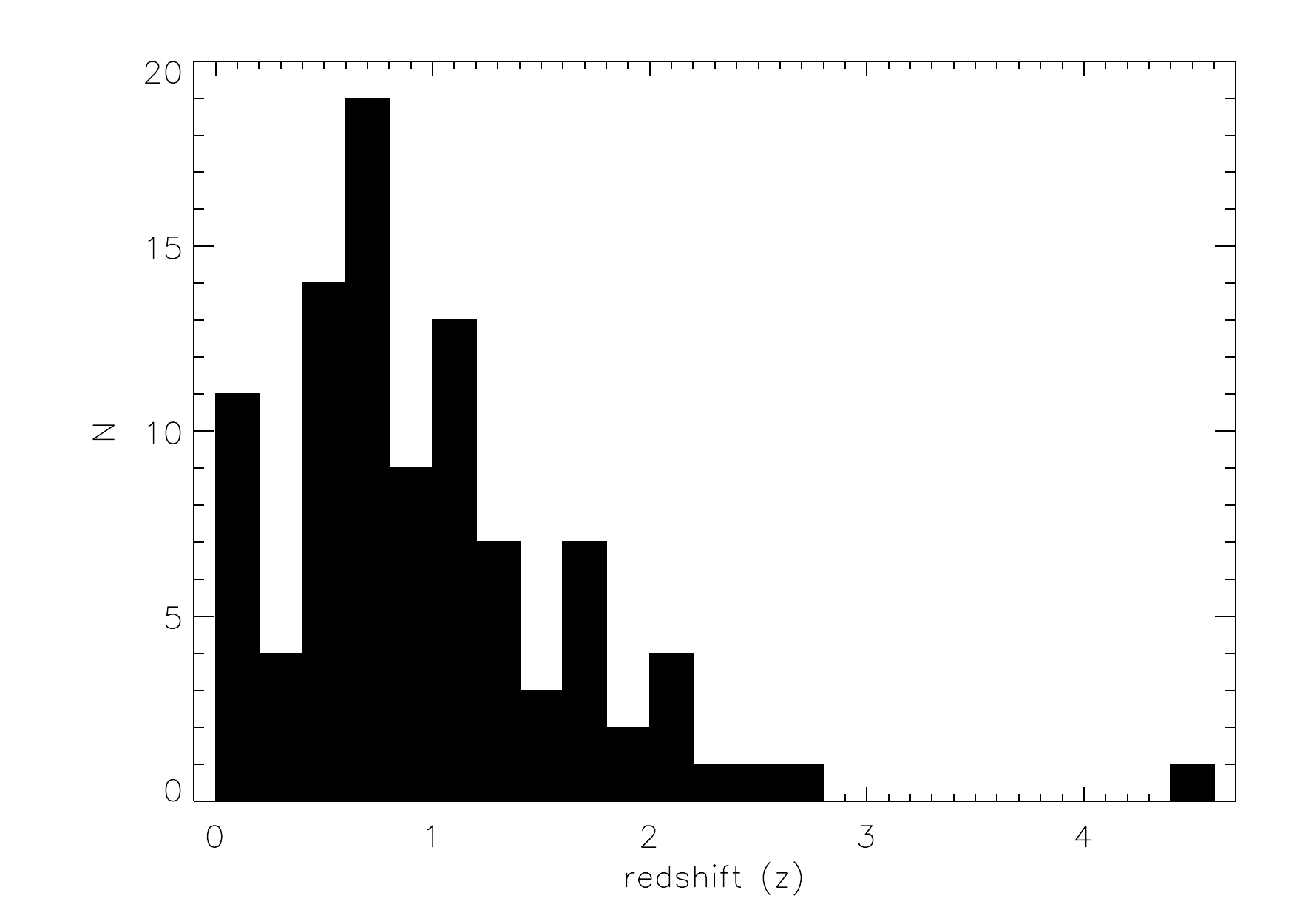}
\caption{LEFT: The 5.5 GHz radio luminosity as a function of redshift.  These are k-corrected assuming the spectral indices measured between 1.4 and 5.5 GHz. RIGHT: The redshift distribution of the radio sources.}
\label{fig:lum55redshift}
\end{figure*}

The radio luminosities and redshift distribution for these sources are shown in Figure \ref{fig:lum55redshift}.
The median redshift of the 5.5 GHz sources is $z=0.87$, and the distribution has a tail to high redshifts. The survey is sensitive to sources with luminosities as faint as $L_{5.5 GHz} $ = 10$^{23}$  W Hz$^{-1}$ up to $z \sim 1$,
indicative of low powered AGN or powerful starbursts. At higher redshifts the sources have AGN-like radio powers.

\subsection{The z-$\alpha$ Relation and Ultra-Steep Spectrum Sources}

Here we further explore the spectral index properties of our sample, with a focus on the extent to which
the observed correlation of radio spectral index with redshift, the $z-\alpha$ relation, persists for the faint
flux-density population.
The $z-\alpha$ relation is one of the most successful tracers of high redshift radio galaxies (HzRGS)
\citep{tielens1979, blundell1999, klamer2006, ker2012}. Although an ultra-steep spectrum (USS, $\alpha < -1$) doesn't guarantee a high redshift  radio source, a higher fraction of high redshift sources can be found amongst USS sources \citep{rottgering1997, debreuck2000, debreuck2001}. Until recently, USS samples have been limited to $S_{1.4 GHz} > 10$ mJy since tens or hundreds of square degrees are required to identify a meaningful sample. A fainter ($S_{610 MHz} > 100$ $\mu$Jy) population of  58 USS sources was found in the Lockman Hole. Optical/IR counterparts for $\sim$60\% of this sample suggests they have redshifts spanning $0.1 < z < 2.8$, and the non-detection of the other $\sim$40\% suggest they lie at even higher redshift \citep{afonso2011}.

\begin{figure*}
\includegraphics[width=0.95\columnwidth]{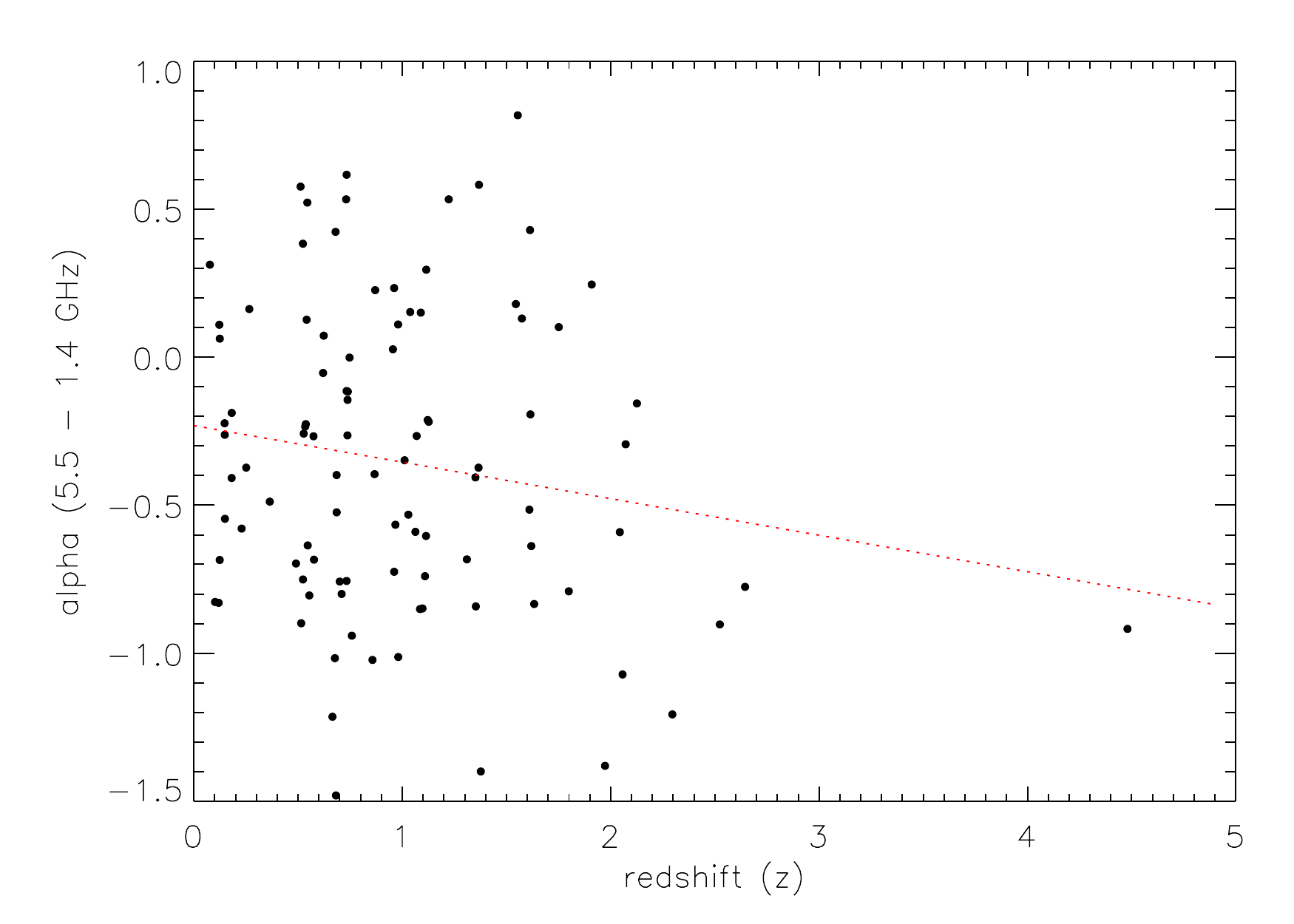}
\includegraphics[width=0.95\columnwidth]{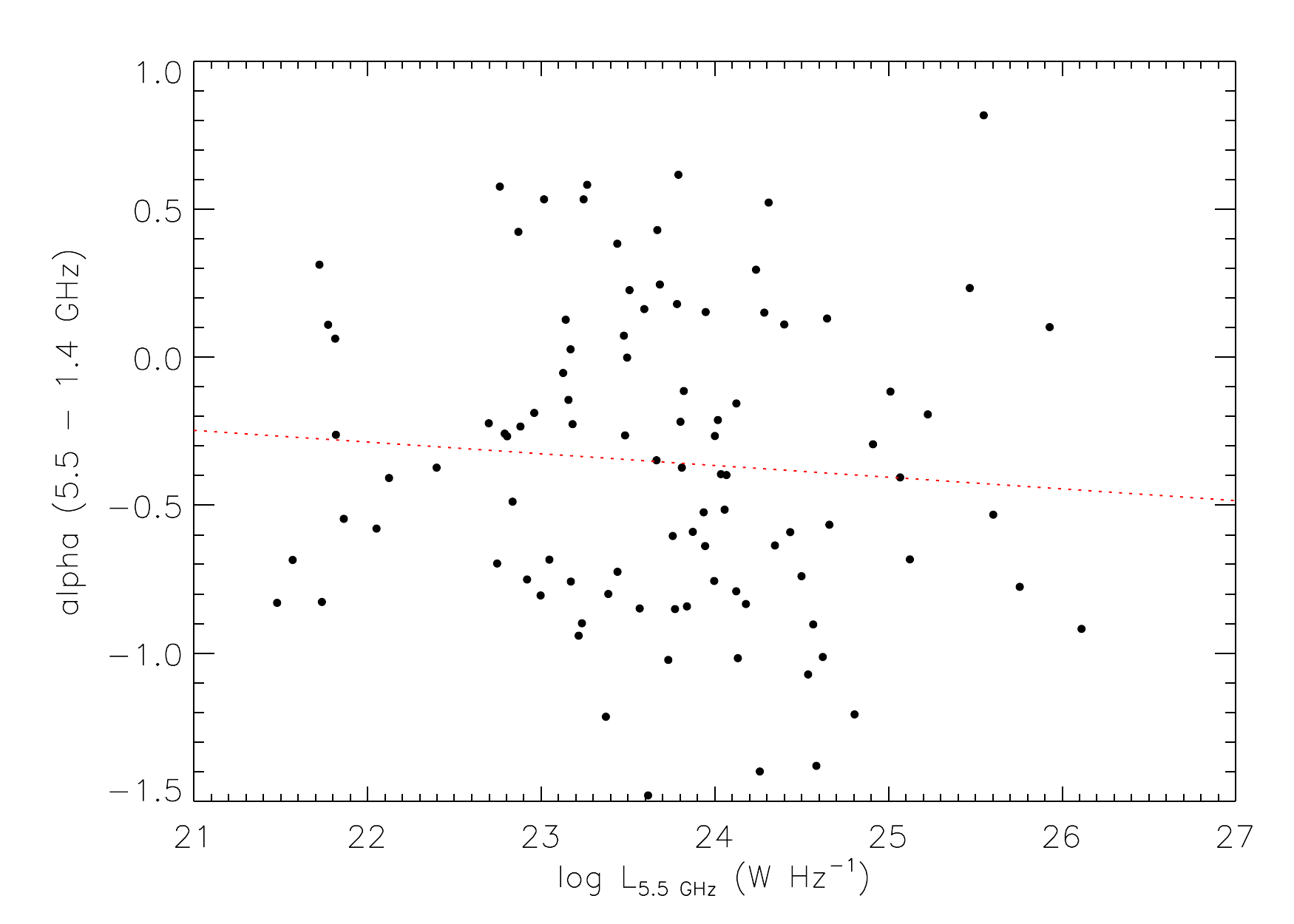}
\caption{LEFT: The spectral index, $\alpha$, as a function of redshift. There is a weak trend of decreasing $\alpha$ with increasing redshift. The dotted line shows the linear best fit.  RIGHT: The spectral index, $\alpha$, as a function of 5.5 GHz radio luminosity. The dotted line shows the linear best fit, but there is no trend of $\alpha$ with radio luminosity.}
\label{fig:alphaz}
\end{figure*}

Figure \ref{fig:alphaz} (left) shows $\alpha$ versus redshift. While a linear fit shows only a weak correlation
(a correlation coefficient of $r = -0.17$, which only becomes weaker if the highest redshift
data point is excluded, $r = -0.13$), there is, however, a clear excess of
steeper spectra at $z>2$. While there are very few systems at $z>2$, reflecting the sparseness of HzRGs on the
sky and the difficulty in identifying them with small area surveys such as this, the fact that 5/8 of the sources
at $z>2$ have $\alpha<-0.7$ is suggestive that the
$z-\alpha$ relation persists even to the faint flux-densities measurable here.
To ensure that this result is not simply a consequence of Malmquist bias, Figure \ref{fig:alphaz} (right) shows
that there is essentially no correlation between luminosity and spectral index (a correlation coefficient of $r = -0.08$),
demonstrating that we are not biased against flat or inverted spectrum systems at high-$z$.

Our results are in agreement with those for much brighter samples \citep{ker2012}. Also consistent with the Lockman Hole
results from \citet{afonso2011}, our steepest spectrum sources have redshifts spanning $0.5 \lesssim z \lesssim 2.7$.
We do, however, have a much higher completeness of optical counterparts for this population ($\sim 80\%$
compared to $\sim 60\%$), even when including a further six 1.4 GHz sources not detected at 5.5 GHz that
have $\alpha < -1.0$. This suggests that, consistent with \cite{afonso2011}, the USS technique for selecting high redshift radio galaxies remains efficient to sub-mJy levels, although the small areas currently accessible
with deep surveys will limit the numbers of such rare high redshift objects able to be identified.

\subsection{Candidate Young AGN Identified Through Spectral Indices}
\label{gps}

One approach to identifying candidate young AGNs is through their size and spectral index properties.
Radio sources with compact sizes and steep spectra (Compact Steep Spectrum, CSS) or peaked spectra
(Gigahertz Peaked Spectrum, GPS), are potential candidates for an early stage of the AGN evolutionary process
\citep{fanti1995, tinti2006}. At bright radio flux-densities such sources have sizes comparable with
or smaller than the optical host galaxy \citep{odea1998}, but still show double-lobed structures
on small scales ($\sim 1-10\,$kpc) \citep{odea1998, snellen1998}.
An alternative hypothesis to explain their small sizes is that, rather than being young,
they are `frustrated' sources, being confined by a dense medium \citep{odea1991}. 
CSS and GPS sources are relatively common in bright source samples, accounting for up to 30\% of sources selected at 2.7 and 5.0 GHz (e.g. \citealp{kapahi1981, peacock1982}). A recent unbiased sample selected at 2.7 GHz
finds more than 10\% of bright (Jy) sources are CSS or GPS sources \citep{randall2011}.
Existing CSS and GPS samples are largely heterogeneous, however, and there are few samples at faint
flux densities \citep[mJy versus Jy level, e.g.][]{randall2012}.

In total, we identify five GPS candidates that are consistent with being either a young AGN or an AGN confined
by a dense medium, based on their size and spectral shape (full details are given in Appendix~\ref{app}).
More complete radio spectral information is required to obtain a complete sample of GPS sources in this field, but
our preliminary analysis shows at least 5/40 sources with $S_{5.5 GHz} > 0.5$ mJy meet the GPS criteria.
This GPS fraction is similar to that found in Jy level samples \citep{randall2011}, suggesting GPS sources remain
an important population to mJy levels.

\section{Concluding Remarks}

We have presented new observations at 5.5 GHz of the extended Chandra Deep Field South. The 0.25 deg$^2$ region was observed with the Australia Telescope Compact Array using a mosaic of 42 pointings. The resultant image reaches an almost uniform noise level of $\sim$12 $\mu$Jy rms  and has a resolving beam of 4.9 $\times$ 2.0 arcsec. Using a false-discovery-rate method, we extracted 123 sources or 142 source components. Ten sources were resolved multiple sources with jet-lobe structure and hence fitted as multiple components.

After carefully correcting for completeness, flux boosting and resolution bias, we derived source counts at 5.5 GHz. These are amongst the deepest source counts at 6cm but from an area 3 to 5 times larger than the previous counts to these depths. The ATLAS 5.5 GHz counts are consistent with the counts derived from other 5 GHz surveys at brighter flux densities, but are lower than counts in the literature by a factor of two for $S_{5.5GHz} < 0.3$ mJy. This discrepancy is attributed to cosmic variance because of the small size of the surveys involved. This fluctuation in the 5.5 GHz source counts at the faint end is similar to that seen at 1.4 GHz for $S_{1.4GHz} < 0.1$ mJy \citep{norris2011}

The 1.4 -- 5.5 GHz spectral index has also been determined for all the sources in this field. 
We find a median spectral index for the ATCA 5.5 GHz sample of $\alpha_{\rm med} = -0.40$.
We find the median spectral index for the faintest flux density levels in our sample, $S_{5.5 GHz} < 0.1$ mJy, is $\alpha_{\rm med} = -0.68$. 
The 1.4 GHz sources start to be missed in the 5.5 GHz image for $S_{1.4 GHz} < 0.5$ mJy so we performed a stacking analysis to examine the average spectral index of the full faint 1.4 GHz selected sample.
From the stacking analysis we find that the faintest 1.4 GHz sources  ($40 < S_{1.4 GHz} < 200$ $\mu$Jy)  also have steep spectra.
This is consistent with the results of \cite{owen2009} who found that radio spectra steepens at the faintest flux density levels ($S_{1.4 GHz} < 100$ $\mu$Jy), after flattening from mJy to sub-mJy flux densities. 
 
The 5.5 GHz sources have been cross-matched to existing optical/NIR data to obtain redshifts and we find,
consistent with earlier work, that the $z-\alpha$ relation seems to persist to these low flux densities.
In addition, several candidate young AGN have been identified using the peak in the spectral energy
distribution between 1.4 GHz and 6 GHz, suggesting that GPS sources are
as common in the mJy population as they are at Jy levels.
 
In future work we will explore the spectral index properties of the ATLAS sources against other attributes, such as polarization and X-ray hardness, with the aim of distinguishing low-luminosity AGN from starforming galaxies in the sub-mJy population and determining how the observed properties of the sub-mJy population are linked to these two physical emission processes. 

\section*{Acknowledgements}

We wish to thank the anonynmous referee for comments which improved this paper. We also thank Christopher Hales for assisting with the C2028 ATCA observations of this project. ATCA is part of the Australia Telescope which is funded by the Commonwealth of Australia for operation as a National Facility managed by CSIRO.

\bibliographystyle{mn2e}
\bibliography{refs}

\begin{appendix}

\section{The GPS Candidates}
\label{app}

\begin{figure*}
\includegraphics[width=0.9\textwidth]{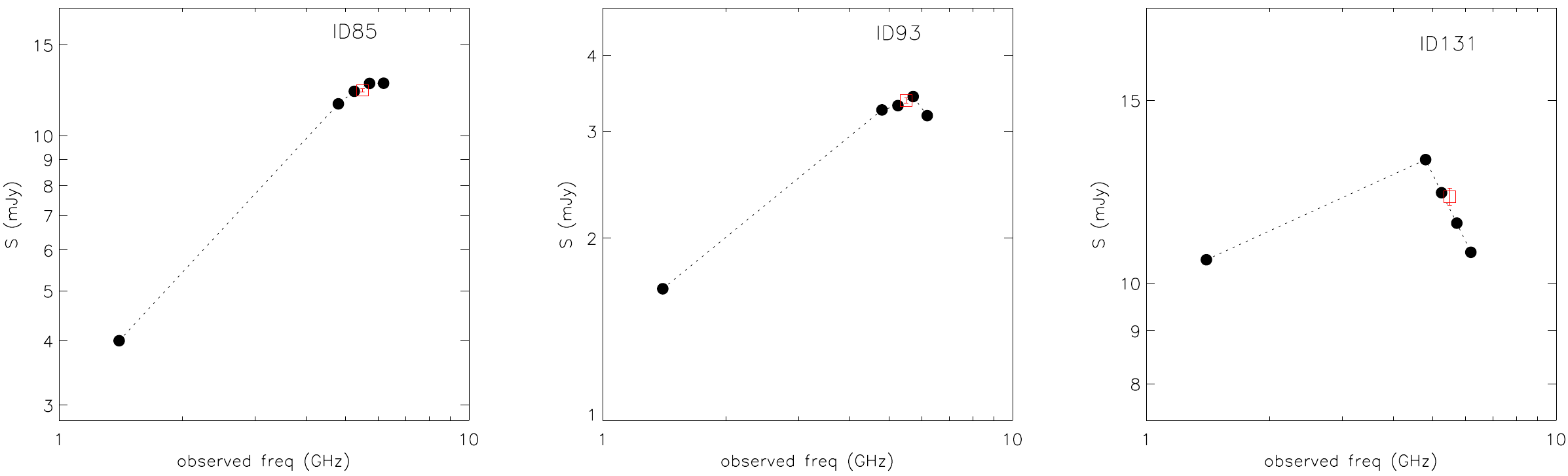}
\caption{The radio spectral energy distributions of the 3 strong GPS candidates. The red square marks the total $S_{5.5 GHz}$ from the full CABB band image and the other 6cm points are from the four CABB sub-band images. The 1.4 GHz point is from Miller et al. 2008.}
\label{fig:gpssed}
\end{figure*}

\begin{figure*}
\includegraphics[width=0.95\textwidth]{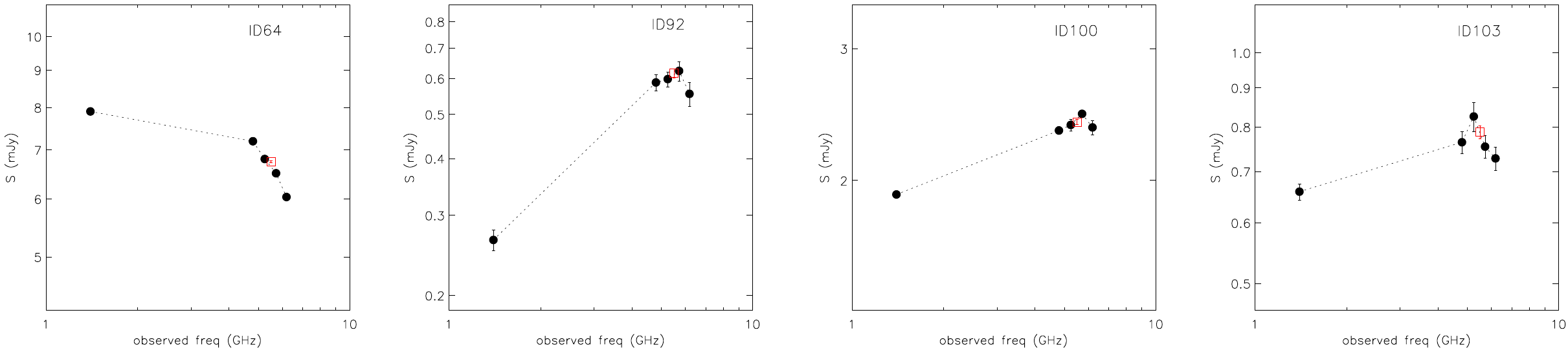}
\caption{The radio spectral energy distributions of the 4 weak GPS candidates. The red square marks the total $S_{5.5 GHz}$ from the full CABB band image and the other 6cm points are from the four CABB sub-band images. The 1.4 GHz point is from Miller et al. 2008.}
\label{fig:gpssed2}
\end{figure*}

\begin{table*}
\renewcommand{\thefootnote}{\alph{footnote}}
\begin{minipage}{\textwidth}
\caption{Summary of the properties of the GPS candidates. The 3 strong candidates are listed before the weak candidates. The radio luminosities have not been k-corrected.} 
\vspace{2mm}
\begin{tabular}{rcccccrrrr} \hline
ID & $S_{1.4 GHz} $ &  $S_{5.5 GHz} $ & $R$ mag &  phot $z$ & spec $z$ & angular-size & linear-size & $L_{1.4 GHz}$ & $L_{5.5 GHz}$ \\
 & (mJy)  &  (mJy) &  & & & (arcsec)& (kpc) & (W Hz$^{-1}$) & (W Hz$^{-1}$) \\ \hline

85 & 4.00 & 12.25 & 18.63\footnotemark[1] & 1.57 $\pm$ 0.09\footnotemark[1]  & 1.5542\footnotemark[2] & $0.7 \times 0.5$ & 6.0 $\times$ 4.3 & 2.48 $\times$ 10$^{25}$ & 7.61 $\times$ 10$^{25}$  \\
93 & 1.65 & 3.37 & 18.84\footnotemark[3] & 0.55 $\pm$ 0.01\footnotemark[3] & 0.5444\footnotemark[2] & $1.2 \times 0.9$ & 10.0 $\times$ 5.7 & 1.25 $\times$ 10$^{24}$ & 2.56 $\times$ 10$^{24}$  \\
131 & 39.5\footnotemark[4] & 12.10 & 22.01\footnotemark[3] & 1.75 $\pm$ 0.01\footnotemark[3] & -- & 2.5\footnotemark[5] & 21.4\footnotemark[5] & 3.06 $\times$ 10$^{26}$ & 9.38 $\times$ 10$^{25}$  \\ \hline

64 & 15.3\footnotemark[6] & 6.74 & 19.00\footnotemark[1] & 0.75 $\pm$ 0.15\footnotemark[1] & 0.7423\footnotemark[2] & $3.2 \times 1.3$ & 23.3 $\times$ 9.5 &  2.21 $\times$ 10$^{25}$ & 9.72 $\times$ 10$^{24}$  \\
92 & 0.265 & 0.616 & 21.63\footnotemark[3] & 0.73 $\pm$ 0.01\footnotemark[3] & 0.733\footnotemark[2] & $1.2 \times 0.6$ & 8.8 $\times$ 4.4 &  3.72 $\times$ 10$^{23}$ & 8.66 $\times$ 10$^{23}$  \\
100 & 1.91 & 2.39 & 22.30\footnotemark[3] & 0.81 $\pm$ 0.01\footnotemark[3] & -- & $0.8 \times 0.3$ & 6.0 $\times$ 2.3 &  3.29 $\times$ 10$^{24}$ & 4.12 $\times$ 10$^{24}$  \\
103 & 0.66 & 0.79 & 23.63\footnotemark[3] & 1.57 $\pm$ 0.01\footnotemark[3] & -- & $0.8 \times 0.4$ & 7.1 $\times$ 3.8 &  4.18 $\times$ 10$^{24}$ & 5.00 $\times$ 10$^{24}$  \\
\hline
\end{tabular}
\footnotetext{$^{\rm a}$ COMBO-17, Wolf et al. 2004, Vega-magnitude}
\footnotetext{$^{\rm b}$ ATLAS, Mao et al. 2012}
\footnotetext{$^{\rm c}$ MUSYC, Cardamone et al. 2010, AB-magnitude}
\footnotetext{$^{\rm d}$ revised using a double Gaussian fit to Miller et al. (2008) 1.4 GHz image}
\footnotetext{$^{\rm e}$ distance between lobes in Miller et al. (2008) 1.4 GHz image}
\footnotetext{$^{\rm f}$ revised using a free Gaussian fit to Miller et al. (2008) 1.4 GHz image}
\label{tab:gps}
\end{minipage}
\end{table*}

We combined the four sub-band images from our 6cm observations with the VLA 1.4 GHz observations to
identify GPS source candidates. Based on the spectral index properties alone, we identify 3 strong
GPS candidates, and another 4 weaker candidates where there is possibly a peak but the large errors
in the flux densities makes the existence of the peak less clear. 
The SEDs for these sources are shown in Figure \ref{fig:gpssed} and \ref{fig:gpssed2}.
ID64 is included in the weak category as the 1.4 GHz flux density is greater than the 6cm sub-band GHz flux densities, and there is a possibility that the SED may not peak at all in between 1.4 and 6.2 GHz.

Subsequent measures of source size show that all but two sources have compact ($<10$ kpc) radio morphology.
Of the two extended sources, ID131 is a classical FRII double with lobes separated by about 21 kpc
and ID64 is a FRI radio galaxy with lobes extending out to about 110 kpc. These two sources are rejected
as GPS candidates on the basis of their extent, and in the case of ID64 also on the spectral shape.

Optical counterparts are taken from the MUSYC and COMBO-17 data, and the properties of the
GPS candidates are summarised in Table \ref{tab:gps}.  The radio luminosities are in the observed frame,
i.e. not $k$-corrected, as the spectral index between 1.4 and 5.5 GHz is uncertain for these sources.

In summary, all our low flux-density GPS candidates are consistent with FRI or FRII radio sources, at redshifts about $z \sim 0.5$ and higher. They are dominated by those with physical extents around 5-10 kpc, well within the characteristic size of massive galaxies at these redshifts. As such, they are likely to be young jets yet to have pushed outside the stellar disks of their host galaxies. This is consistent with the properties of GPS sources identified at the brightest radio flux densities.

We discuss the individual GPS candidates in detail below. 

{\it ID85} 

GPS candidate ID85 is a strong ($\sim$ 12 mJy) point source at 5.5 GHz. It is only marginally resolved in the higher resolution 1.4 GHz data, with a deconvolved size of $0.7 \times 0.5$ arscec. The bright ($R_{\rm Vega} = 18.63$) optical counterpart appears to be a point source and has QSO colours. 
The COMBO-17 photometric redshift of $z = 1.57 \pm 0.09$ is consistent with the  spectroscopic redshift of $z = 1.5542$. At this redshift the deconvolved angular size implies this radio source is about 6 kpc in extent and has a radio luminosity of $L_{1.4 GHz}$  =  2.48 $\times$ 10$^{25}$  W Hz$^{-1}$. Hence this source has a radio power consistent with FRI radio galaxies, or low-powered FRIIs. 

{\it ID93} 

This $S_{5.5 GHz}$ = 3.4 mJy source is not resolved at 5.5 GHz and only marginally resolved at 1.4 GHz. The bright ($R _{\rm AB}= 18.84$) optical counterpart appears to be a point source and has QSO colours. 
The MUSYC photometric redshift of $z = 0.55 \pm 0.01$ is consistent with the  spectroscopic redshift of $z = 0.5444$. At this redshift the deconvolved angular size implies this radio source is about 10 kpc in extent. The radio luminosity of $L_{1.4 GHz}$  =  1.25 $\times$ 10$^{24}$  W Hz$^{-1}$ is consistent with a low luminosity FRI or a very powerful starburst. Higher resolution radio imaging is required to establish whether the extended emission shows a morphology characteristic of jets (most likely) or of diffuse star formation (less likely). If the radio emission was all associated with star formation, it would imply a star formation rate (SFR)  of $\sim$ 100 M$_\odot$ yr$^{-1}$, which is high for galaxies at this redshift, but not extreme.

{\it ID131}

This bright ($S_{5.5 GHz}$ = 12.10 mJy) source is extended at 5.5 GHz with a deconvolved size of 3.5 $\times$ 1.0 arcsec. An inspection of the 1.4 GHz image shows a compact double, so we manually refit the 1.4 GHz source as a double Gaussian, finding a total 1.4 GHz flux density of 39.5 mJy. The optical counterpart in MUSYC is a $R_{\rm AB}$ = 22.01 galaxy with a photometric redshift of $z = 1.75$ $\pm$ 0.01. The lobes in the 1.4 GHz image are $\sim$2.5 arcsec apart, which corresponds to a linear extent of 21 kpc. The radio luminosity of $L_{1.4 GHz}$  =  3.06$\times$ 10$^{26}$  W Hz$^{-1}$  is consistent with this source being a FRII, or high-powered FRI, radio galaxy. 

{\it ID64}

This bright ($S_{5.5 GHz}$ = 12.10 mJy) source is extended with a deconvolved size of 3.4 $\times$ 0.9 arcsec at 5.5 GHz. It is also extended at 1.4 GHz and shows an FRI-like morphology with faint lobes. This source is not easily fit as a multiple Gaussian. A manual fit to the 1.4 GHz core of the source yields a 15 mJy source with deconvolved size of 3.2 $\times$ 1.3 arcsec. The bright ($R_{\rm Vega} = 19.00$) optical counterpart appears to be a point source, has QSO colours, and a possible nearby neighbour 5 arcsec to the east. The COMBO-17 photometric redshift of $z = 0.75 \pm 0.15$ is consistent with the spectroscopic redshift of $z = 0.7386$. The optical source to the east does not have a reliable photometric redshift in either catalogue, possibly due to another nearby faint object to its east which could be contaminating the source's photometry, so it is unclear whether a merger is seen in the optical bands. At $z = 0.7386$ this radio source has a linear extent of about 23 kpc, but the faint FRI lobes are about 15 arcsec in distance or $\sim110$kpc apart. The radio luminosity of $L_{1.4 GHz}$  =  2.21$\times$ 10$^{25}$  W Hz$^{-1}$ is consistent with that of an FRI radio galaxy. 

{\it ID92}

This source, with a 5.5 GHz flux density of 0.6 mJy, is the faintest GPS candidate. It is a point source in the 5.5 GHz image but marginally resolved in the 1.4 GHz image with a deconvolved size of 1.2 $\times$ 0.6 arcsec. The source has an $R_{\rm AB}$ = 21.63 counterpart with a photometric redshift of 0.73 $\pm$ 0.01, consistent with the spectroscopic redshift of $z = 0.733$. At this redshift the deconvolved angular size implies this radio source is about 9 kpc in extent and has a radio luminosity of $L_{1.4 GHz}$  =  3.72 $\times$ 10$^{23}$  W Hz$^{-1}$. This radio power is indicative of a low-luminosity AGN or high star formation (SFR $\sim$ 40 M$_\odot$ yr$^{-1}$). 

{\it ID100}

This $S_{5.5 GHz}$ = 2.4 mJy source is not resolved at 5.5 GHz and only marginally resolved at 1.4 GHz. The $R _{\rm AB}= 22.30$ optical counterpart has a photometric redshift of $z = 0.81 \pm 0.01$ but there is no spectroscopic redshift. At this photometric redshift the radio source is about 6 kpc in extent. The radio luminosity of $L_{1.4 GHz}$  =  3.29 $\times$ 10$^{24}$  W Hz$^{-1}$ is consistent with an FRI  radio galaxy.

{\it ID103}

This source is relatively faint for a GPS candidate ($S_{5.5 GHz}$ = 0.79 mJy). It is not resolved at 5.5 GHz and only marginally resolved at 1.4 GHz. The faint $R _{\rm AB}= 23.63$ optical counterpart has a photometric redshift of $z = 1.57 \pm 0.01$ but there is no spectroscopic redshift. At this photometric redshift the radio source is about 7 kpc in extent. The radio luminosity of $L_{1.4 GHz}$  =  4.18 $\times$ 10$^{24}$  W Hz$^{-1}$ is consistent with an FRI  radio galaxy.

\end{appendix}

\label{lastpage}

\end{document}